\documentclass[prd,aps,floatfix,amsmath,amssymb,nofootinbib,twocolumn]{revtex4-1}

\usepackage{graphicx}
\usepackage{amsmath}
\usepackage{amssymb}
\usepackage{cancel}
\usepackage{mathrsfs}
\usepackage{bm}
\usepackage{booktabs}
\usepackage[colorlinks, bookmarks=true, citecolor=cyan, linkcolor=blue, urlcolor=cyan]{hyperref}
\usepackage{xcolor}
\usepackage{url}
\DeclareMathOperator{\real}{Re}
\newcommand{\R}{{\Bbb R}}
\newcommand{\x}{{x}}
\newcommand{\bx}{{\bf x}}

\newcommand{\new}[1]{{#1}}

\begin{document}

\title{Critical Phenomena in Gravitational Collapse}

\newcommand{\mybeginwidetext}{\begin{widetext}}
\newcommand{\myendwidetext}{\end{widetext}}

\author{Carsten Gundlach}
\affiliation{Mathematical Sciences, University of Southampton,
  Southampton SO17 1BJ, United Kingdom} 
\author{David Hilditch}
\affiliation{CENTRA, Departamento de F\'isica,
          Instituto Superior T\'ecnico - IST, 
          Universidade de Lisboa - UL,
          Avenida Rovisco Pais 1, 1049 Lisboa, Portugal} 
\author{Jos\'e M. Mart\'{\i}n-Garc\'{\i}a}
\affiliation{Wolfram Research, 
        100 Trade Center Drive,
        Champaign, IL 61820-7237, USA} 

\date{\today}

\begin{abstract}
  As first discovered by Choptuik, the black hole threshold in the
  space of initial data for general relativity shows both surprising
  structure and surprising simplicity. Universality, power-law scaling
  of the black hole mass, and scale echoing have given rise to the
  term ``critical phenomena''. They are explained by the existence of
  exact solutions which are attractors within the black hole
  threshold, that is, attractors of codimension one in phase space,
  and which are typically self-similar. Critical phenomena give a
  natural route from smooth initial data to arbitrarily large
  curvatures visible from infinity, and are therefore likely to be
  relevant for cosmic censorship, quantum gravity, astrophysics, and
  our general understanding of the dynamics of general relativity.
  \new{Major additions since the 2010 version of this review are
    numerical simulations beyond spherical symmetry, in particular of
    vacuum critical collapse, and new sections on mathematical results
    in PDE blowup (as a toy model for singularity formation) and on
    naked singularity formation in GR.}  \keywords{Gravitational
    collapse \and Critical phenomena \and Scaling laws \and Numerical
    relativity \and Self-similarity \and Black holes \and Naked
    singularities \and Stability \and Phase space}
\end{abstract}

\maketitle

\setcounter{tocdepth}{3}
\tableofcontents

\section{Introduction}

\subsection{Overview of the subject}

\new{Take a smooth one-parameter family of smooth and asymptotically
  flat initial data in general relativity, where some members of the
  family form a black hole in their time evolutions and others do
  not. Otherwise the family can be generic. Adjust the} parameter $p$
of the initial data to the threshold of black hole formation, and
compare the resulting spacetimes as a function of $p$. In many
situations, the following \emph{critical phenomena} are then observed:

\begin{itemize}

\item Near the threshold, black holes with arbitrarily small masses
  are created, and the black hole mass scales as
  \begin{equation}
    M \propto (p-p_*)^\gamma,
    \label{eq1}
  \end{equation}
  where $p$ parameterises the initial data and black holes form for
  $p>p_*$. 

\item The \emph{critical exponent} $\gamma$ is universal with respect to
  initial data, that is, independent of the particular 1-parameter
  family, although it depends on the type of collapsing matter.

\item In the region of large curvature before black hole formation,
  the spacetime approaches a self-similar solution which is also
  universal with respect to initial data, the \emph{critical solution}. 

\end{itemize}

Critical phenomena were discovered by \citet{Choptuik92} in numerical
simulations of a spherical scalar field. They have been found in
numerous other numerical and analytical studies in spherical symmetry,
and a few in axisymmetry; in particular critical phenomena have been
seen in the collapse of axisymmetric gravitational waves in vacuum,
\new{although this is still under very active investigation, see
  Sec.~\ref{section:axi_vacuum}. Scaling laws similar to the one for
  black hole mass have been calculated, and verified in numerical
  experiments, for black hole charge and angular momentum.}

It is still unclear how universal critical phenomena in collapse are
with respect to matter types and beyond spherical symmetry, in
particular for vacuum collapse. \new{However, progress beyond
  spherical symmetry is now being made with better numerical
  simulations.}

Critical phenomena can be usefully described in dynamical systems
terms. A critical solution is then characterised as an attracting
fixed point within a surface that divides two basins of attraction, a
\emph{critical surface} in phase space. Such a fixed point can be
either a stationary spacetime, or one that is scale-invariant and
self-similar. The latter is relevant for the (``type~II'') critical
phenomena sketched above. We shall also see how the dynamical systems
approach establishes a \new{mathematical} connection with critical
phenomena in statistical mechanics.

Therefore, we could define the field of critical phenomena in
gravitational collapse as the study of the boundaries among the basins
of attraction of different end states of self-gravitating systems,
such as black hole formation or dispersion. In our view, the
main physical motivation for this study is that those critical solutions
which are self-similar provide a way of achieving arbitrarily large
spacetime curvature outside a black hole, and in the limit a naked
singularity, by fine-tuning generic initial data for generic matter
to the black hole threshold. Those solutions are therefore likely
to be important for quantum gravity and cosmic censorship.

\subsection{Plan of this review}

This review focuses on numerical and theoretical work on
phenomena at the threshold of black hole formation in 3+1-dimensional
general relativity. We also include work in higher and lower
spacetime dimensions and non-gravity systems that may be relevant as
toy models for general relativity. We exclude work on critical
phenomena in other areas of physics.

The reader unfamiliar with the topic is advised to begin
with either Sections~\ref{section:universality}-\ref{section:scaling},
which give the key theory of universality, self-similarity and
scaling, or Sections~\ref{section:scalarfieldequations}
and~\ref{section:threshold}, which describe the prototype example, the
massless scalar field in spherical symmetry.

\new{

We have left Secs.~\ref{section:theory} and \ref{section:scalarfield}
relatively unchanged from their 2007 form. We have also maintained
the separation of numerical work in and beyond spherical symmetry, see
Secs.~\ref{section:spherical} and \ref{section:nonspherical}
respectively.

The most exciting developments since the 2007 version of this review
article are, in our opinion, first, the improvement of numerical
simulations of vacuum critical collapse and, second, the first
rigorous mathematical results directly relevant to critical collapse
as a route to naked singularities.

The 2007 version included a small number of numerical studies of
self-similar blowup in nonlinear PDEs, which can serve as toy
models for self-similar naked singularity formation in general
relativity. We have regrouped this and subsequent numerical work on
PDE blowup, together with recent rigorous mathematical results on the
same PDEs, in a new section, Sec.~\ref{section:PDEblowup}.

A second new section, Sec.~\ref{section:mathematicalGR} groups
together rigorous mathematical results on critical collapse and naked
singularity formation in general relativity, including also some
papers on blowup of a perfect fluid coupled to Newtonian gravity.

Fore-runners of this Living Review are its 1999 \citep{Gun99} and 2007
\citep{GunMar07} versions, and our earlier reviews
\citep{Gundlach_critreview_Banach98},
\citep{Gundlach_critreview_ATMP98} \citep{MartinGundlach01} and
\citep{Gundlach_critreview_PhysRep03}.  Review papers by others
include \citet{Horne_MOG}, \citet{Bizon}, \citet{Choptuik_review},
\citet{Choptuik_review2} and \citet{BradyCai}.}
\citet{EggersFontelos} review the role of self-similarity in the
formation of singularities in evolutionary PDEs in general.

\subsection{Notation}

\new{We have changed original notation where that was necessary in
order to keep the notation within this review paper consistent. In the
following, $a,b,c...$ are abstract tensor indices, while $\mu,\nu...$
and $i,j...$ are coordinate indices. We use gravitational units, where
$c=G=1$ (with a few exceptions where $G$ is retained), but
$\hbar$ is not set to one. (In these units black hole mass and charge
have dimensions of length, but particle mass and charge parameters in
the field equations have dimensions of inverse length.) $a:=b$ denotes
$a$ being defined in terms of $b$.

$\Phi$ is a diffeomorphism, while $\phi$ is a scalar (matter) field,
with reduction variable~$\varPhi=\phi_{, r}$, and $\varphi$ is an
angle. $p$ is the parameter of a one-parameter family of initial data,
while $P$ is the fluid pressure, or in other contexts a function of
the initial data. $\gamma$ is a critical exponent, while $\Gamma$ is
the parameter of the ``Gamma-law'' fluid equation of state, the
central world line in spherical symmetry, or (with indices on) a
Christoffel symbol. $k$ is the parameter in the linear fluid equation
of state $P=k\rho$. $x^\mu$ are any spacetime coordinates but $x$ is a
similarity coordinate such as $r/(t_*-t)$. $t$ and $r$ are time and
radius, while $\tau$ is a similarity coordinate such as
$-\ln(t_*-t)$. $L$ stands for a length scale, but also angular
momentum, while $l$ stands for a ratio of length scales, such as
$l_1:=L_1/L_0$. $M$ is the mass of a black hole, while $m$ stands for
the Kodama mass function, or in other contexts, the mass parameter of
a massive scalar field.}

\section{Theory}
\label{section:theory}

In this section we describe the basic theory underlying critical
collapse of the type that forms arbitrarily small black holes (called
type~II, see also Sect.~\ref{section:typeI}). We begin with the
mathematical origin of its three main characteristics, which were
already summarised in the introduction:
\begin{itemize}

\item universality with respect to initial data;

\item scale-invariance of the critical solution;

\item \new{power-law scaling of the black hole mass, charge, angular
    momentum, and the global maximum of the curvature.}

\end{itemize}

\subsection{Universality}
\label{section:universality}

Consider general relativity (from now on, also abbreviated as GR) as
an infinite-dimensional continuous dynamical system. Points in the
phase space are initial data sets (3-metric, extrinsic curvature, and
suitable matter variables, which together obey the Einstein
constraints). We evolve with the Einstein equations in a \new{suitably
fixed} gauge (see Sect.~\ref{section:coordinates}). Solution curves of
the dynamical system are spacetimes obeying the Einstein-matter
equations, sliced by specific Cauchy surfaces of constant time $t$.

An isolated system in GR can end up in qualitatively different stable
end states. Two possibilities are the formation of a single black hole
in collapse, or complete dispersion of the mass-energy to
infinity. \new{In the simple case of a massless scalar field in
  spherical symmetry}, these are the only possible end states (see
Sect.~\ref{section:scalarfield}). Any point in phase space can then be
classified as ending up in one or the other type of end state. The
entire phase space therefore splits into two halves, separated by a
``critical surface''.

A phase space trajectory that starts on a critical surface by
definition never leaves it. A critical surface is therefore a
dynamical system in its own right, with one dimension fewer than the
full system. If it has an attracting fixed point, such a point is
called a critical point. It is an attractor of codimension one in the
full system, and the critical surface is its attracting manifold.
\new{This is also visible in its linear perturbations: in a
finite-dimensional dynamical system, it would have exactly one
unstable linear perturbation mode. Making a mode ansatz for the
infinite-dimensional dynamical system that is general relativity, a
single unstable mode is also found.}

\begin{figure}[htbp]
    \centerline{\includegraphics[width=\linewidth]{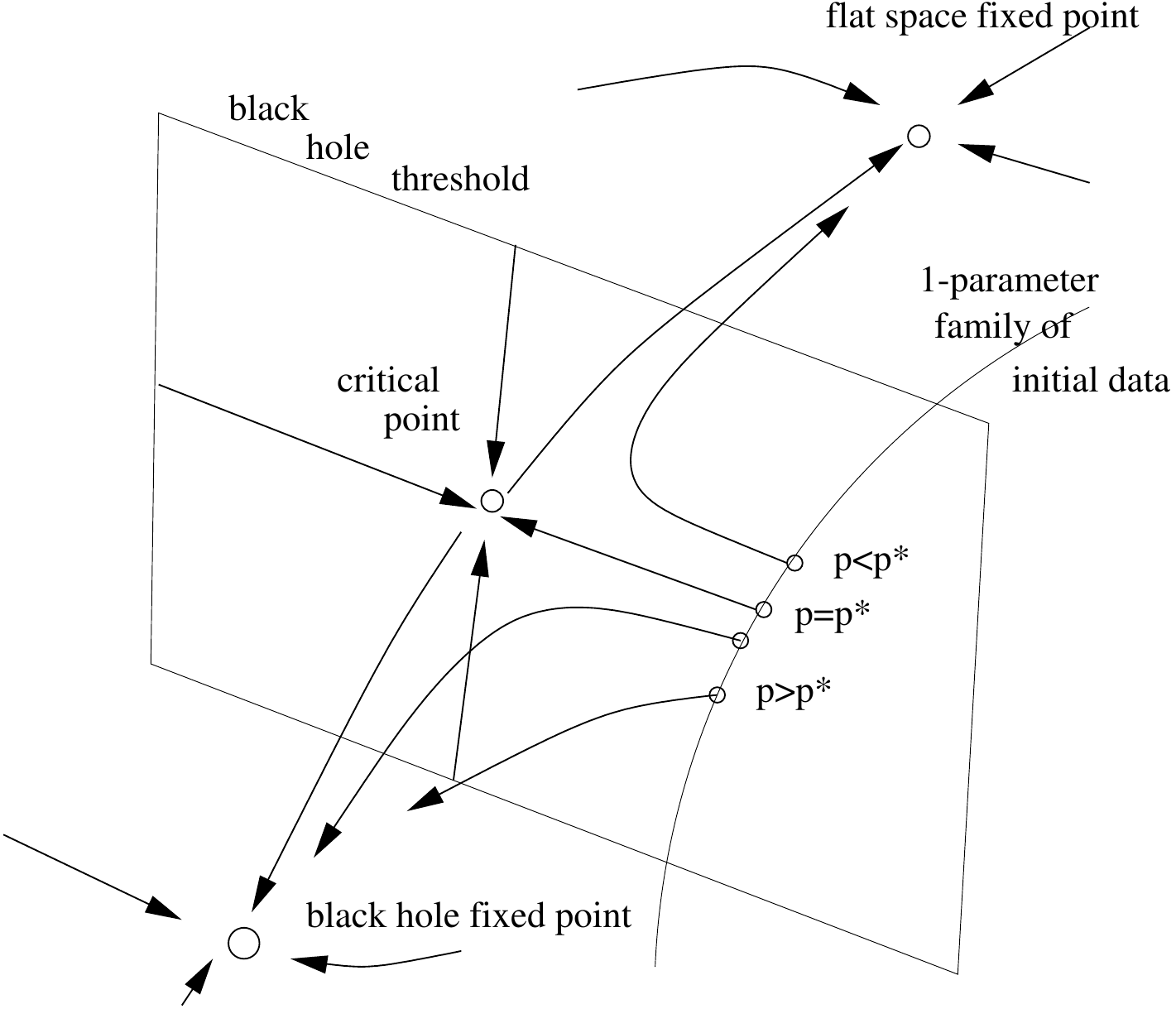}}
    \caption{The phase space picture for the black hole threshold
      in the presence of a critical point. Every point corresponds to
      an initial data set, that is, a 3-metric, extrinsic curvature,
      and matter fields. (In type~II critical collapse these are only
      up to scale). The arrow lines are solution curves, corresponding
      to spacetimes, but the critical solution, which is stationary
      (type~I) or self-similar (type~II) is represented by a
      point. The line without an arrow is not a time evolution, but a
      1-parameter family of initial data that crosses the black hole
      threshold  at $p=p_*$. The 2-dimensional plane represents an
      $(\infty-1)$-dimensional hypersurface, but the third dimension
      represents really only one dimension.}
    \label{figure:dynsim}
  \end{figure}

\begin{figure}[htbp]
    \centerline{\includegraphics[width=\linewidth]{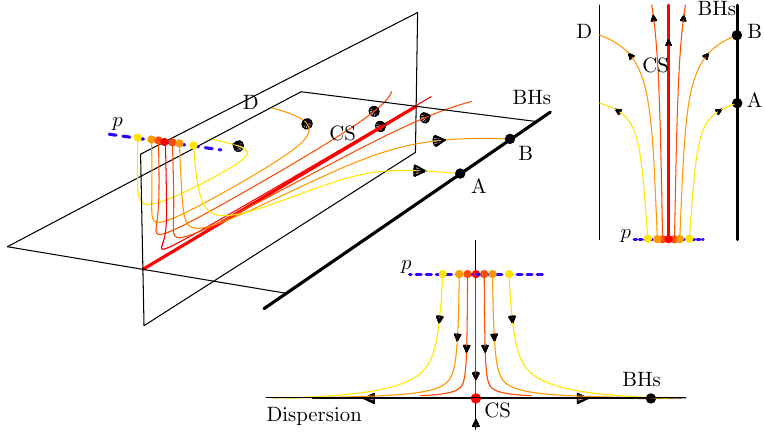}}
    \caption{A different phase space picture, specifically for
      type~II critical collapse, and two 2-dimensional projections of
      the same picture. In contrast with Fig.~\ref{figure:dynsim}, one
      dimension of the two representing the (infinitely many) decaying
      modes has been suppressed. The additional axis now represents a
      global scale which was suppressed in Fig.~\ref{figure:dynsim}, so
      that the scale-invariant critical solution CS is now represented
      as a straight line (in red). Several members of a family of
      initial conditions (in blue) are attracted by the critical
      solution and then depart from it towards black hole formation (A
      or B) or dispersion (D). Perfectly fine-tuned initial data
      asymptote to the critical solution with decreasing
      scale. Initial conditions starting closer to perfect fine tuning
      produce smaller black holes, such that the parameter along the
      line of black hole end states is $-\ln M_{\mathrm{BH}}$. Two
      2-dimensional projections of the same picture are also
      given. The horizontal projection of this figure is the same as
      the vertical projection of Fig.~\ref{figure:dynsim}.}
    \label{figure:phasespace3d}
  \end{figure}

As illustrated in Figs.~\ref{figure:dynsim} and \ref{figure:phasespace3d},
any trajectory beginning near the critical surface, but not
necessarily near the critical point, moves almost parallel to the
critical surface towards the critical point. Near the critical point
the evolution slows down, and eventually moves away from the critical
point in the direction of the growing mode. This is the origin of
universality. All details of the initial data have been forgotten,
except for the distance from the black hole threshold. The closer the
initial phase point is to the critical surface, the more the solution
curve approaches the critical point, and the longer it will remain
close to it.

We should stress that this phase picture is extremely
simplified. Some of the problems associated with this simplification
are discussed in Sect.~\ref{section:coordinates}.


\subsection{Self-similarity}
\label{section:self_similarity}

Fixed points of dynamical systems often have additional symmetries. In
the case of type~II critical phenomena, the critical point is a
spacetime that is self-similar, or scale-invariant. 
The self-similarity can be discrete or continuous. 

\new{

We define a {\bf discretely self-similar (DSS) spacetime} $(M,g,\Phi)$
to be a Lorentzian manifold $(M,g)$ with an invertible smooth map
$\Phi:M\to M$ that obeys \citep{Gundlach_Chop1}
\begin{equation}
\label{DSSdef}
\Phi_*g_{ab}=e^{-2\Delta}g_{ab},
\end{equation}
where $\Phi_*$ is the push-forward under $\Phi$, and $\Delta>0$ a
constant, the log-scale period, or scale-echoing period.  We define a
{\bf DSS-adapted vector field} on a DSS spacetime to be a smooth
vector field $\xi$, such that, along any of its integral curves
$c(\tau)$,
\begin{equation}
\Phi\left(c(\tau)\right)=c(\tau+\Delta).
\end{equation}
Given a DSS-adapted vector field and its integral curves, we can
define {\bf DSS-adapted coordinates} $x^\mu:=(\tau,x^i)$ such that
\begin{equation}
\xi:={\partial\over\partial\tau}.
\end{equation}
In these coordinates, the metric is given by
\begin{equation}
\label{gscaling}
g_{\mu\nu}=e^{-2\tau}\tilde g_{\mu\nu},
\end{equation}
where the rescaled metric coefficients $\tilde g_{\mu\nu}(\tau,x^i)$
are periodic in $\tau$ with period $\Delta$.  It follows from
(\ref{gscaling}) that
\begin{equation}
R_{abcd}R^{abcd}=e^{4\tau}\tilde K,
\end{equation}
where again $\tilde K$ is periodic in $\tau$, and so, unless the
spacetime is flat, $\tau=\infty$ is a curvature singularity.  The
distance (proper distance, proper time, or null affine parameter) to
the singularity, measured along a line of constant $x^i$, is
\begin{equation}
s= e^{-\tau} \tilde S
\end{equation}
where $\tilde{S}$ is periodic [with $S>0$ and
$\tilde{S}_{,\tau}<\tilde{S}$, so that $s_{,\tau}<0$]. Similarly, the
distance between any two spacetime points shrinks by a factor
$e^{-\Delta}$ under an application of $\Phi$.  In this naive sense,
the curvature singularity is a point. If the spacetime exists
everywhere to the past of the future lightcone of this point and this
lightcone extends to infinity, we call the singularity globally naked,
in the sense that it is not hidden inside an event horizon. Its future
lightcone is a Cauchy horizon, beyond which the spacetime does not
have a unique continuation.

DSS-adapted coordinates are not unique, but one such coordinate system
is related to any other one by smooth coordinate transformations of
the form
\begin{subequations}
\begin{align}
  \label{xtauchange}
  x^i&=x^i(\hat\tau,\hat x^j), \\
  \label{tauchange}
  \tau&=\hat\tau+h(\hat\tau,\hat x^j),
\end{align}
\end{subequations}
where the functions $\x^i$ and $h$ are periodic in $\hat\tau$ with
period $\Delta$. Hence the most general DSS-adapted vector field takes
the form
\begin{equation}
\label{generalxi}
\xi:=\partial_{\hat\tau}=(1+h_{,\hat\tau})\partial_\tau+x^i_{,\hat\tau}\partial_i.
=:(1+T)\partial_\tau+X^i\partial_i,
\end{equation}
where $T$ and $X^i$ are periodic functions of $(\tau,x^i)$ with period
$\Delta$ and {\em zero average} over one period. One can further
partition the similarity coordinates $x^i$ as $(x,\theta^A)$ and use
this residual gauge freedom to fix the past light cone of the
singularity at some $x=x_p$. If the singularity admits a future
light-cone, one can also fix this at some $x=x_f$. Finally, if the
spacetime is spherically symmetric, or otherwise has a preferred
worldline running into the singularity, one can independently fix this
central worldline at $x=x_c$.

We define a {\bf continuously self-similar (CSS)
  spacetime} $(M,g,\xi)$ to be a Lorentzian manifold $(M,g)$ with a
smooth vector field $\xi$ such that
\begin{equation}
\label{homothetic}
{\cal L}_\xi g=-2g,
\end{equation}
where ${\cal L}_\xi$ denotes the Lie derivative along $\xi$
\citep{CahillTaub}.  Such a vector field and spacetime are also called
homothetic. CSS is a special case of DSS, where the vector field $\xi$
is unique (up to the addition of constant multiples of any Killing
vectors the spacetime may have), the metric coefficients
$\tilde g_{\mu\nu}$ are independent of $\tau$ altogether, and so are
the functions $x^i$ and $h$ that parameterise residual coordinate
transformations.

}

Using the gauge freedom of general relativity, the lapse and shift in
the ADM formalism can be chosen (non-uniquely) so that the coordinates
become adapted coordinates if and when the solution becomes
self-similar (see Sect.~\ref{section:coordinates}). $\tau$ is then
both a time coordinate (in the usual sense that surfaces of constant
time are Cauchy surfaces), and \new{minus} the logarithm of overall
scale at constant $x^i$. The time parameter used in
Fig.~\ref{figure:phasespace3d} is of this type.

\subsection{Mass scaling}
\label{section:scaling}

Let $Z$ stand for a set of scale-invariant variables of the problem,
such as $\tilde g_{\mu\nu}$ and suitably rescaled matter variables. If
the dynamics is scale-invariant (this is the case exactly for example
for the scalar field, and approximately for other systems, see
Sect.~\ref{section:universalityclasses}), then $Z(x)$ is an element of
the phase space factored by overall scale, and $Z(x,\tau)$ a
solution. Note that $Z(x)$ is an initial data set for GR only up to
scale. The overall scale is supplied by \new{$e^{-\tau}$}.

For simplicity, assume that the critical solution is CSS. It can then
be written as $Z(x,\tau)=Z_*(x)$. One can then make a mode ansatz, in
which the linear perturbation depends on $\tau$
exponentially. Assuming there is a countable set of such modes and no
continuous spectrum, we arrive at the linear approximation
\begin{equation}
  \label{perturbations}
  Z(x,\tau) \simeq
  Z_*(x) + \sum_{i=1}^\infty C_i(p) \, e^{\lambda_i \tau} Z_i(x).
\end{equation}
\new{(Any self-similar solution admits an unstable perturbation with
$\lambda=1$ that corresponds to an infinitesimal time shift of the
background solution, and so is unphysical.)}  The perturbation
amplitudes $C_i$ depend on the initial data, and hence on $p$. As
$Z_*$ is a critical solution, by definition there is exactly one
$\lambda_i$ with positive real part (in fact it is purely real), say
$\lambda_0$. As $\tau\to\infty$, all other perturbations vanish. In
the following we consider this limit, and retain only the one growing
perturbation.

From our phase space picture, the evolution ends at the critical
solution for $p=p_*$, so we must
have $C_0(p_*)=0$. Linearising in $p$ around $p_*$, we obtain
\begin{equation}
  \label{echoing_region}
  Z(x,\tau) \simeq Z_*(x) + {dC_0\over dp} (p-p_*)
  e^{\lambda_0\tau} Z_0(x).
\end{equation}

For $p\ne p_*$, but close to it, the solution has the approximate form
(\ref{echoing_region}) over a range of $\tau$. Now we extract Cauchy
data at one particular $p$-dependent value of $\tau$ within that
range, namely $\tau_*$ defined by
\begin{equation}
  {dC_0\over dp} (p-p_*) \new{e^{\lambda_0\tau_*}} \equiv {\new \pm\epsilon},
  \label{taustardef}
\end{equation}
where \new{$0<\epsilon\ll 1$ is a constant} such that at this $\tau$ the
linear approximation is still valid. At sufficiently large $\tau$, the
linear perturbation has grown so much that the linear approximation
breaks down, and for $C_0(p)>0$ a black hole forms while for $C_0(p)<0$ the
solution disperses. The crucial point is that we need not follow this
evolution in detail, nor does the precise value of $\epsilon$
matter. It is sufficient to note that the Cauchy data at $\tau=\tau_*$
are 
\begin{equation}
  \label{pdata}
  Z(x,\tau_*) \simeq Z_*(x) \new{\pm\epsilon} Z_0(x).
\end{equation}
Due to the funnelling effect of the critical solution, the data at
$\tau_*$ is always the same, except for \new{the sign denoted above by
$\pm$ and} an overall scale, which is \new{not defined by $Z$ but is}
given by $e^{-\tau_*}$. For example, the physical spacetime metric,
with dimension $(\mathrm{length})^2$ is given by
$g_{\mu\nu}=e^{-2\tau} \tilde g_{\mu\nu}$, and similar scalings hold
for the matter variables according to their dimension. In particular,
as $e^{-\tau_*}$ is the only length scale in the initial data~(\ref{pdata}),
the mass of the final black hole must be proportional to that
scale. \new{(See also \citet{CaoCaiYan16} for scalings of this type for
intersections of the apparent horizon with a specific time slicing.)}
Therefore
\begin{equation}
  \label{massscaling}
  M \propto e^{-\tau_*} \propto (p-p_*)^{1 / \lambda_0},
\end{equation}
and, comparing with Equation~(\ref{eq1}), we have found the critical exponent
$\gamma = 1/\lambda_0$.

\new{A similar dimensional analysis applies to all dimensionful geometric
constants, for example the maximum of the Ricci scalar inn
dispersing solutions, which has dimension
(length)${}^{-2}$ \citep{GarfinkleDuncan}:
\begin{equation}
\label{curvaturescaling}
R_\text{max} \propto e^{2\tau_*} \propto (p_*-p)^{-2/ \lambda_0}.
\end{equation}
}

When the critical solution is DSS, a periodic fine structure of small
amplitude \new{(colloquially also called a ``wiggle'')} is
superimposed on this basic power law \citep{Gundlach_Chop2,
  HodPiran_wiggle,GunBauHil24}: \new{
\begin{align}
  \ln M &= A + \gamma \ln (p-p_*) + f_M\left(\gamma \ln (p-p_*) +
  B\right),\label{fM} \\
  -{1\over 2}\ln R_\text{max} &= A + \gamma \ln (p_*-p) 
  + f_R\left(\gamma \ln (p_*-p) + B\right) \label{eqn:wiggles_R}
\end{align}
where $f_M$ and $f_R$ have period $\Delta$, and $A$ and $B$ depend on
the one-parameter family of initial data. $f_R$ is universal,
gauge-independent, and continuous with discontinuous derivative. If $M$
is interpreted as the mass of the first marginally outer-trapped
surface (from now on, MOTS) to be found, it is clearly
slicing-dependent and could be discontinuous\footnote{\new{We reserve
    the term ``apparent horizon'' for the world tube (normally
    spacelike and foliated by MOTSs) that bounds the trapped
    region}}. If $M$ is defined \new{as the mass at the intersection
  of the event horizon with future null infinity}, it becomes
slicing-independent, but the critical scaling, which is essentially
local, may be covered up by later infall. However, where these
distinctions are not important we may just refer to the ``black hole
mass''.} As the critical solution is periodic in $\tau$ with period
$\Delta$, the number $N$ of scaling ``echos'' \new{before black hole
  formation or dispersion} is approximated by
\begin{equation}
  N \simeq \Delta^{-1} \gamma \ln|p-p_*| + \mathrm{constant}.
\end{equation}
Note that this holds for both supercritical and subcritical solutions.

\subsection{Type~I}
\label{section:typeI}

In type~I critical phenomena, the same phase space picture as in
Fig.~\ref{figure:dynsim} and Sect.~\ref{section:universality} applies,
but the critical solution is now stationary or time-periodic instead
of self-similar or scale-periodic. It also has a finite mass and can
be thought of as a metastable star. (Type~I and~II are so named
after first and second order phase transitions in statistical
mechanics, in which the order parameter is discontinuous and
continuous, respectively.)  Universality in this context implies that
the black hole mass near the threshold is independent of the initial
data, namely a certain fraction of the mass of the stationary critical
solution. The dimensionful quantity that scales is not the black hole
mass, but the lifetime $t_p$ of the intermediate state where the
solution is approximated by the critical solution. This is clearly
\begin{equation}
  \label{typeIscaling}
  t_p = - {1\over \lambda_0} \ln|p-p_*| + \mathrm{constant}.
\end{equation}

Type~I critical phenomena occur when a mass scale in the field
equations becomes dynamically relevant. (This scale does not
necessarily set the mass of the critical solution absolutely: There
could be a family of critical solutions selected by the initial
conditions.) Conversely, as the type~II power law is scale-invariant,
type~II phenomena occur in situations where either the field equations
do not contain a scale, or this scale is dynamically irrelevant. Many
systems, such as the massive scalar field, show both type~I and
type~II critical phenomena, in different regions of the space of
initial data \citep{BradyChambersGoncalves}.

\subsection{Coordinate choices for the dynamical systems picture}
\label{section:coordinates}

The time evolution of Cauchy data in GR can only be considered as a
dynamical system if the ADM evolution equations are complemented by a
prescription for the lapse and shift. To realise the phase space
picture of Figs.~\ref{figure:dynsim} and \ref{figure:phasespace3d} and
Sect.~\ref{section:universality}, the critical solution must be a
fixed point (CSS) or limit cycle (DSS)
\new{in these coordinates}. We have seen how coordinates adapted to an
exact self-similarity can be constructed, but is there a prescription
of the lapse and shift for \emph{arbitrary} initial data, such that,
given initial data for the critical solution, the resulting time
evolution actively drives the metric to a form~(\ref{gscaling}) that
explicitly displays the self-similarity?

\citet{GarfinkleGundlach} have suggested several
combinations of lapse and shift conditions that leave CSS spacetimes
invariant and turn the Choptuik DSS spacetime into a limit cycle
(see \citet{GarfinkleMeyer,Garfinkle2} for partial successes). Among
these, the combination of maximal slicing
with minimal strain shift has been suggested in a different context
but for related reasons \citep{SmarrYork}. Maximal slicing requires the
initial data slice to be maximal ($K_a{}^a=0$), but other
prescriptions, such as freezing the trace of $K$ together with minimal
distortion, allow for an arbitrary initial slice with arbitrary
spatial coordinates.

All these coordinate conditions are elliptic equations that require
boundary conditions, and will turn CSS spacetimes into fixed points
(or DSS into limit cycles) only given correct boundary
conditions. Roughly speaking, these boundary conditions require a
guess of how far the slice is from the accumulation point $t=t_*$, and
answers to this problem only exist in spherical symmetry. Appropriate
boundary conditions are also needed if the dynamical system is
extended to include the lapse and shift as evolved variables, turning
the elliptic equations for the lapse and shift into hyperbolic or
parabolic equations.

Turning a CSS or stationary spacetime into a fixed point of the
dynamical system also requires an appropriate choice of the phase
space variables $Z(x^i)$. To capture CSS (or DSS) solutions, one needs
scale-invariant variables. Essentially, these can be constructed by
dimensional analysis. The coordinates $x^i$ and $\tau$ are
dimensionless, \new{
but when dimensions are put back all occurrences of $e^{-\tau}$ become
$Le^{-\tau}$ where $L$ is an {\em arbitrary} fixed length scale. In
particular $g_{\mu\nu}$ in similarity coordinates then has dimension
$L^2$.} The scaling for the ADM and any matter variables follows.

Even with a prescription for the lapse and shift in place, a given
spacetime does not correspond to a unique trajectory in phase
space. Rather, for each initial slice through the same spacetime one
obtains a different slicing of the entire spacetime. A possibility for
avoiding this ambiguity would be to restrict the phase space further,
for example by restricting possible data sets to maximal or constant
extrinsic curvature slices.

Another open problem is that in order to talk about attractors and
repellers on the phase space we need to define a norm on a suitable
function space which includes both asymptotically flat data and data for
the exact critical solution. The norm itself must favour the central region
and ignore what is further out (and asymptotically flat) if all black
holes of the same mass are to be considered as the same
end state. 

\subsection{Approximate self-similarity and universality classes}
\label{section:universalityclasses}

The field equations for the massless scalar field coupled to the
Einstein equations are scale-free. Realistic matter models introduce
length scales, and the field equations then do not allow for
exactly self-similar solutions. They may however admit solutions which
are CSS or DSS asymptotically on small spacetime scales as the
dimensionful parameters become irrelevant, including type~II critical
solutions
\citep{Choptuik94,BradyChambersGoncalves,ChoptuikChmajBizon}.

\new{
As the prototype example, consider the massive scalar field. Its
stress energy tensor contains the terms $\nabla_a\phi\nabla_b\phi$ and
$m^2\phi^2$. (Note that $m$ has dimension of inverse length in
geometric units $c=G=1$). In an (approximately) self-similar solution,
which echos on a length scale $\sim e^{-\tau}$, we have
$\nabla_a\phi\sim e^{\tau}$ but $\phi\sim 1$. One can make a formal
expansion in the small parameter $me^{-\tau}$.} The zeroth order of
the expansion is the self-similar critical solution of the system with
$m=0$. A similar ansatz can be made for the linear perturbations of
the resulting background. The zeroth order of the background expansion
determines $\Delta$ exactly and independently of $m$, and the
zeroth order term of the linear perturbation expansion determines the
critical exponent $1/\lambda_0$ exactly, so that there is no need in
practice to calculate any higher orders in $m$ to make predictions for
type~II critical phenomena where they occur. (With $m\ne 0$, the basin
of attraction of the type~II critical solution will depend on $m$, and
type~I critical phenomena may also occur; see
Sect.~\ref{section:typeI}.) A priori, there could also be more than
one type~II critical solution for $m=0$, although this has not been
observed.)

This procedure has been carried out for the Einstein--Yang--Mills
system \citep{Gundlach_EYM} and for massless scalar electrodynamics
\citep{GundlachMartin}. Both systems have a single length scale $1/e$
(in geometric units $c=G=1$), where $e$ is the gauge coupling
constant. All values of $e$ can be said to form one universality class
of field equations \citep{HaraKoikeAdachi} represented by $e=0$,
\new{see also Sec.~\ref{section:scalarcharge}.} This notion of
universality classes is fundamentally the same as in statistical
mechanics. Any scalar field potential $V(\phi)$ becomes dynamically
irrelevant compared to the kinetic energy $|\nabla\phi|^2$ in a
self-similar solution \citep{Choptuik94}, so that all scalar fields
with potentials are in the universality class of the free massless
scalar field. Surprisingly, even two different models like the SU(2)
Yang--Mills and SU(2) Skyrme models in spherical symmetry are members
of the same universality class \citep{BizonChmajTabor}. Other examples
include modifications to the perfect fluid equation of state (EOS)
that do not affect the limit of high density \citep{NeilsenChoptuik}.

If there are several length scales $L_0$, $L_1$, $L_2$ etc.\ present
in the \new{field equations, a possible approach is to make a single
  formal expansion in $L_0^{-1}e^{-\tau}$, using one of these scales},
and define the dimensionless constants $l_i=L_i/L_0$ from the
others. The scope of the universality classes depends on where the
$l_i$ appear in the field equations. If a particular $L_i$ appears in
the field equations only in \new{negative} powers, the corresponding
$l_i$ appears multiplied by $e^{-\tau}$, and will be irrelevant in the
scaling limit. All values of this $l_i$ therefore belong to the same
universality class. From the example above, adding a quartic
self-interaction $\lambda\phi^4$ to the massive scalar field gives
rise to the dimensionless number \new{$\lambda^{-1} m^2$} but its
value is an irrelevant parameter (in the language of renormalisation
group theory).

\new{\citet{GundlachMartin} conjecture that} massive
scalar electrodynamics, for any values of $e$ and $m$, is in the
universality class of the massless uncharged scalar field in a region
of phase space where type~II critical phenomena occur. Examples of
dimensionless parameters which do change the universality class are
\new{$k$ in the perfect fluid equation $P=k\rho$}, the $\kappa$ of the
2-dimensional sigma model or, probably, a conformal coupling of the
scalar field \citep{Choptuik91} (the numerical evidence is weak but a
dependence should be expected).

\subsection{The analogy with critical phase transitions}
\label{section:analogy}

Some basic aspects of critical phenomena in gravitational collapse,
such as fine-tuning, universality, scale-invariant physics, and
critical exponents for dimensionful quantities, can also be identified
in critical phase transitions in statistical mechanics
(see \citet{Yeomans} for an introductory textbook).

From an abstract point of view, the objective of statistical mechanics
is to derive relations between macroscopic observables $A$ of the
system and macroscopic external forces $f$ acting on it, by
considering ensembles of microscopic states of the system. The
expectation values $\langle A\rangle$ can be generated as partial
derivatives of the partition function
\begin{equation}
  Z(\mu, f) = \!\!\!\!\!\!\! \sum_\mathrm{microstates}
  \!\!\!\!\!\!\! e^{- H(\mathrm{microstate}, \mu, f)}
\end{equation}
Here the $\mu$ are parameters of the Hamiltonian such as the strength
of intermolecular forces, and $f$ are macroscopic quantities which
are being controlled, such as the temperature or magnetic field.

Phase transitions in thermodynamics are thresholds in the space of
external forces $f$ at which the macroscopic observables $A$, or one
of their derivatives, change discontinuously. We consider two
examples: the liquid-gas transition in a fluid, and the ferromagnetic
phase transition.

The liquid-gas phase transition in a fluid occurs at the boiling curve
$p=p_{\mathrm{b}}(T)$. In crossing this curve, the fluid density changes
discontinuously. However, with increasing temperature, the difference
between the liquid and gas density on the boiling curve decreases, and
at the critical point $(p_*=p_{\mathrm{b}}(T_*),T_*)$ it vanishes as
a non-integer power:
\begin{equation}
  \rho_\mathrm{liquid} - \rho_\mathrm{gas} \sim (T_*-T)^\gamma.
\end{equation}

At the critical point an otherwise clear fluid becomes opaque, due to
density fluctuations appearing on all scales up to scales much larger
than the underlying atomic scale, and including the wavelength of
light. This indicates that the fluid near its critical point is
approximately scale-invariant (for some range of scales between the
size of molecules and the size of the container). 

In a ferromagnetic material at high temperatures, the magnetisation
$\bm{m}$ of the material (alignment of atomic spins) is determined by
the external magnetic field $\bm{B}$. At low temperatures, the material
shows a spontaneous magnetisation even at zero external field. In the
absence of an external field this breaks rotational symmetry: The
system makes a random choice of direction. With increasing
temperature, the spontaneous magnetisation $\bm{m}$ decreases and
vanishes at the Curie temperature $T_*$ as
\begin{equation}
  |\bm{m}|\sim (T_*-T)^\gamma.
\end{equation}
Again, the correlation length, or length scale of a typical
fluctuation, diverges at the critical point, indicating
scale-invariant physics. 

Quantities such as $|\bm{m}|$ or
$\rho_\mathrm{liquid}-\rho_\mathrm{gas}$ are called order
parameters. In statistical mechanics, one distinguishes between
first-order phase transitions, where the order parameter changes
discontinuously, and second-order, or critical, ones, where it goes to
zero continuously. One should think of a critical phase transition as
the critical point where a line of first-order phase transitions ends
as the order parameter vanishes. This is already clear in the fluid
example. In the ferromagnet example, at first one seems to have only
the one parameter $T$ to adjust. But in the presence of a very weak
external field, the spontaneous magnetisation aligns itself with the
external field $\bm{B}$, while its strength is to leading order
independent of $\bm{B}$. The function $\bm{m}(\bm{B},T)$ therefore
changes discontinuously at $\bm{B}=0$. The line $\bm{B}=0$ for $T<T_*$
is therefore a line of first order phase transitions between
directions (if we consider one spatial dimension only, between
$\bm{m}$ up and $\bm{m}$ down). This line ends at the critical point
$(\bm{B}=0,T=T_*)$ where the order parameter $|\bm{m}|$ vanishes. The
critical value $\bm{B}=0$ of $\bm{B}$ is determined by symmetry; by
contrast $p_*$ depends on microscopic properties of the material.

\citet{Gundlach_scalingfunctions} has suggested that the angular
momentum of the initial data can play the role of $\bm{B}$, and the
final black hole angular momentum the role of $\bm{m}$. Like the
magnetic field, angular momentum is a vector, with a critical value
that must be zero because all other values break rotational
symmetry. \new{See also Sec.~\ref{section:nonsphericalfluid} below for
numerical tests of this conjecture.}

We have already stated that a critical phase transition involves
scale-invariant physics. In particular, the atomic scale, and any
dimensionful parameters associated with that scale, must become
irrelevant at the critical point. This is taken as the starting point
for obtaining properties of the system at the critical point.

One first defines a semi-group acting on micro-states: the
renormalisation group. Its action is to group together a small number
of adjacent particles as a single particle of a fictitious new system
by using some averaging procedure. This can also be done in a more
abstract way in Fourier space. One then defines a dual action of the
renormalisation group on the space of Hamiltonians by demanding that
the partition function is invariant under the renormalisation group
action:
\begin{equation}
  \sum_\mathrm{microstates} \!\!\!\!\!\!\! e^{-H} = \!\!\!\!\!\!\!\!\!
  \sum_\mathrm{microstates'} \!\!\!\!\!\!\! e^{-H'}. 
\end{equation}
The renormalised Hamiltonian is in general more complicated than the
original one, but it can be approximated by the same Hamiltonian with
new values of the parameters $\mu$ and external forces $f$. (At this
stage it is common to drop the distinction between $\mu$ and $f$, as
the new $\mu'$ and $f'$ depend on both $\mu$ and $f$.) Fixed points of
the renormalisation group correspond to Hamiltonians with the
parameters at their critical values. The critical values of many of
these parameters will be $0$ or $\infty$, meaning that the
dimensionful parameters they were originally associated with are
irrelevant. Because a fixed point of the renormalisation group cannot
have a preferred length scale, the only parameters that can have
nontrivial values are dimensionless.

The behaviour of thermodynamical quantities at the critical point is
in general not trivial to calculate. But the action of the
renormalisation group on length scales is given by its definition. The
blowup of the correlation length $\xi$ at the critical point is
therefore the easiest critical exponent to calculate. \new{In critical
phenomena in gravitational collapse,} the same is true for the black
hole mass, which is just a length scale. We can immediately
reinterpret the mathematics of Sect.~\ref{section:scaling} as a
calculation of the critical exponent for $\xi$, by substituting the
correlation length $\xi$ for the black hole mass $M$, $T_*-T$ for
$p-p_*$, and taking into account that the $\tau$-evolution in critical
collapse is towards smaller scales, while the renormalisation group
flow goes towards larger scales: $\xi$ therefore diverges at the
critical point, while $M$ vanishes.

\new{

\section{The spherically symmetric scalar field (in 3+1)}
\label{section:scalarfield}

}

Critical phenomena in gravitational collapse were first discovered by
Choptuik \citep{Choptuik91, Choptuik92, Choptuik94} in the model of a
spherically symmetric, massless scalar field $\phi$ minimally coupled
to general relativity. The scalar field matter is both simple, and
acts as a toy model in spherical symmetry for the effects of
gravitational radiation.  Given that it is still the best-studied
model in spherical symmetry, we review it here as a case study. For
other numerical work on this model, see \citet{GPP2, Garfinkle,
  HamadeStewart, FrolovPen, Puerrer, ZiprickKunstatter0812}. \new{For
mathematical results see Sec.~\ref{section:mathscalarGR}.}

We first review the field equations and Choptuik's observations at the
black hole threshold, mainly as a concrete example for the general
ideas discussed above. \new{The critical solution in the spherical
Einstein-scalar system (the Choptuik solution) is also the only
situation whose complete spacetime structure is known.  This throws
an interesting light on cosmic censorship.} In particular, the exact
critical solution \new{is real-analytic \citep{ReiTru12} and contains
a curvature singularity that is locally and globally naked
\citep{critcont}}, and \new{therefore} any critical solution
obtained in the limit of perfect fine-tuning of asymptotically flat
initial data is at least locally naked.

\subsection{Field equations in \new{polar-radial coordinates}}
\label{section:scalarfieldequations}

The Einstein equations with a massless scalar field as the
matter are
\begin{equation}
  \label{scalar_stress_energy}
  G_{ab} = 8 \pi \left(\nabla_a \phi \nabla_b \phi - {1\over 2} g_{ab}
  \nabla_c \phi \nabla^c \phi\right),
\end{equation}
and the wave equation obeyed by the matter is
\begin{equation}
  \nabla_a \nabla^a \phi = 0.
\end{equation}
Note that the matter equation of motion is contained within the
contracted Bianchi identities. Choptuik chose Schwarzschild-like
coordinates, 
\begin{equation}
  \label{tr_metric}
  ds^2 = - \alpha^2(t,r) \, dt^2 + a^2(t,r) \, dr^2 + r^2 \, d\Omega^2,
\end{equation}
where $d\Omega^2 = d\theta^2 + \sin^2\theta \, d\varphi^2$ is the
metric on the unit 2-sphere. This choice of coordinates is defined by
the radius $r$ giving the surface area of 2-spheres as $4\pi r^2$, and by
$t$ being orthogonal to $r$ (polar-radial coordinates). One more
condition is required to fix the coordinate completely. Choptuik chose
$\alpha=1$ at $r=0$, so that $t$ is the proper time of the central
observer.

In the auxiliary variables
\begin{equation}
  \varPhi = \phi_{, r},
  \qquad
  \Pi={a\over\alpha} \phi_{, t},
\end{equation}
the wave equation becomes a first-order system, 
\begin{align}
  \varPhi_{, t} &= \left({\alpha\over a}\Pi\right)_{\!\!,r},
  \label{wave_1}
  \\
  \Pi_{, t} &= \frac{1}{r^2} \left(r^2{\alpha\over a} \varPhi\right)_{\!\!,r}.
  \label{wave_2}
\end{align}
In spherical symmetry there are four algebraically independent
components of the Einstein equations. Of these, in polar-radial
coordinates one is a linear combination of derivatives of the
others. The other three contain only first derivatives of the metric,
namely $a_{,t}$, $a_{,r}$, and $\alpha_{,r}$, and are
\begin{align}
  \label{da_dr}
  {a_{, r}\over a}  + {a^2 -1 \over 2r} &= 2\pi r \, (\Pi^2 + \varPhi^2),
  \\
  \label{dalpha_dr}
  {\alpha_{, r}\over \alpha} - {a_{, r}\over a}  - {a^2 -1 \over r} &= 0,
  \\
  \label{da_dt}
  {a_{, t}\over \alpha} &= 4\pi r \, \varPhi \, \Pi.
\end{align}
Because of spherical symmetry, the only dynamics is in the scalar
field equations~(\ref{wave_1}, \ref{wave_2}). The metric can be found
by integrating the ODEs~(\ref{da_dr}) and~(\ref{dalpha_dr}) for $a$
and $\alpha$ at any fixed $t$, given $\phi$ and
$\Pi$. Equation~(\ref{da_dt}) can be ignored in this ``fully
constrained'' evolution scheme.

\subsection{The black hole threshold}
\label{section:threshold}

The free data for the system are the two functions $\Pi(0,r)$ and
$\varPhi(0,r)$. Choptuik investigated several 1-parameter families of
such data by evolving the data for many different values of the
parameter. Simple examples of such families are $\Pi(0,r)=0$ and a
Gaussian for $\varPhi(0,r)$, with the parameter $p$ taken to be either
the amplitude of the Gaussian, with the width and centre fixed, or the
width, with position and amplitude fixed, or the position, with width
and amplitude fixed. For sufficiently small amplitude, for example,
the scalar field will then disperse, and for sufficiently large
amplitude it will form a black hole.

Generic 1-parameter families behave in this way, but this is difficult
to prove in generality. Christodoulou showed for the spherically
symmetric scalar field system that data sufficiently weak in a
well-defined way evolve to a Minkowski-like spacetime
\citep{Christodoulou0a, Christodoulou3}, and that a class of
sufficiently strong data forms a black hole \citep{Christodoulou2}.

Choptuik found that in all 1-parameter families of initial data he
investigated he could make arbitrarily small black holes by
fine-tuning the parameter $p$ close to the black hole threshold. An
important fact is that there is nothing visibly special to the black
hole threshold. One cannot tell that one given data set will form a
black hole and another one infinitesimally close will not, short of
evolving both for a sufficiently long time. 

As $p\to p_*$ along the family, the spacetime varies on ever smaller
scales. Choptuik developed numerical techniques that recursively
refine the numerical grid in spacetime regions where details arise on
scales too small to be resolved properly. In the end, he could
determine $p_*$ up to a relative precision of $10^{-15}$, and make
black holes as small as $10^{-6}$ times the ADM mass of the
spacetime. The power-law scaling~(\ref{massscaling}) was obeyed from
those smallest masses up to black hole masses of, for some families,
$0.9$ of the ADM mass, that is, over six orders of magnitude
\citep{Choptuik94}. There were no families of initial data which did
not show the universal critical solution and critical
exponent. Choptuik therefore conjectured that $\gamma$ is the same for
all 1-parameter families of smooth, asymptotically flat initial data
that depend smoothly on the parameter, and that the approximate
scaling law holds ever better for arbitrarily small $p-p_*$.

Typical 1-parameter families cross the threshold only once, so that
there is every indication that it is a smooth submanifold, as we
assumed in the phase space picture. Taking into account the discussion
of mass scaling above, we can formally write the black hole mass as a
functional of the initial data $Z:=\left(\phi(0,r),\Pi(0,r)\right)$
\emph{exactly} as
\begin{equation}
\label{PQ}
  M[Z] = Q[Z] H(P[Z])\,(P[Z])^\gamma, 
\end{equation}
where $P$ and $Q$ are \emph{smooth} functions on phase space and $H$ is
the Heaviside function. $Q$ could be absorbed into $P$.

In hindsight, polar-radial gauge is well-adapted to
self-similarity. In this gauge, DSS corresponds to
\begin{equation}
  \label{tr_scaling}
  Z(t,r) = Z\left(e^{n\Delta}t, e^{n\Delta}r\right)
\end{equation}
for any integer $n$, where $Z$ stands for any one of the
dimensionless quantities $a$, $\alpha$ or
$\phi$ (and therefore also for $r\Pi$ or $r\varPhi$). With 
\begin{equation}
  \label{x_tau}
  x:= -{r \over t-t_*},
  \qquad
  \tau:= - \ln\left(-{t-t_*\over L}\right),
  \qquad
  t<t_*,
\end{equation}
\new{for some $t_*$, DSS with respect to the point $(t=t_*,r=0)$
  corresponds to}
\begin{equation}
  Z(x,\tau+\Delta) = Z(x,\tau).
\end{equation}
The dimensionful constants $t_*$ and $L$ depend on the particular
1-parameter family of solutions, but the dimensionless critical fields
$Z_*$, and in particular their dimensionless period $\Delta$, are
universal. \new{$L$ here has two roles: it makes $\tau$ in
  (\ref{x_tau}) dimensionless, and its value can be adjusted to shift
  the origin of $\tau$ in a family-dependent way so that
  $Z(x,\tau)\simeq Z_*(x,\tau)$ universally for all families of
  initial data.}  Empirically, $\Delta\simeq 3.44$ for the scalar
field in numerical time evolutions, and $\Delta=3.445452402(3)$ from a
numerical construction of the critical solution based on exact
self-similarity and analyticity \citep{critcont}. \new{The latter
  agrees with the leading digits of the same quantity as computed in
  the existence proof of \citet{ReiTru12}.}

\new{The curvature scaling function $f_R$ defined above in
  (\ref{eqn:wiggles_R}) has been found to be universal,
  coordinate-independent and continuous (with a discontinuity in the
  derivative once per period)
  \citep{GarfinkleDuncan,Bau18,GunBauHil24}. The mass scaling function
  $f_M$ defined in (\ref{fM}) is slicing-dependent, and discontinuous
  once per period in some but not all slicings
  \cite{Puerrer,CreDeOWin19,Rin20,ZiprickKunstatter0812,GunBauHil24}.}

\subsection{Global structure of the critical solution}
\label{section:globalstructure}

In adapted coordinates, the metric of the critical spacetime is of the
form $e^{-2\tau}$ times a regular metric. From this general form
alone, one can conclude that $\tau=\infty$ is a curvature singularity,
where, \new{for example, $R^{abcd}R_{abcd}$ blows up like $e^{4\tau}$}
(unless the spacetime is flat), and which is at finite proper time
from regular points in its past.  The Weyl tensor with index position
$C^a{}_{bcd}$ is conformally invariant, so that components with this
index position remain finite as $\tau\to\infty$. This type of
singularity is called ``conformally compactifiable'' \citep{Tod_pc}
or ``isotropic'' \citep{Goodeetal}. For a
classification of all possible \emph{global} structures of spherically
symmetric self-similar spacetimes see \citet{GundlachMartinCSS}.

The global structure of the scalar field critical solution was
determined accurately in \citep{critcont} by assuming analyticity at
the centre of spherical symmetry and at the past light cone of the
singularity (the self-similarity horizon, or SSH). The critical
solution is then analytic up to the future lightcone of the
singularity (the Cauchy horizon, or CH). Global adapted coordinates
$x$ and $\tau$ can be chosen so that the regular centre $r=0$, the SSH
and the CH are all lines of constant $x$, and surfaces of constant
$\tau$ are never tangent to $x$ lines. (A global $\tau$ is no longer a
global time coordinate.) This is illustrated in
Fig.~\ref{figure:redshiftmatched}. \new{See also
Sec.~\ref{section:ReitererTrubowitz} for an existence proof based on
this construction.}

Approaching the CH, the scalar field oscillates an
infinite number of times but with the amplitude of the oscillations
decaying to zero. The scalar
field in regular adapted coordinates $(x,\tau)$ is of the form
\begin{equation}
  \label{phicritcont}
  \phi(x,\tau) \simeq F_\mathrm{reg}(\tau) +
  |x|^\epsilon F_\mathrm{sing} \left[ \tau + H(\tau) + K \ln |x| \right],
\end{equation}
where $F_\mathrm{reg}(\tau)$, $F_\mathrm{sing}(\tau)$ and $H(\tau)$
are periodic with period $\Delta$, and the SSH is at $x=0$. These
functions have been computed numerically to high accuracy, together
with the constants $K$ and $\epsilon$. The scalar
field itself is smooth with respect to $\tau$, and as $\epsilon>0$, it
is continuous but not
differentiable with respect to $x$ on the CH itself. The same is true
for the metric and the curvature. Surprisingly, the ratio $m/r$
of the Hawking mass over the area radius on the CH is of
order $10^{-6}$ but not zero (the value is known to eight significant
figures). 

\new{Even though the CH itself is regular with smooth null data except
for only a singular point at its base, that point singularity makes
 the continuation non-unique. An illustration of this} is given in
\citet{critcont}, where all \emph{DSS} continuations are
considered. Within a DSS ansatz, the solution just to the future of
the CH has the same form as Equation~(\ref{phicritcont}).
$F_\mathrm{reg}(\tau)$ is the same on both sides, but
$F_\mathrm{sing}(\tau)$ can be chosen freely on the future side of the
CH. Within the restriction to DSS this function can be taken to
parameterise the information that comes out of the naked singularity.

There is precisely one choice of $F_\mathrm{sing}(\tau)$ on the future
side that gives a regular centre to the future of the CH, with the
exception of the naked singularity itself, which is then a point. This
continuation was calculated numerically, and is almost but not quite
Minkowski in the sense that $m/r$ remains small everywhere to the
future of the SSH. 

\begin{figure}[htbp]
    \centerline{\includegraphics[width=\linewidth]{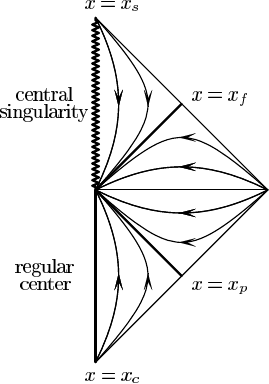}}
    \caption{The spacetime diagram of all generic \emph{DSS}
      continuations of the scalar field critical solution, from
      \citep{critcont}. The naked singularity is timelike, central,
      strong, and has negative mass. There is also a unique
      continuation where the singularity is replaced by a regular
      centre except at the spacetime point at the base of the CH,
      which is still a strong curvature singularity. No other
      spacetime diagram is possible if the continuation is DSS. The
      lines with arrows are lines of constant adapted coordinate $x$,
      with the arrow indicating the direction of
      $\partial/\partial\tau$ towards larger curvature.}
    \label{figure:critcont}
  \end{figure}

All other DSS continuations have a naked, timelike central curvature
singularity with negative mass. More exotic continuations including
further CHs would be allowed kinematically \citep{CarrGundlach} but
are not achieved dynamically if we assume that the continuation is
DSS. The spacetime diagram of the generic DSS continued solution is
given in Fig.~\ref{figure:critcont}.


\subsection{Near-critical spacetimes and naked singularities}

Choptuik's results have an obvious bearing on the issue of cosmic
censorship (see \citet{Wald_censorship} for a general review of cosmic
censorship). Roughly speaking, fine-tuning to the black hole threshold
provides a set of data which is codimension one in the space of
all smooth, asymptotically flat initial data, and whose time
evolution contains at least the point singularity of the critical
solution. The cosmic censorship hypothesis must therefore be
formulated as ``\emph{generic} smooth initial data for reasonable
matter do not form naked singularities''. Here we look at the relation
between fine-tuning and naked singularities in more detail. 

\citet{Christodoulou5} proves rigorously that naked singularity
formation is not generic, but in a rather larger function space,
functions of bounded variation, than one would naturally consider. In
particular, the instability of the naked singularity found by
Christodoulou is not differentiable on the past light cone. This
is unnatural in the context of critical collapse, where the naked
singularity can arise from generic (up to fine-tuning) smooth initial
data, and the intersection of the past light cone of the singularity
with the initial data surface is as smooth as the initial data
elsewhere. It is therefore not clear how this theorem relates to the
numerical and analytical results strongly indicating that naked
singularities are codimension-1 generic within the space of
smooth initial data. 

\begin{figure}[htbp]
    \centerline{\includegraphics[width=\linewidth]{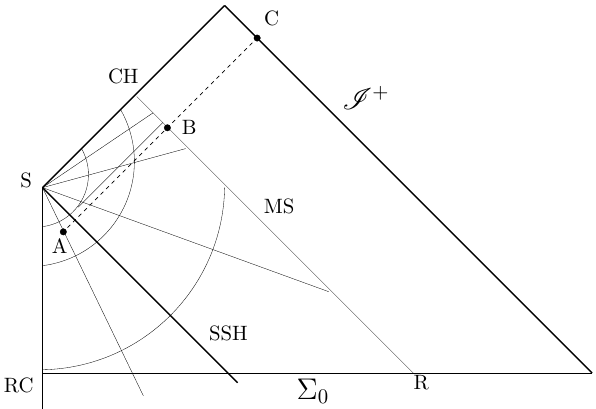}}
    \caption{Conformal diagram of the critical solution matched to
      an asymptotically flat one (RC: regular centre, S: singularity,
      SSH: self-similarity horizon). Curved lines are lines of
      constant coordinate $\tau$, while converging straight lines are
      lines of constant coordinate $x$. Let the initial data on the
      Cauchy surface $\Sigma_0$ be those for the exact critical solution out
      to the 2-sphere R, and let these data be smoothly extended to
      some data that are asymptotically flat, so that the future null
      infinity $\mathscr{I}^+$ exists. To the past of the matching
      surface MS the solution coincides with the critical
      solution. The spacetime cannot be uniquely continued beyond the
      Cauchy horizon CH. The redshift from point A to point B is
      finite by self-similarity, and the redshift from B to C is
      finite by asymptotic flatness.}
    \label{figure:redshiftmatched}
  \end{figure}

First, consider the exact critical solution. The lapse $\alpha$
defined by Equation~(\ref{tr_metric}) is bounded above and below in the
critical solution. Therefore the redshift measured between constant
$r$ observers located at any two points on an outgoing radial null
geodesic in the critical spacetime to the past of the CH is bounded
above and below. Within the exact critical solution, a point with
arbitrarily high curvature can therefore be observed from a point with
arbitrarily low curvature. Next consider a spacetime where the critical
solution in a central region is smoothly matched to an asymptotically
flat outer region such that the resulting asymptotically
flat spacetime contains a part of the critical solution that includes the
singularity and a part of the CH. In this spacetime, a point of
arbitrarily large curvature can be seen from $\mathscr{I}^+$ with finite
redshift. This is illustrated in Fig.~\ref{figure:redshiftmatched}.

Now consider the evolution of asymptotically flat initial data that
have been fine-tuned to the black hole threshold. The global structure
of such spacetimes has been investigated numerically in
\citet{HamadeStewart,FrolovPen,Garfinkle,Puerrer}. Empirically, these
spacetimes can be approximated near the singularity by
Equation~(\ref{perturbations}). Almost all perturbations decay as the
singularity is approached and the approximation becomes better, until
the one growing perturbation (which by the assumption of fine-tuning
starts out small) becomes significant. A maximal value of the
curvature is then reached which \new{scales as in (\ref{curvaturescaling}).}

Finally, consider the limit of perfect fine-tuning. The growing mode
is then absent, and all other modes decay as the naked singularity is
approached. Note that Equation~(\ref{perturbations}) suggests this is
true regardless of the direction (future, past, or spacelike,
depending on the value of $x$) in which the singularity is
approached. This seems to be in tension with causality: the
issue is sensitively connected to the completeness of the modes $Z_i$
and to the stability of the CH, and requires more investigation.


\begin{figure}[htbp]
    \centerline{\includegraphics[width=\linewidth]{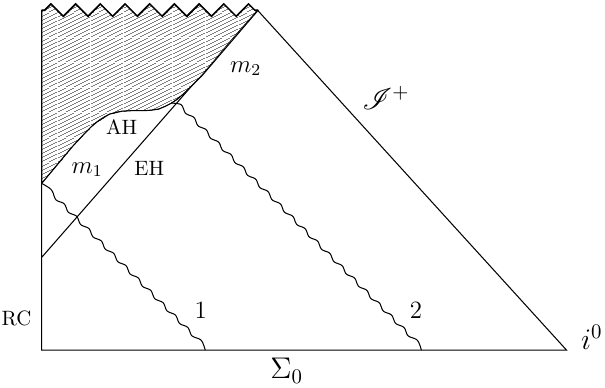}}
    \caption{The final event horizon of a black hole is only known
      when the infall of matter has stopped. Radiation at 1 collapses
      to form a small black hole which settles down, but later more
      radiation at 2 falls in to give rise to a larger final
      mass. Fine-tuning of a parameter may result in $m_1\sim
      (p-p_*)^\gamma$, but the final mass $m_2$ would be approximately
      independent of $p$.}
    \label{figure:smallBHlargeBH}
  \end{figure}

A complication in the supercritical case has been pointed out
in \citet{Puerrer}. In the literature on supercritical evolutions,
what is quoted as the black hole mass is in fact the mass of the
first \new{MOTS} that appears in the time slicing used by the code
(spacelike or null). The black hole mass can and generically will be
larger than \new{the first MOTS mass} because of matter falling in
later, (see Fig.~\ref{figure:smallBHlargeBH} for an illustration.)
The true black hole mass can only be measured at $\mathscr{I}^+$,
where it is defined to be the limit of the Bondi mass $m_{\mathrm{B}}$
as the Bondi time $u_{\mathrm{B}}\to\infty$. This was implemented
in \citet{Puerrer}. Only one family of initial data was investigated,
but in this family it was found that $m_{\mathrm{B}}$ converges to
$10^{-4}$ of the initial Bondi mass in the fine tuning limit. More
numerical evidence would be helpful, but the result \new{can be
understood as follows}. As the underlying physics is perfectly
scale-invariant in the massless scalar field model, the minimum mass
must be determined by the family of initial data through the infall of
matter into the black hole. Simulations of critical collapse of a
perfect fluid in a cosmological context show a similar lower
bound \citep{HawkeStewart} due to matter falling back after shock
formation, but this may not be true for all initial
data \citep{MuscoMiller}.


\subsection{Electric charge}
\label{section:scalarcharge}

Given the scaling power law for the black hole mass in critical
collapse, one would like to know what happens if one takes a generic
1-parameter family of initial data with both electric charge and
angular momentum (for suitable matter), and fine-tunes the parameter
$p$ to the black hole threshold. A simple model for charged matter is
a complex scalar field coupled to electromagnetism with the
substitution $\nabla_a\to\nabla_a+ieA_a$, or scalar
electrodynamics. (Note that in geometric units $c=G=1$, black hole
charge $Q$ has dimension of length, but the charge parameter $e$ has
dimension 1/length. \new{In classical physics there is no motivation for
  setting $\hbar=1$.})

\citet{GundlachMartin} have studied
scalar massless electrodynamics in spherical symmetry
perturbatively. Clearly, the real scalar field critical solution of
Choptuik is a solution of this system too. In fact, it remains
a critical solution within massless scalar electrodynamics in the
sense that it still has only one growing perturbation mode within the
enlarged solution space. Some of its perturbations carry electric
charge, but as they are all decaying, electric charge is a subdominant
effect. The charge of the black hole in the critical limit is
dominated by the most slowly decaying of the charged modes. From this
analysis, a universal power-law scaling of the black hole charge
\begin{equation}
Q\sim (p-p_*)^\delta
\end{equation}
was predicted. The predicted value $\delta\simeq 0.88$ of the critical
exponent (in scalar electrodynamics) was subsequently verified in
collapse simulations by \citet{HodPiran_charge} and later again by
\citet{Petryk}. (The mass scales with $\gamma\simeq 0.37$ as for the
uncharged scalar field.)

General considerations similar to those in
Sect.~\ref{section:universalityclasses} led Gundlach and Mart\'\i
n-Garc\'\i a to the general prediction that \new{in type~II critical
  collapse, governed by a self-similar critical solution,} the two
critical exponents are always related, for any matter model, by the
inequality
\begin{equation}
\delta\ge2\gamma
\end{equation}
(with the equality holding if the critical solution is charged), so
that black hole charge can always be treated perturbatively at the
black hole threshold. \new{To the best of our knowledge,} this has not
yet been verified in any other matter model. \new{(See also
  Sec.~\ref{section:extremalcriticalcollapse} for possibility that the
  critical solution is an extremal black hole.)}

\subsection{Self-interaction potential}
\label{section:self-interaction}

An example of the richer phenomenology in the presence of a scale in
the field equations is the spherical \emph{massive} scalar field with a
potential $m^2\phi^2$ \citep{BradyChambersGoncalves} coupled to
gravity: In one region of phase space, with characteristic scales
smaller than $1/m$, the black hole threshold is dominated by the
Choptuik solution and type~II critical phenomena occur. In another it
is dominated by metastable oscillating boson stars (whose mass is of
order $1/m$ in geometric units) and type~I critical phenomena
occur. (For the real scalar field, the type~I critical solution
is an (unstable) oscillating boson star \citep{SeidelSuen} while for
the complex scalar field it can be a static (unstable) boson
star \citep{HawleyChoptuik} \new{and \citep{JimAlc22}}).
\new{\citet{KelBel20,KelBel21} study spacetimes in which two scalar
fields, one of which is massive, interact through
gravity.}

When the scalar field with a potential is coupled to electromagnetism
type~II criticality is still controlled by a solution which
asymptotically resembles the uncharged Choptuik spacetime, but type~I
criticality is now controlled by charged boson stars
\citep{Petryk}. There are indications that subcritical type~I
evolutions lead to slow, large amplitude oscillations of stable boson
stars \citep{Lai, LaiChoptuik, Petryk} and not to dispersion to
infinity, as had been conjectured in \citet{HawleyChoptuik}.

Another interesting extension is the study of the dynamics of a real
scalar field with a symmetric double-well potential, in which the
system displays type~I criticality between the two possible vacua
\citep{HondaChoptuik}. \new{Extending this to consider an
asymmetric double-well potential, \citet{Hon10} finds that the
boundary of the basins of attraction between the two vacua are fractal
in nature. \citet{IkeYoo16} consider a spherically symmetric real
scalar field with a double-well potential $V(\phi)=C(\phi^2-\sigma)^2$
and $\phi(r)=\sigma\tanh(r-r_0)$ static initial data interpolating
between the two minima (a spherical domain wall). They find the same
type-II critical phenomena as for the massless scalar field.}

\new{

\subsection{The Choptuik solution as a testbed}
\label{section:Choptuik_Testbed}

Choptuik's original approach to the spherical scalar field was to work
in polar-radial coordinates with lapse~$\alpha=1$ at the origin. In
these coordinates ever finer features appear closer to the critical
solution. To overcome this difficulty he implemented an adaptive mesh
algorithm. An alternative that beautifully avoids the need for mesh
refinement was given by~\citet{Garfinkle}, who suggested to work with
null coordinates adapted to the DSS.  Both of these approaches allow
perfect numerical tuning in double-precision arithmetic in spherical
symmetry. However, scalar field critical collapse still plays the role
of a testbed for the development of other formulations and
discretizations.

There are a number of works that employ free-evolution Cauchy
formulations of GR. Such systems are popular in numerical relativity
for the treatment of compact binaries. \citet{CorRenRut23} used a
first order reduction of the generalised harmonic system, and
investigated constraint damping and various choices for the gauge
source functions. \citet{AkbCho15} used a spherical reduction of BSSN
with moving-puncture gauge conditions. \citet{WerEtiAbd21} presented
similar results with the BSSN formulation with open source
tools. \citet{Rin20} successfully used a bespoke spherical Cauchy
formulation to investigate the use of spatial coordinates that zoom in
towards the center of collapse, in an approach that resembles that
of~\citet{Garfinkle} with null coordinates.

There have also been developments in simulations using null
coordinates other than \citep{Garfinkle}. \citet{BarOliRod17}
presented a mesh refinement scheme in Bondi gauge. \citet{CreDeOWin19}
presented results in affine-null gauge, with an adaptive mesh Galerkin
numerical method and compactification to future null infinity.

}

\section{Other systems in spherical symmetry or
cohomogeneity 2}
\label{section:spherical}

The pioneering work of Choptuik on the spherical massless scalar field
has been followed by a plethora of further investigations. These could
be organised under many different criteria. We have chosen the
following rough categories:
\begin{itemize}

\item Systems in which the field equations \new{for the matter},
  \emph{when reduced to spherical symmetry}, \new{form one or more}
  wave-like equations \new{in radius and time only}, typically with
  explicit $r$-dependence in its coefficients. \new{We shall call
    these cohomogeneity-2 (in spacetime).} This includes Yang--Mills
  fields, sigma models, vector and spinor fields, scalar fields in 2+1
  or in 4+1 and higher spacetime dimensions, \new{a vacuum gravity
    ansatz in 4+1 dimensions that reduces to an effective scalar
    field}, and scalar fields in a semi-classical approximation to
  quantum gravity.

\item Perfect fluid matter \new{in spherical symmetry}, either in an
  asymptotically flat or a cosmological context. The linearised Euler
  equations are in fact wave-like, but the full non-linear equations
  admit shock heating and are therefore not even time-reversal
  symmetric.

\item Collisionless matter described by the Vlasov equation is a
  partial differential equation on particle phase space as well as
  spacetime. Therefore even in spherical symmetry, the matter equation
  is a partial differential equation in four independent variables
  (rather than two). Intuitively speaking, there are infinitely more
  matter degrees of freedom \new{in Vlasov than in scalar field or
    perfect fluid matter, or in (non-spherical) vacuum gravity}.

\end{itemize}

Some of these examples were constructed because they may have
intrinsic physical relevance (semiclassical gravity, primordial black
holes), others as toy models for 3+1-dimensional gravity, and others
mostly out of a purely mathematical
interest. Table~\ref{table:mattermodels} gives an overview of these
models. \new{Many of the matter models coupled here to Einstein
  gravity also admit blowup solutions on Minkowski spacetime, but
  these are discussed in Sec.~\ref{section:PDEblowup}.}

\subsection{Matter fields obeying wave equations}
\label{section:matterobeyingwaveequations}

\subsubsection{Scalar field collapse in 2+1}

Spacetime in 2+1 dimensions is flat everywhere where there is no
matter, so that gravity is not acting at a distance in the usual
way. There are no gravitational waves, and black holes can only be
formed in the presence of a negative cosmological constant (see
\citet{Carlip} for a review). \new{Collapse in 2+1 is an interesting
  toy model for 3+1, as both rotating and non-rotating spacetimes can
  have circular symmetry, meaning all fields depend only on time and
  radius (whereas rotating solutions in 3+1 are at best
  axisymmetric).}

Scalar field collapse in circular symmetry was investigated
numerically by \citet{PretoriusChoptuik}, and
\citet{HusainOlivier}. In a regime where the cosmological constant is
small compared to spacetime curvature they find type~II critical
phenomena with a universal CSS critical solution, and
$\gamma=1.20\pm 0.05$ \citep{PretoriusChoptuik}. The value
$\gamma\simeq 0.81$ \citep{HusainOlivier} appears to be less
accurate. \new{Because of $\Lambda<0$ the spacetime is not
asymptotically flat but asymptotically anti-de~Sitter, which is
similar to a box with reflecting walls. Therefore, if a black hole
forms one expects all mass to fall into it eventually, so one would
expect scaling only for the mass of the first MOTS.}

Looking for the critical solution in closed form,
\citet{Garfinkle2+1} found a countable family of exact
spherically symmetric CSS solutions for a massless scalar field with
$\Lambda=0$, but his results remain inconclusive. The $q=4$ solution
appears to match the numerical evolutions inside the past light cone,
but its past light cone is also an apparent horizon. The $q=4$
solution has three growing modes although the top one would give
$\gamma=8/7\simeq 1.14$ if only the other two could be ruled
out \citep{GarfinkleGundlach2+1}. An attempt at
this \citep{HirschmannWangWu} seems unmotivated. At the same time, it
is possible to embed the $\Lambda=0$ solutions into a family of
$\Lambda<0$ ones \citep{ClementFabbri1,ClementFabbri2,Cavaglia}, which
can be constructed along the lines of
Sect.~\ref{section:universalityclasses}, so that Garfinkle's
solution could be the leading term in an expansion in
$e^{-\tau}(-\Lambda)^{-1/2}$.

\new{

\citet{JalGun15} confirmed previous numerical results of collapse
simulations and proposed a partial theory: the critical solution is
well approximated by the $q=4$ Garfinkle solution in the interior of
the past lightcone, glued to a Vaidya exterior that is marginally
trapped everywhere. This has the same perturbation spectrum as the
original Garfinkle solution. Taking the effect of $\Lambda<0$ into
account perturbatively untraps the entire solution. Taking into
account only the dominant growing mode then predicts critical scaling
with a curvature critical exponent of $7/8$ and a mass critical
exponent of $8/23$, consistent with numerical experiment. They
conjecture that there is a smooth approximately CSS critical solution
that approaches the gluing solution on scales $\ll
1/\sqrt{-\Lambda})$, with the other two growing modes suppressed by
smoothness.

\citet{JalGun17} have numerically investigated the collapse of a
rotating scalar field, with an ansatz
$\psi(t,r,\varphi)=f(t,r)e^{im\varphi}$. They find mass scaling but it
stops at small scales, and the critical exponent is not
universal. They find no scaling of the angular momentum of the first
MOTS.

}

\subsubsection{Scalar field collapse in higher dimensions}
\label{subsubsection:Higher_Dimensions}

Critical collapse of a massless scalar field in spherical symmetry in
5+1 spacetime dimensions was investigated in
\citet{GarfinkleCutlerDuncan}. Results are similar to 3+1 dimensions,
with a DSS critical solution and mass scaling with
$\gamma\simeq 0.424$. \citet{Birukou, Birukou2} have developed a code
for arbitrary spacetime dimension. They confirm known results in 3+1
($\gamma\simeq 0.36$) and 5+1 ($\gamma\simeq 0.44$) dimensions, and
investigate 4+1 dimensions. Without a cosmological constant they find
mass scaling with $\gamma\simeq 0.41$ for one family of initial data
and $\gamma\simeq 0.52$ for another. They see wiggles in the $\ln M$
versus $\ln(p-p_*)$ plot that indicate a DSS critical solution, but
have not investigated the critical solution directly. With a negative
cosmological constant and the second family, they find
$\gamma=0.49$. \citet{BlaKun07} have made a more precise
determination: $\gamma=0.4131\pm 0.0001$. This was motivated by an
attempt to explain this exponent using a holographic duality between
the strong coupling regime of 4+1 gravity and the weak coupling regime
of 3+1 QCD \citep{Alvarez-Gaume}, which had predicted
$\gamma=0.409552$. \new{See \citep{Bla07} for more details.}

\citet{Kol05} \new{presents a transformation between two classical
  solutions, one that is related to the Choptuik solution and one that
  is related to the critical solution in the black-string black hole
  transition}, and claims to obtain analytic estimates for $\gamma$
and $\Delta$ \new{in arbitrary dimension}. This has motivated a
numerical determination of $\gamma$ and $\Delta$ for the spherical
massless scalar field in noninteger dimension up to 14, \new{
see \citet{SorkinOren,BlaPreBec05,TavKun11}. See also \citet{PorGun22}
for results in 8+1 and a critical discussion of the methods
of \citep{BlaPreBec05}.

\citet{GolPir12} and \citet{DepLeoTav12} investigate collapse of
the spherical scalar field in Einstein-Gauss-Bonnet gravity. The
additional terms in the action are topological in 3+1 dimensions, so
they consider 4+1 (and also 5+1 in the latter work). The dimensionful
parameter strongly affects near-critical solutions, preventing
self-similarity.

\citet{AnZhang2018} note that in a Kaluza-Klein ansatz the warped
product of a 4-dimensional spacetime with $S^n$ or $T^n$ is equivalent
to a massless scalar field with an exponential potential minimally
coupled to Einstein gravity in the 4-dimensional spacetime. They
numerically construct spherically symmetric CSS solutions and show
that there are naked singularities in both the 4-dimensional spacetime
with scalar field and the higher-dimensional spacetime, for the
example of $n+4=6$ with $S^2$ compactification. There is no discussion
of stability and so no indication that these are critical solutions.

}

\subsubsection{Nonlinear sigma models (or wave maps)}
\label{section:wavemapGR}

We have already discussed in Sect.~\ref{section:self-interaction}
the effects of adding a potential to the evolution of the scalar field.
An alternative generalisation is a modification of the kinetic term in
the Lagrangian, with the general form
\begin{equation}
\frac{1}{2} g^{\mu\nu} G_{IJ}(\phi^K)\nabla_\mu\phi^I \nabla_\nu \phi^J
\end{equation}
for an $N$-dimensional vector field $\phi^I(x)$ with $I=1,\dots,N$,
and where $G_{IJ}(\phi^K)$ is a fixed nonlinear function acting as a
metric on the so-called \emph{target space} of the fields $\phi^I$.
Such a system is called \new{a nonlinear sigma model (in the physics
  literature), or harmonic map or wave map (in the mathematics
  literature)}. The fields $\phi^I$ and metric $G_{IJ}$ are
dimensionless, and this allows the introduction of dimensionless
parameters in the system, which cannot be asymptotically neglected
using the arguments of Sect.~\ref{section:universalityclasses}.
(Compare with the potential $V(\phi^I)$, which has dimensions
$(\mathrm{length})^{-2}$, and hence requires dimensionful parameters.)

The case $N=1$ gives nothing new, and so Hirschmann and Eardley (HE
from now on) studied the $N=2$ case with a target manifold with
constant curvature \citep{HE3}, proportional to a real dimensionless
constant $\kappa$. Using a single complex coordinate $\phi$ the action
of the system can be written as
\begin{equation}
\label{N=2wavemapEinstein}
  \int d^4x \, \sqrt{g}
  \left[ \frac{R}{2} - \frac{|\nabla\phi|^2}{(1 - \kappa|\phi|^2)^2} \right].
\end{equation}
For $\kappa\ge 0$ this system is equivalent to the problem of a real
massless scalar field coupled to Brans--Dicke (BD) gravity (with the
BD coupling constant given by
$8\omega_{\mathrm{BD}}=-12+ \kappa^{-1}$).  \citet{LieblingChoptuik}
have shown that there is a smooth transition in the BD system from DSS
criticality for low $\kappa$ to CSS criticality for larger $\kappa$.
The flat target space case $\kappa=0$ is equivalent to a
self-gravitating complex massless scalar field, whose critical
solution is the original DSS spacetime found by Choptuik. The case
$\kappa=1$ is equivalent to the axion-dilaton system, and has been
shown to display CSS criticality in \citet{HamadeHorneStewart}.
\new{Spherical CSS solutions and their perturbations of the
axion-dilaton system have been constructed in \citet{AlvHat12,AlvHat13}.}

Generalising their previous results for $\kappa=0$ \citep{HE1,HE2}, HE
constructed for each $\kappa$ a CSS solution based on the ansatz
$\phi(\tau,x)=e^{i\omega\tau}\phi(x)$. Studying its perturbations HE
concluded that this solution is critical for $\kappa>0.0754$, but has
three unstable modes for $\kappa<0.0754$ and even more for
$\kappa<-0.28$. Below 0.0754 a DSS solution takes over, as shown in
the simulations of
\citet{LieblingChoptuik}, and HE conjectured that the
transition is a Hopf bifurcation, such that the DSS cycle smoothly
shrinks with growing $\kappa$, collapsing onto the CSS solution at the
transition and then disappearing while the echoing
period $\Delta$ remains finite.

\new{We note that in terms of $\tilde\phi:=\sqrt{\kappa}\phi$ and
  $\eta:=1/\kappa$, the action (\ref{N=2wavemapEinstein}) becomes
\begin{equation}
\label{N=2wavemapEinsteinbis}
  \int d^4x \, \sqrt{g} \left[ \frac{R}{2}
  - \eta\frac{|\nabla\tilde\phi|^2}{(1 -|\tilde\phi|^2)^2} \right].
\end{equation}
In gravitational units $c=G=1$, the constant $\eta$ in
(\ref{S3wavemapGRaction}) and (\ref{N=2wavemapEinsteinbis}) is
dimensionless. Hence $\eta\to 0$ corresponds to weak coupling to
gravity. Also, for $\eta=0$, that is on Minkowski spacetime, the
equations admit CSS blowup solutions, see
Sec.~\ref{section:sphericalwavemap}. These two facts may explain why the
naked singularity solutions with gravity are CSS for small $\eta$ but
DSS for large $\eta$.

The close relation between the CSS and DSS critical solutions is also 
manifest in the construction of their global structure. In particular,
the results of \citet{HE2} and \citet{EardleyHirschmannHorne} for the CSS
$\kappa=0$ and $\kappa=1$ solutions, respectively, show that the Cauchy
horizon of the singularity is almost but not quite flat, exactly as
was the case with the Choptuik DSS spacetime (see
Sect.~\ref{section:globalstructure}).

\new{The $N=3$ case where the target manifold is $S^3$ with the round
(constant curvature) metric was investigated by
\citet{Aichelburg1}}. In the corotational reduction to spherical
symmetry, the effective action is
\begin{equation}
\label{S3wavemapGRaction}
  \int d^4x \, \sqrt{g} \left[\frac{R}{2} - \eta
  \left( |\nabla\phi|^2 + \frac{2\sin^2\phi}{r^2} \right) \right],
\end{equation}
where $r$ is the area radius, \new{$\phi$ is real}, and the coupling
constant $\eta$ is dimensionless. \new{(The flat space case of this
  ansatz gives Eq.~(\ref{corotationalwavemap}) below, with $d=3$.)}
\new{\citet{Aichelburg1} find} a transition between CSS and DSS
criticality, but this is a totally different type of transition from
the $N=2$ case, in particular showing a divergence in the echoing
period $\Delta$.

It has been shown (numerically in \citet{BizonWassermann} and then
analytically in \citet{BizonWassermann2}) that for $0\le\eta<1/2$,
there is an infinite sequence of CSS solutions $\phi_n$ labelled by a
nodal number $n$, and having $n$ growing modes.  The $n$-th solution
is always regular in the past light cone of the singularity, but is
regular up to the future light cone only for $\eta<\eta_n$ where
$\eta_0\simeq 0.0688$, $\eta_1\simeq 0.1518$, and
$\eta_n<\eta_{n+1}<1/2$. For larger couplings an apparent horizon
develops and the solution cannot be smoothly continued.  These results
suggest that $\phi_0$ is a stable naked singularity for $\eta<\eta_0$,
and $\phi_1$ acts as a critical solution between naked singularity
formation and dispersal for $\eta<\eta_0$ and between black hole
formation and dispersal for $\eta_0<\eta<\eta_1$. The numerical
experiments agree with this scenario in the range $0\le\eta <
0.14$. Other CSS solutions of this system are investigated
in \citet{BizonSzybkaWassermann}, and the possibility of chaos
in \citet{Szybka}.

\citet{Aichelburg1,LechnerPhD} have shown that for $\eta\gtrsim 0.2$
there is clear DSS type~II criticality at the black hole
threshold. The period $\Delta$ depends on $\eta$, monotonically
decreasing towards an asymptotic value for $\eta\to\infty$.
Interesting new behaviour occurs in the intermediate range
$0.14<\eta<0.2$ that lies between clear CSS and clear DSS.  With
decreasing $\eta$ the overall DSS includes episodes of approximate CSS
\citep{Aichelburg2}, of increasing length (measured in the log-scale
time $\tau$). As $\eta\to\eta_{\mathrm{c}}\simeq 0.170$ from above the
duration of the CSS epochs, and hence the overall DSS period $\Delta$
diverges. For $0.14<\eta<0.17$ time evolutions of initial data near
the black hole threshold no longer show overall DSS, but they still
show CSS episodes. Black hole mass scaling is unclear in this regime.

It has been conjectured that this transition from CSS to DSS can be
interpreted, in the language of the theory of dynamical systems, as
the infinite-dimensional analogue of a 3-dimensional Shil'nikov
bifurcation \citep{Lechner}. High-precision numerics in
\citet{Aichelburg4} further supports this picture: For
$\eta>\eta_{\mathrm{c}}$ a codimension-1 CSS solution coexists in
phase space with a codimension-1 DSS attractor such that the
(1-dimensional) unstable manifold of the DSS solution lies on the
stable manifold of the CSS solution.  For $\eta$ close to
$\eta_{\mathrm{c}}$ the two solutions are close and the orbits around
the DSS solution become slower because they spend more time in the
neighbourhood of the CSS attractor. A linear stability analysis
predicts a law
$\Delta\simeq -\frac{2}{\lambda} \log(\eta-\eta_{\mathrm{c}})+b$ for
some constant $b$, where $\lambda$ is the Lyapunov exponent of the CSS
solution. For $\eta=\eta_{\mathrm{c}}$ both solutions touch and the
DSS cycle disappears.

}

\subsubsection{$SU(2)$ (or $SO(3)$) Yang-Mills}

\citet{ChoptuikChmajBizon} have found both type~I and type~II critical
collapse in the spherical Einstein--Yang--Mills (EYM) system with
$SU(2)$ \new{(or $SO(3)$)} gauge potential, restricting to the
purely magnetic case, in which the matter is described by a single
real scalar field. The situation is very similar to that of the
massive scalar field, and now the critical solutions are the
well-known static $n=1$ Bartnik--McKinnon solution
\citep{BartnikMcKinnon} for type~I and \new{an asymptotically DSS
  critical solution for type~II. The DSS} critical solution was later
constructed in \citet{Gundlach_EYM}. In both cases the black holes
produced in the supercritical regime are Schwarzschild black holes
with zero Yang--Mills field strength, but the final states (and the
dynamics leading to them) can be distinguished by the value of the
Yang--Mills final gauge potential at infinity, which can take two
values, corresponding to two distinct vacuum states.

\citet{ChoptuikHirschmannMarsa} have investigated the boundary in
phase space between formation of those two types of black holes, using
a code that can follow the time evolutions for long after the black
hole has formed. This is a new ``type~III'' phase transition whose
critical solution is an unstable static black hole with Yang--Mills
hair \citep{Bizon0,VolkovGaltsov}, which collapses to a hairless
Schwarzschild black hole with either vacuum state of the Yang--Mills
field, depending on the sign of its one growing perturbation
mode. This ``coloured'' black hole is actually a member of a
1-parameter family parameterised by its apparent horizon radius and
outside the horizon it approaches the corresponding Bartnik-McKinnon
solution. When the horizon radius approaches zero the three critical
solutions meet at a ``triple point''. What happens there deserves
further investigation.

\new{

Working again in the purely magnetic case, and using a novel
numerical setup that includes future null infinity within the
domain, \citet{Rin14} has confirmed these findings. Observing
furthermore that the Einstein-Yang-Mills equations are invariant
under changing the sign of the Yang-Mills potential, he investigated
a two-parameter family of initial data in which unstable coloured
black holes of both signs arise as a codimension-one critical
solution. He found that the codimension-two critical solution at
their intersection is the Reissner-Nordstr\"om solution.

\citet{MalRin17} then dropped the magnetic ansatz to consider the
general spherically symmetric SU(2) Einstein-Yang-Mills
system. Given that in this more general setting the
Bartnik--McKinnon solution has an additional perturbative unstable
mode, they anticipated that the critical behaviour would differ
greatly. Working with two independent numerical codes they in fact
found considerable simplification in the critical behaviour, with
both type~I and type~III absent. Their numerical evidence suggests
that the type~II critical solution that survives is similar to, but
distinct from the magnetic type~II critical solution. In agreement
with the PhD thesis of~\citep{Jac18} they found evidence against the
existence of a universal critical solution at the threshold.

}

\citet{Millward} have further coupled a Higgs field to the
Einstein--Yang--Mills system, \new{with the Higgs field in the
fundamental representation}. New possible end states appear: regular
static solutions, and stable hairy black holes (different from the
coloured black holes referred to above).  Again there are type~I or
type~II critical phenomena depending on the initial conditions.
\new{\citet{Kai18,Kai19}
works instead with a Higgs field in the adjoint representation, in
which a similar phenomenology occurs, with possible end-states
including a regular monopole solution, (multiple) static monopole
black holes, and the Reissner-Nordstr\"om solution. In the first work,
he tuned initial data to examine the transition between a monopole
black hole and the Reissner-Nordstr\"om end-states, both of which
appear stable. He found that an unstable monopole black hole solution
manifests as the critical solution between these two end-states. In
the second study he examined the threshold of black hole formation,
finding on the subcritical side that solutions settle down to a static
monopole configuration. Interestingly, he found that there are two
type II critical solutions, with distinct scaling and echoing
parameters, that manifest depending on whether the Higgs or Yang-Mills
fields drives the collapse. In the latter case there appear to be
deviations from pure type II behaviour.}

\new{

Motivated by the observation that beyond spherical symmetry,
gravitational waves will be excited by the presence of dynamical
matter and may compete as the dominant driver towards
collapse, \citet{GunBauHil19} considered the competition between two
spherically symmetric massless matter fields, with the first
satisfying the wave equation, and the second an SU(2) Yang-Mills field
under the magnetic ansatz. In that it contains multiple dynamical
fields, this system is qualitatively similar both to generic spherical
SU(2) Yang-Mills and Yang-Mills-Higgs models. Surprisingly they found
that in this combined system there is a single combined approximately
DSS critical solution that transitions from the pure Yang-Mills
critical solution at large scales to the Choptuik solution at small
scales, so that the scalar field eventually always
dominates. Interpreting the results of~\citet{MalRin17}, they suggest
that the general spherically SU(2) critical solution may eventually be
dominated by the purely magnetic critical solution, with the
additional degrees of freedom eventually subdominant. This would
explain the differences between the two generic and magnetic critical
solutions observed by~\citet{MalRin17} at finite scale.

It is interesting to note that the ansatz of \citep{Gundlach_EYM} for
an (asymptotically) DSS solution of EYM in 3+1 is $w=1-rS$ with
$S(e^\Delta t,e^\Delta r)\simeq S(t,r)$, whereas the ansatz of
\citep{BizBie15} for (exactly) CSS solutions of pure YM in 5+1 and
higher is $w=w(r/t)$. Here $w$ of \citep{Gundlach_EYM} on Minkowski
spacetime is the same as $w$ of \citep{BizBie15}, that is both obey
(\ref{sphericalYM}). This is consistent with the fact that in 3+1
dimensions, pure YM (without gravity) is energy-subcritical and
globally regular, so the blowup is due to gravity, and the DSS
solution with this ansatz has no non-gravitating counterpart.

}

\subsubsection{Vacuum Bianchi-IX ansatz in 4+1 dimensions}
\label{section:VacuumBianchiIX}

\citet{BizonChmajSchmidt} have found a way of constructing
asymptotically flat vacuum spacetimes in 4+1 dimensions \new{which
  have nontrivial dynamics while all fields depend only the
  coordinates $(t,r)$. This is possible because} Birkhoff's theorem
does not hold in more than 3+1 dimensions. Recall that in (3+1
dimensional) Bianchi~IX cosmology the manifold is $M^1\times S^3$
where the $S^3$ is equipped with an SU(2) invariant (homogeneous but
anisotropic) metric
\begin{equation}
  ds^2 = L_1^2 \sigma_1^2 + L_2^2 \sigma_3^2 + L_3^2 \sigma_3^2,
\end{equation}
where the $\sigma_i$ are the SU(2) left-invariant 1-forms, and the
$L_i$ are functions of time only. Similarly, the spacetimes of
\citet{BizonChmajSchmidt} are of the product form $M^2\times S^3$
where the $L_i$ now depend only on $r$ and $t$. This gives rise to
nontrivial dynamics, including a threshold between dispersion and
black hole formation. With the additional U(1) symmetry $L_1=L_2$
(biaxial solutions) there is only one dynamical degree of freedom,
\new{whose equation of motion looks much like a spherical wave
  equation in 9+1 dimensions}. At the black hole threshold, type~II
critical phenomena are seen with $\Delta\simeq 0.49$ and
$\gamma\simeq 0.3289$. \new{While the DSS critical solution is known
  only numerically, \citet{BizWas10} have  proved that regular CSS solutions
  cannot exist.

\citet{PorGun22} have added a massless scalar field,
motivating this as a toy model for matter coupled to gravitational
waves, and find type-II critical collapse with a codimension-two DSS
solution that decays into the known critical solutions for pure
gravitational waves or pure scalar field, both of which are still
codimension-one in this system. Compare also \citet{GunBauHil19},
where a different phase space diagram is found for a massless scalar
field and a Yang-Mills field in spherical symmetry are coupled to
gravity: a single, approximately DSS, critical solution links the
(exactly DSS) Yang-Mills critical solution with codimension two to the
(exactly DSS) Choptuik solution with codimension one.}

In evolutions with the general ansatz where all $L_i$ are different
(triaxial solutions) \citep{BizonChmajSchmidt2}, the U(1) symmetry is
recovered dynamically in the approach to the critical
surface. However, each biaxial solution, and in particular the
critical solution, exists in three copies obtained by permutation of
the $\sigma_i$. Therefore, in the triaxial case, the critical surface
contains three critical solutions. The boundaries, within the critical
surface, between their basins of attraction contain in turn
codimension-two DSS attractors. It is conjectured that there is in
fact a countable family of DSS solutions with $n$ unstable modes.
\citet{SzybkaChmaj} give numerical evidence that these boundaries
\emph{within} the critical surface are fractal (in contrast to the
critical surface itself, which is smooth, as it is in all other known
systems.) \new{\citet{CheWay19} present a holographic interpretation
  of the analogous results with asymptotically adS boundaries.}

A similar ansatz can be made in other odd spacetime dimensions, and in
8+1 dimensions type~II critical behaviour is again observed
\citep{Bizonetal2005}.

\subsubsection{Anti-de Sitter and other confined models}
\label{section:adSConfined}

\new{

\citet{BizRos11} have studied the wave equation minimally coupled to
gravity in 3+1 dimensions but, in contrast to Choptuik's setup, with
negative cosmological constant. The vacuum solution is then locally
anti-de~Sitter (adS) spacetime, rather than Minkowski. This renders
null-infinity timelike, and one needs to impose reflecting outer
boundary conditions. They find that data near the origin which are
close to collapsing give rise to solutions which are locally identical
to that of Choptuik, and that with each passing reflection from the
outer boundary the data become ever more compact, and so closer to
collapsing. Their numerical evidence suggests that after enough
reflections even arbitrarily small generic initial data
collapse. These results were backed up by a multiscale perturbative
analysis whose predictions were carefully compared with the numerical
data. This result, the first numerical demonstration of the
instability of adS, has proven highly influential and has inspired a
large number of follow-up works. We restrict our attention here just
to those most closely related to critical collapse. For an
introductory exposition, see~\citep{MalRos13}. For a more detailed
review, see~\citep{Mar17}. Complementary to the numerical results in
spherical Einstein-scalar, \citet{Mos18} has given a proof of the
instability of adS for massless collisionless matter (Einstein-Vlasov)
in spherical symmetry.

In a direct follow-up, \citet{SanSop15} studied in detail the
transition that occurs as the collapse is induced by one further
reflection of the data. They corroborated the earlier results but
additionally find that, as the amplitude of the initial data is
continuously reduced below the threshold $p_{*n}$ for collapse after
$n$ bounces, the threshold apparent horizon mass after $n+1$ bounces
is finite and that this mass is approached as a universal power law (a
result that has been reproduced in \citet{CaiJiYan17}). Details of
their numerical method, and an extension to arbitrary precision
arithmetic are presented in~\citet{SanSop16} and \citet{SanSop18}
respectively.

Working still with adS asymptotics,
\citet{BizJal13} study a configuration analogous to that
of~\citet{BizRos11}, but in $2+1$ spacetime dimensions. This is an
interesting case because although turbulent phenomena again present
themselves, sufficiently weak data can not form a black
hole. Numerical evidence suggests that, whilst various norms of the
solution do rapidly grow over the time evolution, solutions
nevertheless remain globally regular. Working analytically within
the same model, \citet{BaiStrTaa14} find families of self-similar
solutions which can be tuned to the threshold of apparent horizon
formation.

Similarly~\citet{Dep16} investigates from 3+1
to $9+1$ dimensions, finding results compatible with those
of~\citep{BizRos11}. \citet{BizRos17} work in
spherical symmetry with adS asymptotics but in 4+1 dimensions under
the Bianchi-IX ansatz discussed in
Section~\ref{section:VacuumBianchiIX}, observing similar turbulent
dynamics.

Several studies have been performed with the aim of disaggregating the
role of adS asymptotics, reflecting boundary conditions, and the
details of the field content. \citet{OkaCarPao14} point out that the
inclusion of confining interaction potentials alone for scalar fields
in the asymptotically flat setting are not sufficient to induce the
turbulent behaviour observed in adS, since such potentials still allow
dispersion, and may even permit alternative end-states without a black
hole. \citet{Mal12} has studied spherical spacetimes with a vanishing
cosmological constant, with a massless scalar field but with perfectly
reflecting outer boundary conditions imposed on a timelike
world-tube. He finds result analogous to \citep{BizRos11}, indicating
that reflecting boundary conditions, rather than the specific form of
the metric, are responsible for the turbulent instability. Similarly,
\citet{BucLehLie12} impose reflecting boundary conditions on a finite
world-tube, but now also with a nonvanishing cosmological
constant~$\Lambda<0$, and with a complex scalar field, and find
analogous results.

Studying a semilinear wave equation in the fixed adS
background \citet{Lie12} again sees the same phenomenology, in this
context meaning blow-up of the solution. Imposing boundary
conditions on a finite timelike world-tube, he finds that
sufficiently small data do not blow up because of dispersion
associated with the boundary condition. This is compatible with the
observation of \citet{Fri14}, in a more general context, that
stability may depend upon the choice of boundary conditions.
Working with a vanishing cosmological constant, reflecting boundary
conditions and a charged scalar field, \citet{CaiYan16} observe
similar behavior, including the power-law associated with the jump
in the apparent horizon mass as the number of reflections increase.}

\subsubsection{Other \new{matter models} obeying wave equations}
\label{section:other_matter}

\citet{ChopHirschLieb} have presented perturbative evidence that
the static solutions found by \citet{vanPutten} in the vacuum
Brans--Dicke system are critical solutions. They have also performed
full numerical simulations, but only starting from small deviations
with respect to those solutions.

\new{

\citet{JimAlc22} work in scalar-tensor theory with a massless
scalar field model. In the Jordan frame this results in a scalar field
coupled to GR non-minimally. They work with a carefully chosen radial
coordinate and shock-avoiding slicing slicing condition \citep{Alc96}
see also~\citep{BauHil22}. The system exhibits type II behavior. The
power-law exponent and echoing-period decrease as the degree of
non-minimal coupling increases, and the universal periodic
function~$f_R$, see \eqref{eqn:wiggles_R}, takes a more complicated
form than in GR.

As mentioned above in section~\ref{subsubsection:Higher_Dimensions},
when included in the action in 3+1 dimensions the Gauss-Bonnet term
has no affect on the gravitational field equations. This can be
overcome by coupling it to a scalar field. The resulting
theory is called Einstein-dilaton-Gauss-Bonnet
gravity. \citet{ZhaChiLiu21} have studied the transition from
nonscalarized to scalarized black hole end-states in this
theory. They find type I critical behaviour.

\citet{Bae22} studies the spherical critical collapse of a massless
scalar field within the Starobinsky~$R^2$ model, a particular form
of~$f(R)$ gravity, in the Einstein frame, where the theory is
equivalent to Einstein gravity with a dilaton field, finding power-law
scaling with an exponent similar to that in GR.}

\citet{VentrellaChoptuik} have performed numerical simulations of
collapse of a massless Dirac field in a special state: an incoherent
sum of two independent left-handed zero-spin fields having opposite
orbital angular momentum. This is prepared so that the total
distribution of energy-momentum is spherically symmetric. The freedom
in the system is then contained in a single complex scalar field
obeying a modified linear wave equation in spherical symmetry. There
are clear signs of CSS criticality in the metric variables, and the
critical complex field exhibits a phase of the form $e^{i\omega\tau}$
for a definite $\omega$, \new{as in the ansatz of \citep{HE1} for the
  complex scalar field critical solutions.}

\citet{GarfinkleMannVuille} have found coexistence of types I and II
criticality in the spherical collapse of a massive vector field (the
Proca system), the scenario being almost identical to that of a
massive scalar field. In the self-similar phase the collapse amplifies
the longitudinal mode of the Proca field with respect to its
transverse modes, which become negligible, and the critical solution
is simply the gradient of the Choptuik DSS spacetime.

\citet{SarbachLehner} find type~I critical behaviour in
$q+3$-dimensional spacetimes with
$\mathrm{U}(1) \times \mathrm{SO}(q+1)$ symmetry in Einstein--Maxwell
theory at the threshold between dispersion and formation of a black
string.

\begin{table*}
  \renewcommand{\arraystretch}{1.2}
  {\small 
  \begin{tabular}{l@{\quad}ccccc}
    \toprule
    \hline\hline
    Matter &
    Type &
    Collapse &
    Critical solution &
    Perturbations of \\ [-0.3 em]
    & &
    simulations & &
    critical solution \\
    \hline\hline
    \midrule
    Perfect fluid, $P=k\rho$ &
    II &
    \citep{EvansColeman,NeilsenChoptuik, Novak, Noble, NobleChoptuik, NobCho16} &
    CSS \citep{EvansColeman,Maison,NeilsenChoptuik} &
    \citep{Maison,KoikeHaraAdachi2,Gundlach_nonspherical,Gundlach_critfluid2}
    \\
    -- in 2+1 & I & \citep{BouGun21a, BouGun21b}  & \cite{GunBou20}

    &
    \\
    -- in 4+1, 5+1, 6+1 & II &  ~ &
    CSS \citep{Alvarez-Gaume0804}
    & \citep{Alvarez-Gaume0804}
    \\
    \hline
    Collisionless matter:
    \\
    -- massive particles &
    I &
    \citep{ReinRendallSchaeffer,OlabarrietaChoptuik} &
    & ~
    \\ 
    -- massless particles &
    I &
    \citep{AkbCho14} &
    \citep{vlasov1,Gun16} & \citep{Gun17}
    \\ 
    \midrule\hline
    Real scalar field:
    \\
    -- massless, minimally coupled &
    II &
         \citep{Choptuik91,Choptuik92,Choptuik94, GPP2, Garfinkle,
         HamadeStewart, FrolovPen, Puerrer} &
    DSS \citep{Gundlach_Chop1, ReiTru12} &
    \citep{GarfinkleDuncan, Gundlach_Chop2, HodPiran_wiggle, GunBauHil24,Gundlach_Chop2, 
                                           Bau18, MartinGundlach}
    \\
           & &
               \citep{ZiprickKunstatter0812, CreDeOWin19, Rin20, 
               AkbCho15, WerEtiAbd21, BarOliRod17, CorRenRut23} &
                      &
    \\
    -- massive &
    I &
    \citep{BradyChambersGoncalves, KelBel20, KelBel21} &
    oscillating \citep{SeidelSuen} & ~
    \\
    &
    II &
    \citep{Choptuik94, KelBel20, KelBel21} &
    DSS \citep{HaraKoikeAdachi,GundlachMartin} &
    \citep{HaraKoikeAdachi,GundlachMartin}
    \\
    -- double-well &
    I (vacua)/II & \citep{HondaChoptuik, Hon10, IkeYoo16}
     &
    \\
    -- conformally coupled &
    II &
    \citep{Choptuik94} &
    DSS
    \\
    -- 2+1 & II & \citep{PretoriusChoptuik, HusainOlivier, JalGun15, JalGun17}
     &  CSS~\citep{Garfinkle2+1, JalGun15} &  
    \\
    -- 4+1 &
    II &
    \citep{Birukou, Birukou2, BlaKun07} & DSS & ~
    \\
    -- 5+1 &
    II &
    \citep{GarfinkleCutlerDuncan} & DSS & ~
    \\ 
    \midrule\hline
    Massive complex scalar field &
    I, II &
    \citep{HawleyChoptuik,JimAlc22} &
    \citep{SeidelSuen} &
    \citep{HawleyChoptuik}
    \\
    \hline
    Massless scalar electrodynamics &
    II &
    \citep{HodPiran_charge, Lai, LaiChoptuik} &
    DSS \citep{GundlachMartin} &
    \citep{GundlachMartin}
    \\ 
    \midrule\hline
    Massive vector field &
    II &
    \citep{GarfinkleMannVuille} &
    DSS \citep{GarfinkleMannVuille} &
    \citep{GarfinkleMannVuille}
    \\ 
    \midrule\hline
    Massless Dirac &
    II &
    \citep{VentrellaChoptuik} &
    CSS \citep{VentrellaChoptuik}
    \\
    \hline
    Vacuum Brans--Dicke &
    I &
    \citep{ChopHirschLieb} &
    static \citep{vanPutten} &
    \citep{ChopHirschLieb}
    \\ 
    \midrule\hline
    2-d sigma model:
    \\
    -- complex scalar ($\kappa=0$) &
    II &
    \citep{Choptuik_pc} &
    DSS \citep{Gundlach_Chop2} &
    \citep{Gundlach_Chop2}
    \\
    -- axion-dilaton ($\kappa=1$) &
    II & 
    \citep{HamadeHorneStewart} & 
    CSS \citep{EardleyHirschmannHorne,HamadeHorneStewart} & 
    \citep{HamadeHorneStewart}
    \\
    -- scalar-Brans--Dicke ($\kappa>0$) &
    II &
    \citep{LieblingChoptuik,Liebling} &
    CSS, DSS \citep{HE1,HE2}
    \\
    -- general $\kappa$ including $\kappa<0$ &
    II &
    &
    CSS, DSS \citep{HE1,HE2, HE3} &
    \citep{HE3} \\ 
    \midrule\hline
    3-d sigma model
    & II
    & \citep{Aichelburg1,BizonWassermann,LechnerPhD} 
    & CSS, DSS \citep{BizonWassermann2} &\\
    & 
    & \citep{Aichelburg2,Aichelburg4} 
    &  &\\
    \midrule\hline
    SU(2) Yang--Mills &
    I &
    \citep{ChoptuikChmajBizon,Rin14} &
    static \citep{BartnikMcKinnon} &
    \citep{LavrelashviliMaison}
    \\
    &
    II &
    \citep{ChoptuikChmajBizon,Rin14, MalRin17, Jac18} &
    DSS \citep{Gundlach_EYM} &
    \citep{Gundlach_EYM}
    \\
    & ``III'' &
    \citep{ChoptuikHirschmannMarsa,Rin14} &
    coloured BH \citep{Bizon0,VolkovGaltsov} &
    \citep{StraumannZhou,VolkovBrodbeckLavrelashviliStraumann,BizonChmaj3}
    \\
    \hline
    SU(2) Yang--Mills--Higgs &
    (idem) &
    \citep{Millward, Kai18, Kai19} &
    (idem) & ~
    \\
    \hline
    SU(2) Skyrme model &
    I &
    \citep{BizonChmaj} &
    static \citep{BizonChmaj} &
    \citep{BizonChmaj}
    \\
    &
    II &
    \citep{BizonChmajTabor} &
    DSS \citep{BizonChmajTabor}
    \\
    \hline
    SO(3) Mexican hat &
    II &
    \citep{Liebling2} &
    DSS & ~
    \\
    \hline
    Bianchi-IX ansatz & II
     & \citep{BizonChmajSchmidt, Bizonetal2005, PorGun22}
     & DSS (\citep{BizWas10}; no CSS)
     & ~
    \\
    \bottomrule\hline\hline
  \end{tabular}
}
\vskip 4mm
\centering
\caption{\new{An overview of critical collapse in asymptotically
    flat general relativity in spherical symmetry. For discussion of
    confined models, see section \ref{section:adSConfined}. For work
    on more exotic configurations, see \ref{section:other_matter}, and
    for cosmological
    applications~\ref{section:cosmological_applications}.}}
\label{table:mattermodels}
\renewcommand{\arraystretch}{1.0}
\end{table*}

\subsection{Perfect fluid matter}
\label{section:perfectfluid}

\subsubsection{The asymptotically flat case}
\label{section:perfectfluidasymflat}

\citet{EvansColeman} performed the first simulations of critical
collapse with a perfect fluid with \new{the equation of state (EOS)
  $P=k\rho$} (where $\rho$ is the \new{total} energy density and $P$
the pressure) for $k=1/3$ (radiation), and found a CSS critical
solution with a mass-scaling critical exponent $\gamma\simeq 0.36$.
\citet{KoikeHaraAdachi,KoikeHaraAdachi2} constructed that critical
solution and its linear perturbations from a CSS ansatz as an
eigenvalue problem, computing the critical exponent to high precision.
Independently, \citet{Maison} constructed the regular CSS solutions
and their linear perturbations for a large number of values of $k$,
showing for the first time that the critical exponents were
model-dependent. As \citet{OriPiran,OriPiran2} before, he claimed that
there are no regular CSS solutions for $k>0.89$, but
\citet{NeilsenChoptuik,NeilsenChoptuik2} have found CSS critical
solutions for all values of $k$ right up to 1, both in collapse
simulations and by making a CSS ansatz. The difficulty comes from a
change in character of the sonic point, which becomes a nodal point
for $k>0.89$, rather than a focal point. \citet{Harada2} has also
found that the critical solution becomes unstable to a ``kink''
(discontinuous at the sonic point of the background solution) mode for
$k>0.89$, but because it is not smooth it does not seem to have any
influence on the numerical simulations of collapse. On the other hand,
the limit $k\to 1$ leading to the stiff EOS $P=\rho$ is singular in
that during evolution the fluid 4-velocity can become spacelike and
the density $\rho$ negative. The stiff fluid equations of motion are
in fact equivalent to the massless-scalar field, but the critical
solutions can differ, depending on how one deals with the issue of
negative density \citep{BradyChoptuikGundlachNeilsen}. Summarising, it
is possible to construct the Evans--Coleman CSS critical
(codimension-1) solution for all values $0<k<1$. This solution can be
identified in the general classification of CSS perfect-fluid
solutions as the unique spacetime that is analytic at the center and
at the sound cone, is ingoing near the center, and outgoing everywhere
else \citep{CarrColey2,CarrColeyGoliathNilssonUggla,CarrGundlach}.
There is even a Newtonian counterpart of the critical solution: the
Hunter~(a) solution \citep{HaradaMaeda}.  \citet{Alvarez-Gaume0804}
have calculated $\gamma$ using perturbation theory for the spherically
symmetric perfect fluid with $P=k\rho$ for $k\lesssim 1/4$ in
$d=5,6,7$ dimensions.

$P=k\rho$ is the only EOS compatible with exact CSS
(homothetic) solutions for perfect fluid collapse \citep{CahillTaub}
and therefore we might think that other equations of state would not
display critical phenomena, at least of type~II. 
\citet{NeilsenChoptuik} have given evidence that for the ideal
gas EOS $P=k\rho_0\epsilon$ (where
$\rho_0$ is the rest mass density and $\epsilon$ is the internal
energy per rest mass unit) the black hole threshold also contains a
CSS attractor, and that it coincides with the CSS exact critical
solution of the ultrarelativistic case with the same $k$. This is
interpreted a posteriori as a sign that the critical CSS solution is
highly ultrarelativistic, $\rho:=(1+\epsilon)\rho_0 \simeq
\epsilon\rho_0\gg \rho_0$, and hence rest mass is
irrelevant. \citet{Novak} has also shown in the case $k=1$, or
even with a more general tabulated EOS, that type~II critical
phenomena can be found by velocity-induced perturbations of static TOV
solutions. A thorough and much more precise analysis by 
\citet{Noble, NobleChoptuik} of the possible collapse scenarios of the stiff
$k=1$ ideal gas has confirmed this surprising result, and again the
critical solution (and hence the critical exponent) is that of the
ultrarelativistic limit problem.  Parametrizing, as usual, the TOV
solutions by the central density $\rho_{\mathrm{c}}$, they find that for
low-density initial stars it is not possible to form a black hole by
velocity-induced collapse; for intermediate initial values of
$\rho_{\mathrm{c}}$, it is possible to induce type~II criticality for large
enough velocity perturbations; for large initial central densities
they always get type~I criticality, as we might have anticipated.

\citet{Noble} has also investigated the evolution of a perfect fluid
interacting with a massless scalar field indirectly through
gravity. By tuning of the amplitude of the pulse it is possible to
drive a fluid star to collapse.  For massive stars type~I criticality
is found, in which the critical solution oscillates around a member of
the unstable TOV branch. For less massive stars a large scalar
amplitude is required to induce collapse, and the black hole threshold
is always dominated by the scalar field DSS critical solution, with
the fluid evolving passively. \new{\citet{NobCho16} perturb a static
  stiff fluid star by both an initial velocity and a scalar field, and
  find both type~I and type~II critical collapse in this wider
  parameter space.}

\subsubsection{Cosmological applications}
\label{section:cosmological_applications}

In the early universe, quantum fluctuations of the metric and matter
can be important, for example providing the seeds of galaxy
formation. Large enough fluctuations will collapse to form primordial
black holes. \citet{NiemeyerJedamzik} first noted that, as large
quantum fluctuations are exponentially more unlikely than small ones,
$P(\delta)\sim e^{-\delta^2}$, where $\delta$ is the density contrast
of the fluctuation, one would expect the spectrum of primordial black
holes to be sharply peaked at the minimal $\delta$ that leads to black
hole formation, giving rise to critical phenomena. See
also \citep{GreenLiddle,Yokoyama,KuhRamSan16}.

An approximation to primordial black hole formation is a spherically
symmetric distribution of a radiation gas ($P=\rho/3$) with
cosmological rather than asymptotically flat boundary conditions. In
\citet{NiemeyerJedamzik, NiemeyerJedamzik2} type~II critical phenomena
were found, which would imply that the mass of primordial black holes
formed are much smaller than the naively expected value of the mass
contained within the Hubble horizon at the time of collapse. The
boundary conditions and initial data were refined in
\citet{HawkeStewart, MuscoMiller}, and a minimum black hole mass of
$\sim 10^{-4}$ of the horizon mass was found, due to matter accreting
onto the black hole after strong shock formation. However, when the
initial data are constructed more realistically from only the growing
cosmological perturbation mode, no minimum mass is found
\citep{PolnarevMusco, MuscoMillerPolnarev}. \new{Other values of $k$
  in the ultrarelativistic equation of state $P=k\rho$ are considered
  in \citep{Mus12}.}

\subsubsection{Very soft equations of state}
\label{section:verysoft}

\citet{OriPiran,OriPiran2} have pointed out that there exists a
different CSS perfect fluid solution for $0<k<0.036$ (generalising the
Larston-Penston solution of Newtonian fluid collapse), which has a
naked singularity for $0<k<0.0105$. \new{More specifically, the
  Penrose diagram given in Fig.~16 of \citet{OriPiran2} shows a Cauchy
  horizon followed by an event horizon (both outgoing null), an
  apparent horizon and a curvature singularity (both spacelike). The
  regular centre, Cauchy horizon, event horizon, apparent horizon and
  future spacelike branch of the singularity are linked by an ingoing
  null branch of the singularity. Radial light rays starting at this
  null singularity between the Cauchy horizon and the event horizon
  reach infinity (if the CSS solution is suitably matched to an
  asymptotically flat one), so this part of the null singularity is
  globally naked, while radial light rays emerging from the null
  singularity inside the event horizon end again at the future
  spacelike part of singularity. The same spacetime diagram had
  previously been given in \citet{EarSma79}. All possible Penrose
  diagrams of CSS fluid solutions are listed in \citet{CarrGundlach}.
  Existence of this solution has been proved much more recently by
  \citet{GuoHadJang21b}, see also Sec.~\ref{section:mathfluidGR}.

  \citet{Harada} and \citet{HaradaMaeda} have shown numerically that
  this solution is an attractor, and that it} has no growing
perturbative modes in spherical symmetry. That this solution is an
attractor has also been confirmed in more accurate numerical time
evolutions by \citet{Snajdr}. This seems to violate cosmic censorship,
as generic spherical initial data would create a naked singularity.

Naked singularities were already known to be generic in spherical dust
collapse, see \citet{GriPol09} and \citet{OrtSar11} for
reviews. \new{Dust collapse is not generally considered relevant to}
the cosmic censorship conjecture as spherical dust generically reaches
infinite density even in flat spacetime, so the singularity can be
ascribed to the matter. For $0<k\lesssim 0.105$ this objection does
not apply. However, in spherical symmetry(!) the formal limit $k\to 0$
of the field equations in polar-radial coordinates is not dust coupled
to GR but an isothermal fluid coupled to Newtonian gravity
\citep{HaradaMaeda2, HaradaMaeda3}. One might therefore formulate
cosmic censorship to also exclude matter models that become singular
when coupled to Newtonian gravity.

\new{However, \citet{Gundlach_critfluid2} has shown that {\em any} regular
CSS spherical perfect fluid solution with $0<k<1/9$ has an
analytically known, unstable $l=1$ odd-parity mode, see also
Sec.~\ref{section:nonsphericalfluid}. Therefore the Larston-Penston
spherically symmetric naked singularity solution must be unstable to
rotating perturbations. This does of course not imply that such
perturbed solutions may not have another, nonspherical, naked
singularity structure.

\subsubsection{Perfect fluid in 2+1 dimensions}
\label{section:2+1fluid}

\citet{GunBou20} constructed all rigidly rotating perfect fluid stars
in 2+1 dimensions with negative cosmological constant, for arbitrary
barotropic equation of state $P=P(\rho)$, mass and spin. These
families contain stable and unstable branches, just as in 3+1.
\citet{GunBou21} noted that in circular symmetry in 2+1 not only the
mass $M$ but also the angular momentum $J$ are derived from a
conserved current, in analogy to the Kodama mass\footnote{\new{A
    spherically symmetric spacetime, in any dimension, admits a
    conserved mass function that we call the Kodama mass, see
    \citet{Kod79,Kin21}. The Kodama mass in 3+1 is the same as the
    Misner-Sharp mass or the Hawking mass (on symmetry 2-spheres).}}
$m$ in spherical symmetry in any dimensions. This considerably
simplifies numerical time evolutions with angular momentum. Bourg and
Gundlach investigated the collapse threshold in time evolutions with
the linear equation of state $P=k\rho$, without rotation in
\citep{BouGun21a} and with rotation in \citep{BouGun21b}.  For
$k\lesssim 0.42$ they found type-I critical collapse with lifetime
scaling. The critical solution is a static star on the unstable
branch found in \citep{GunBou20}. For $k\gtrsim 0.43$ they found
type-II critical phenomena with scaling of the mass $M$ and spin $J$
of the first MOTS. Here the critical solution is a star on the
unstable branch that contracts quasistatically. If angular momentum is
present, the critical exponent for $|J|$ is smaller than for $M$, so
$|J|/M$ increases as $p\to p_*$. As $|J|\to M$ with increasing
fine-tuning, scalings stops, so only subcritical black holes are
formed in critical collapse. Dynamically, the critical solution stops
contracting as it reaches extremality.

}

\subsection{Collisionless matter}

A cloud of collisionless particles can be described by the Vlasov
equation, i.e., the Boltzmann equation without collision term.
This matter model differs from field theories by having a much larger
number of matter degrees of freedom: The matter content is described
by a statistical distribution $f(x^\mu, p^\mu)$ on the point particle
phase space, instead of a finite number of fields $\phi(x^\mu)$.
When restricted to spherical symmetry, individual particles move
tangentially as well as radially, and so individually have angular
momentum, but the stress-energy tensor averages out to a spherically
symmetric one, with zero total angular momentum. The distribution
$f$ is then a function $f(t,r,p^r,L^2)$ of radius, time, radial
momentum and (conserved) angular momentum $L$.

Several numerical simulations of critical collapse of collisionless
matter in spherical symmetry have been published to date.
\citet{ReinRendallSchaeffer} find that black hole formation turns on
with a mass gap that is a large part of the ADM mass of the initial
data, and this gap depends on the initial matter condition.  No
critical behaviour of either type~I or type~II was
observed. \citet{OlabarrietaChoptuik} find evidence of a metastable
static solution at the black hole threshold, with type~I scaling of
its lifetime as in (\ref{typeIscaling}). However, the critical
exponent depends weakly on the family of initial data, ranging from
5.0 to 5.9, with a quoted uncertainty of 0.2. Furthermore, the matter
distribution does not appear to be universal, while the metric seems
to be universal up to an overall rescaling, so that there appears to
be no universal critical solution. More precise computations by
\citet{Stevenson}, using finite volume methods, have confirmed the
existence of static intermediate solutions and non-universal scaling
with exponents ranging now from 5.27 to 11.65.  Remarkably no type~II
scaling phenomena has been discovered.  The investigation of
\citet{AndreassonRein} with massive particles has confirmed again the
existence of a mass gap and the existence of metastable static
solutions at the black hole threshold, though there is no estimation
of the scaling of their life-times. More interestingly, they show that
the subcritical regime can lead to either dispersion or an oscillating
steady state depending on the binding energy of the system. They also
conclude, based on perturbative arguments, that there cannot be an
isolated universal critical solution.
 
\new{

Motivated by the non-observation of type-II critical collapse with
massive particles, \citet{vlasov1} have constructed CSS spherically
symmetric solutions of Einstein-Vlasov with massless particles that
are generic by function counting. For massless particles, the
stress-energy and hence spacetime depend only on an integral of the
Vlasov distribution function over the absolute value of the momentum,
thus reducing the effective momentum variables to one (which in
spherical symmetry one can take to be $p^r/L$).  

\citet{RenVel10} have
given an existence proof of ``dustlike'' CSS solutions of massless
Einstein-Vlasov in spherical symmetry, defined by $f(x^\mu,p^\mu)$
having support only on a hypersurface. In \citet{RenVel16} these
solutions are truncated to make them asymptotically flat, and the
resulting causal structure is examined. The conformal diagram takes
the form of a triangle, bounded by an initial Cauchy surface $t=0$, a
regular centre $r=0$, and future null infinity $r=\infty$. The only
singular point is the top vertex where the regular centre and future
null infinity meet: this is not future timelike infinity but a
singular point. It can be reached by timelike geodesics, such as
$r=0$, in finite time, but not by any null geodesic. Future null
infinity extends only to finite affine parameter, truncated not by a
Cauchy horizon but by the singularity itself. The authors call this a
``veiled singularity''.

\citet{AkbCho14} carried out numerical time evolutions in spherical
symmetry with massless particles. They found type~I critical phenomena
with and approximately universal critical solution, but also found
that all static solutions are codimension-one stable and sit on the
threshold of collapse.
To understand this apparent contradiction, Gundlach wrote down generic
static solutions \citep{Gun16}, and their perturbation equations in
Hamiltonian form \citep{Gun17}, and calculated specific solutions and
their perturbation spectrum numerically. In this framework he
conjectured that the apparently universal critical solution
of \citet{AkbCho14} is singled out by the way the threshold solution
is approached in the evolution of critical {\em smooth} initial data.
}

\subsection{Quantum effects}

Type~II critical phenomena provide a relatively natural way of
producing arbitrarily high curvatures, where quantum gravity effects
should become important, from generic initial data. Approaching the
Planck scale from above, one would expect to be able to write down a
critical solution that is the classical critical solution
asymptotically at large scales, as an expansion in inverse powers of
the Planck length (see Sect.~\ref{section:universalityclasses}).

Black hole evolution in semiclassical gravity has been investigated in
1+1 dimensional models which serve as toy models for spherical
symmetry (see \citet{Giddings_BH} for a review). The black hole
threshold in such models has been investigated in \citet{ChibaSiino,
  AyalPiran, StromingerThorlacius, Kiem, ZhouKirstenYang,
  PelegBoseParker, BoseParkerPeleg} \new{and
  \citet{AshPreRam2010}}. In some of these models, the critical
exponent is $1/2$ for kinematical reasons.

In \citet{BradyOttewill} a 3+1-dimensional but perturbative approach
is taken. The quantum effects then give rise to an additional unstable mode
with $\lambda=2$. If this is larger than the classical positive Lyapunov
exponent $\lambda_0$, it will become the dominant perturbation for
sufficiently good fine-tuning, and therefore sufficiently good
fine-tuning will reveal a mass gap. The mass gap is found also in
numerical evolutions of a spherical scalar field in 3+1 dimensions
with the semiclassical equations obtained in the framework of
``singularity resolution'' in loop quantum gravity \citep{Husain,
  ZiprickKunstatter0902}.
  
\new{\citet{BenGamLeh20} propose an effective semiclassical theory
that, at least in spherical symmetry, does {\em not introduce} a
length scale into the field equations, and indeed find type~II scaling
with $\gamma\simeq 0.37$. \citet{BerSadZho21} propose another
semiclassical theory, equivalent to the classical theory with five
additional minimally coupled massive scalar fields, three of which
have the opposite sign for the kinetic and mass term in the Lagrangian
(ghosts). The initial data chosen remain unclear. They find type~II
scaling with $\gamma\simeq 0.38$, and little difference to the purely
classical results. This is consistent with a mass term becoming
irrelevant on small scales, but it is not clear if and why the ghost
fields remain stable.}

\subsection{Toy models}

\new{A number of papers have studied specific one-parameter families of
exact solutions, say of the spherical Einstein-scalar system, that
include dispersing and black-hole solutions, and may show scaling
along the family. What they have in common is that the threshold
solution of the one-parameter family is not a critical solution in the
full infinite-dimensional phase space.}

The Roberts 1-parameter family of 3+1 solutions \citep{Roberts} has
been analysed along this line in \citet{Oshiro_Roberts, Brady_Roberts,
WangOliveira, Kiem}. This family contains black holes whose masses
(with a suitable matching to an asymptotically flat solution) scale as
$(p-1)^{1/2}$ for $p\gtrsim 1$, but such a special family of solutions
has no direct relevance for collapse from generic data. Its
generalisation to other dimensions has been considered in
\citet{Frolov2}. A fully analytic construction of all (spherical and
nonspherical) linear perturbations of the Roberts solution by
\citet{Frolov, Frolov3} has shown that there is a continuum of unstable
spherical modes filling a sector of the complex plane with $\real
\lambda\ge 1$, so that it cannot be a critical
solution. Interestingly, all nonspherical perturbations decay.
\citet{Frolov4} has speculated that the critical ($p=1$) Roberts
solution, which has an outgoing null singularity, plus its most
rapidly growing (spherical) perturbation mode would evolve into the
Choptuik solution, which would inherit the oscillation in $\tau$ with
a period $4.44$ of that mode.

A similar transition within a single 1-parameter family of
solutions has been pointed out in \citet{Oliveira} for the Wyman
solution \citep{Wyman}.

\new{\citet{HusMarNun94} consider another family of solutions, with a
timelike conformal Killing vector.}

\citet{Hayward,Haywarderratum} and 
\citet{ClementFabbri1,ClementFabbri2} have also proposed
critical solutions with a null singularity, and have attempted to
construct black hole solutions from their linear perturbations. This
is probably irrelevant to critical collapse, as the critical spacetime
does not have an outgoing null singularity. Rather, the singularity is
naked but first appears in a point. The future light cone of that
point is not a null singularity but a CH with finite curvature.

Other authors have attempted analytic approximations to the Choptuik
solution.  \citet{Pullin_Chop} has suggested describing critical
collapse approximately as a perturbation of the Schwarzschild
spacetime. \citet{PricePullin} have approximated the
Choptuik solution by two flat space solutions of the scalar wave
equation that are matched at a ``transition edge'' at constant
self-similarity coordinate $x$. The nonlinearity of the gravitational
field comes in through the matching procedure, and its details are
claimed to provide an estimate of the echoing period $\Delta$.

\new{Other self-gravitating systems that have been studied as toy
models for critical collapse can be characterised as showing a
transition between dispersion and black hole formation, but in a setup
that has only a finite number of degrees of freedom, in contrast to a
field theory. It is characteristic for such toy models that the
critical exponent $\gamma$, suitably defined, is $1/2$.}

\citet{PelegSteif} have investigated the collapse of a dust ring. In
both cases the mass of the black holes is a known function of the
parameters of the initial condition, giving a ``critical exponent''
1/2, but no underlying self-similar solution is involved.

\citet{MahajanHaradaJoshiNakao} expand the initial data
for Einstein clusters in powers of the radius and, assuming that there
are no shell crossings, find a mass scaling exponent of 3/2 for two
such expansion coefficients. Universality is not demonstrated, and so
the connection with a CSS solution discussed by 
\citet{HaradaMahajan} is unclear.

\citet{FrolovV} and \citet{FrolovVetal} consider a stationary
2+1-dimensional Nambu--Goto membrane held fixed at infinity in a
stationary 3+1-dimensional black hole background spacetime. The
induced 2+1 metric on the membrane can have wormhole, black hole, or
Minkowski topology. The critical solution between Minkowski and black
hole topology has 2+1 CSS. The mass of induced
black hole metrics scales with $\gamma=2/3$, superimposed with a
wiggle of period $3\pi/\sqrt{7}$ in $\ln p$. The mass scaling is
universal with respect to different background black hole metrics, as
they can be approximated by Rindler space in the mass scaling limit.

\citet{HorowitzHubeny} and \citet{BirminghamAdS} have attempted to
calculate the critical exponent in toy models from the adS-CFT
correspondence. {\'A}lvarez-Gaum{\'e} et al.\ have attempted to use
the adS-CFT correspondence for calculating $\gamma$ from the QCD side
for the spherical massless scalar field in 5 dimensions
\citep{Alvarez-Gaume}, and the spherical perfect fluid with $P=k\rho$
for $k=1/(d-1)$ in $d=5,6,7$ dimensions \citep{Alvarez-Gaume0804}.

\new{

Spherical dust collapse has been widely used as as a toy model for
naked singularity formation, even though, as mentioned above, dust
forms singularities already without gravity. However, the solution is
known in closed form, see \citet{GriPol09} and \citet{OrtSar11} for
reviews. \citet{WatNol09,DufNol11a, DufNol11b} investigated
(nonspherical) linear perturbations of CSS spherical dust collapse,
and found that the Cauchy horizon is unstable to even-parity
perturbations.

\citet{AlvGomSab08} study the formation of trapped surfaces in the
head-on solution of two gravitational shock waves in axisymmetry in
4+1 and 5+1, with vanishing and negative cosmological constant. Here a
shock wave means that the metric contains $\delta(u)$ terms, where $u$
is retarded time. They find that in 4+1 the trapped surface first
appears with finite size, while in 5+1 it appears with zero size and
scales with $\gamma=1/2$ in 4+1 and with $\gamma=1$ in 5+1.

}  

\section{Beyond spherical symmetry}
\label{section:nonspherical}

Studies of critical collapse should go beyond spherical
symmetry (and in the first instance to axisymmetry) for three reasons:

\begin{itemize}

\item Weak gravitational waves in vacuum general relativity can focus
  and collapse. The black hole threshold in this process shows what in
  critical phenomena in gravitational collapse is intrinsic to gravity
  rather than the matter model. 

\item Black holes are characterised by charge and angular momentum as
  well as mass. Angular momentum is the more interesting of the two
  because it is again independent of matter, but cannot be studied in
  spherical symmetry.

\item Angular momentum resists collapse, but angular momentum in the
  initial data is needed to make a black hole with angular
  momentum. Therefore it is an interesting question to ask what
  happens to the dimensionless ratio $J/M^2$ at the black hole
  threshold. 

\end{itemize}

In the following we review what has been done so far. 

\subsection{Nonspherical scalar field}
\label{section:nonsphscalar}

\subsubsection{Real scalar field in axisymmetry}

\citet{MartinGundlach} have analysed all nonspherical mode
perturbations of the scalar field critical solution by solving a
linear eigenvalue problem with an ansatz of regularity at the centre
and the SSH. They find that the only growing mode is the known
spherical one, while all other spherical modes and all non-spherical
modes decay. This \new{mode stability result suggests that the
  Choptuik solution remains the critical solution in a finite
  neighbourhood of spherical symmetry, including also the presence of
  weak gravitational waves.}

\citet{Choptuik_axisymmetry} carried out axisymmetric time evolutions
for the massless scalar field. They found that in the limit of
fine-tuning generic axisymmetric initial data the spherically
symmetric critical solution is approached at first but then deviates
from spherical symmetry and eventually develops two centres. \new{
This suggested that the critical solution had an unstable $l=2$ mode.

The apparent contradiction was resolved by further numerical time
evolutions. \citet{Bau18}, working in spherical polar coordinates,
demonstrated decay of sufficiently small nonspherical perturbations in
quantitative agreement with \citet{MartinGundlach}, but for more
non-spherical initial data also confirmed the observations
of \citet{Choptuik_axisymmetry} for more non-spherical families of
initial data. \citet{MarCorRut24} and \citet{CorRenRut23}, with an
adaptive mesh and pseudospectral code, investigated critical collapse
of a complex massless scalar field in axisymmetry. They therefore
evolved different families, but agree with the qualitative picture
just described. In particular, the critical solution is again the real
Choptuik solution (up to a constant complex
phase). Finally, \citet{GunBauHil24} investigated axisymmetric (real)
scalar field critical collapse using null coordinates (and so imposing
initial data on an outgoing null cone), but were not able to examine
the black hole threshold for families non-spherical enough to
bifurcate.

The results of \citep{Choptuik_axisymmetry,Bau18,MarCorRut24} can be
described heuristically in terms of {\em effective} values of $\Delta$
and $\gamma$ that decrease with increasing non-sphericity parameter
$\epsilon$, and a $\lambda$ for the dominant non-spherical
perturbation mode that increases with $\epsilon$, eventually becoming
positive. The families considered in these papers can be sorted by the
changes in $\Delta$ and $\gamma$ in the same order as by the
non-sphericity of $\psi$. By contrast \citep{GunBauHil24} sees
significantly much smaller changes in $\Delta$ and $\gamma$ at the
same non-sphericity.

A direct comparison of non-sphericity between different families of
initial data is provided by the difference between the scalar field
$\psi$ at the pole and equator, in the region of a near-critical
evolution where it is approximated by the Choptuik solution.  This
diagnostic was recorded in \citep{Bau18,MarCorRut24,GunBauHil24} but
not \citep{Choptuik_axisymmetry}. The families of \citep{GunBauHil24}
would, by this ranking, not be expected to show bifurcation (and do
not).

\citet{DepKidSch19} examined massless scalar field critical collapse
without symmetry assumptions, using a spectral discretization in
generalised harmonic gauge and adaptive mesh refinement. They found
the spherical critical solution with the expected scaling and echoing
behaviour, but not the bifurcation observed by
\citep{Choptuik_axisymmetry,Bau18,MarCorRut24}.  This may be because
of the level of fine-tuning achieved: $p-p_*\sim 10^{-15}$
in \citet{Choptuik_axisymmetry} and \citet{GunBauHil24}, $10^{-12}$
in \citet{Bau18}, $10^{-10}$ in \citet{MarCorRut24} and $10^{-6}$
in \citet{DepKidSch19}.

In earlier work, \citet{HeaLag13} had also examined massless scalar
field critical collapse without symmetry assumptions, using the BSSN
formulation. They found $\gamma\simeq 0.37$ but no clear evidence of
echoing. \citet{CloLim16} (see also \citep{Clo17}) examined critical
collapse with a double-well potential, using the BSSN formulation and
adaptive mesh refinement. They found $\gamma\simeq 0.37$ but achieved
fine-tuning only to $p-p_*\sim 10^{-1}$.

}

\subsubsection{Pseudo-rotating scalar fields}

Perturbing the scalar field around spherical symmetry, angular
momentum comes in to second order in perturbation theory. All angular
momentum perturbations were found to decay, and a critical exponent
$\mu\simeq 0.76$ for the angular momentum was derived for the massless
scalar field in \citet{GarfinkleGundlachMartin}. This prediction has
not yet been tested in nonlinear collapse simulations, \new{
and would require going beyond axisymmetry: a real scalar field has
vanishing angular momentum current in axisymmetry and so needs to have
a non-axisymmetric structure going around to admit angular
momentum. However, axisymmetric and spherically symmetric toy models
of a ``rotating'' scalar field have been explored in axisymmetry and
spherical symmetry. We may call these ``pseudo-rotating'', as the
scalar field matter appears to experience a centrifugal force but
there is no spacetime angular momentum.}

\citet{ChoptuikHirschmannLieblingPretorius} considered a complex
scalar field giving rise to an axisymmetric spacetime with angular
momentum, by making the ansatz, in cylindrical polar coordinates,
\begin{equation} \label{m1ansatz}
  \Psi(\rho,z,t,\varphi)=\phi(\rho,z,t)e^{im\varphi}
\end{equation}
where $\phi$ is complex.  This ansatz allows the ratio of
energy density to angular momentum density in the initial data to be
adjusted arbitrarily, including to zero angular momentum for a $\phi$
that is real (up to a trivial constant phase). On the other hand, even
in the absence of angular momentum a purely real $\phi$ obeys a wave
equation with an explicit $m^2/\rho^2$ centrifugal potential. For all
initial data in numerical evolutions with quantum number $m=1$ a real,
non-rotating, DSS critical solution with $\Delta\simeq 0.42$ is
approached, and a Ricci scaling exponent of $\gamma\simeq 0.11$ is
observed in subcritical evolutions.

Far from the black hole threshold $J\sim M^2$ in the final black hole,
but nearer the black hole threshold, $J\sim M^6$, where $J$ and $M$
are measured on the \new{first MOTS}. Note $J/M^2\to 0$ at the
black-hole threshold is compatible with a non-rotating critical solution. \new{A
  simple calculation shows that} $J\sim M^6$ implies $\lambda= -4$ for
the angular momentum mode.

\new{We note in this context that}
\citet{OlabarrietaVentrellaChoptuikUnruh} study a similar system in
spherical symmetry, by arranging $2l+1$ separate scalar fields given
by
\begin{equation}
\phi_{lm}=\psi(t,r)Y_{lm}(\theta,\varphi)
\end{equation}
for $m=-l,\dots l$ with the \emph{same} $\psi(t,r)$ for all values of
$m$ so that the sum of their stress-energy tensors and the spacetime
are spherically symmetric. $\psi$ again sees a centrifugal barrier
$\sim l/r^2$. DSS critical behaviour is found, where the logarithmic
echoing period $\Delta$ and mass scaling exponent $\gamma$ both
decrease approximately exponentially with $l$.

\citet{Lai} has studied type~I critical phenomena for boson (massive
complex scalar field) stars in axisymmetry, the first study of type~I
in axisymmetry. He finds that the subcritical end state is a boson
star with a large amplitude fundamental mode oscillation.

\new{

\subsection{Nonspherical perfect fluid}
\label{section:nonsphericalfluid}

Perfect fluid matter is interesting for the study of critical collapse
because it allows for rotation and the formation of a rotating black
hole in collapse already in axisymmetry, in contrast to the scalar
field and vacuum cases.

The nonspherical perturbations of the spherically symmetric critical
solution for $P=k\rho$ were investigated numerically in a mode ansatz
by \citet{Gundlach_nonspherical}. For $0.49\lesssim k<1$, there are
multiple unstable modes. For $1/9<k\lesssim 0.49$, all nonspherical
perturbations decay. For $0<k<1/9$, precisely one $l=1$ odd-parity
mode that carries angular momentum is unstable, and this mode and the
known spherical mode are the only two unstable modes of the spherical
critical solution.

The axial (odd-parity) $l=1$ perturbations are special in that they
admit an analytic calculation of the entire mode spectrum, and a
closed form expression for the corresponding modes in terms of the
spherically symmetric CSS background solution (which itself is not
known in closed form). In particular one can show
\citep{Gundlach_critfluid2} that the most rapidly growing or least
damped of these modes has the growth rate
\begin{equation}
  \label{lambda1}
  \lambda_1(k)={1-9k\over 3k+3}.
\end{equation}
(A similar result had previously been found by \citet{HanNak97} for
the non-relativistic isothermal fluid coupled to Newtonian
gravity. This corresponds to the limit $k=0$ with $\lambda_1=1/3$.)

With rotation, in the regime $0<k\lesssim 0.49$, for evolutions going
through an approximately CSS phase the mode stability results suggest
the approximation
\begin{equation}
Z\simeq Z_*+Pe^{\lambda_0\tau}Z_0+Qe^{\lambda_1\tau}Z_1+\text{decaying modes},
\end{equation}
where $Z$ represents the entire solution, $Z_*$ the critical solution,
$Z_0$ the unstable spherically symmetric mode, and $Z_1$ the dominant
$l=1$ axial mode (unstable for $0<k<1/9$). $P$ and $Q$ are the
amplitudes of the two modes in this approximately CSS phase and depend
on the initial data. Note that this is a generalisation of
Eq.~(\ref{echoing_region}) above, with $C_0$ now written as $P$ and
including one more perturbation mode.  The theory for the case
$0<k<1/9$, where $\lambda_1>0$, was developed
in \citet{GunBau17}. Here we note only two important features: first,
the black-hole threshold is no longer given exactly by $P=0$. Second,
sufficiently close to the black hole threshold one would generically
expect scaling to break down, that is, one would not expect $M$ and
$J$ to both go to zero at the black hole threshold. Third, one would
expect the scaling laws for the black hole mass and spin to display
universal scaling functions as well as universal critical exponents.

Non-rotating axisymmetric critical collapse of the perfect fluid was
investigated numerically in \citet{BauMon15} for $k=1/3$, and for values of
$k$ between $0.2$ and $0.6$ in \citet{CelBau18}. The complex frequency
of the least damped or most rapidly growing $l=2$ even-parity
perturbation of the critical solution was found to be in agreement
with \citet{Gundlach_nonspherical}. 

Rotating axisymmetric collapse with $k=1/3$ was considered in
\citet{BauGun16}, \citet{GunBau16}. Specifically, the initial data
depended on two parameters $p$ and $q$, where $p$ controls strength
and $q$ initial angular momentum. The scaling laws are in agreement
with the predictions of \citet{Gundlach_angmom,Gundlach_critfluid2}.
This was generalised to the equations of state $k=0.08$, $0.1$, $0.2$,
$0.5$ in \citet{GunBau17}. For $k=0.2$ and $0.5$ there was a again
good agreement with the simple power laws.

For the soft equations of state $k=0.1$ and $0.08$, however, where the
angular momentum mode is unstable, the expected breakdown of scaling
was not observed for the level of fine-tuning to the black hole
threshold that could be achieved. Nor could any non-trivial scaling
functions be fitted to the data. Rather, a purely heuristic
description of the observed functions $M(p,q)$ and $J(p,q)$ could be
given by making $\lambda_1$ depend weakly on $q$ and becoming negative
for large $q$.  A possible explanation for the observed numerical
results could be that at large $q$ there no longer is a phase where
both $Z_0$ and $Z_1$ can be treated as linear perturbations. If so,
then better fine-tuning of initial data with small angular momentum
would be required to see the universal scaling functions.

\subsection{Axisymmetric collisionless matter}
\label{section:axi_Vlasov}

\citet{AmeAndRin20} investigate the threshold of collapse in massive
Einstein-Vlasov in axisymmetry and find type~I lifetime scaling.

\subsection{Axisymmetric vacuum gravity}
\label{section:axi_vacuum}

The simplest context in which to study asymptotically flat vacuum
collapse is to start from regular single-ended axisymmetric Cauchy
data. Symmetry reduction under the Killing
vector~$\xi=\partial/\partial\varphi$ can then be used to suppress one
spatial dimension. In this setting angular momentum moreover
necessarily vanishes, and so generic members of the Kerr family
are excluded as possible end-states of the time evolution. The
geometric subclass of these spacetimes most studied at the time of
writing is defined by the condition that the twist
vector~$\omega_a=\epsilon_{abcd}\xi^b\nabla^c\xi^d$ vanishes, and that
there is a reflection symmetry $z\to-z$ through the equatorial
plane. A spacetime with $\xi=\partial/\partial\varphi$ is twistfree if
and only if $g_{\mu\varphi}=0$ for~$\mu\ne\varphi$.

Two strategies have been used to build initial data for vacuum
critical collapse. In the first, solutions of the linearised vacuum
Einstein equations, often called Teukolsky waves, are used as seed
data within a general constraint solution procedure. The resulting
initial data are often referred to as nonlinear Teukolsky waves or
just Teukolsky waves. The second approach is to use Brill wave
initial data, which are time symmetric and result in a linear
equation for a spatial conformal factor. In the linearised setting
Brill waves reduce to initial data that give rise to time-symmetric
Teukolsky waves.

Vacuum critical collapse was first studied numerically
by~\citet{AbrahamsEvans}, who employed a partially constrained
formulation of GR, together with maximal slicing and a
quasi-isotropic spatial gauge in explicit axisymmetry. They
considered one family of Teukolsky wave initial data consisting of a
mostly incoming pulse with pure~$l=2$ angular dependence. Limited
numerical resolution allowed for tuning to find black holes with
masses only down to $0.2$ of the ADM mass. Even this far from the
threshold, they found power-law scaling of the black hole mass, with
exponent $\gamma\simeq 0.36$. The black hole mass was determined
from the apparent horizon surface area, and the frequencies of the
lowest quasi-normal modes of the black hole. They found tentative
evidence for scale echoing in the time evolution, with
$\Delta\simeq 0.6$, with about three echos seen. In a subsequent
paper \citep{AbrahamsEvans2}, some evidence for universality of the
critical solution, echoing period and critical exponent was given in
the evolution of a second family of Teukolsky wave initial data. In
this family, black hole masses down to $0.06$ of the ADM mass were
achieved. The bounds on $p_*$ of these two papers correspond to fine-tuning
to $|p-p_*|\sim 10^{-4}$ and~$10^{-5}$, respectively.

Building on these early results has proved
difficult. \citet{Alcubierre_Brill} performed evolutions using a
family of Brill waves in a fully 3-dimensional code using the
free-evolution BSSN formulation with maximal slicing and vanishing
shift, but were unable to achieve the same degree of tuning as in
the study of \citep{AbrahamsEvans}. From a slightly different
physical motivation (looking for evidence of cosmic censorship
violation), \citet{GarDun01} used a partially constrained
formulation with maximal slicing and an elliptic condition on the
shift vector to evolve the same family of Brill waves, this time in
explicit axisymmetry. They established coarse, though compatible,
bounds for the black hole threshold for this family. They
additionally investigated prolate Brill waves.

Employing the Geroch decomposition to reduce to axisymmetry,
\citet{Rin05} worked with the same gauge condition
as~\citep{GarDun01} with different partially constrained evolution
schemes. He generalised Brill wave data to include twist, and found
bounds broadly compatible with those of earlier work when treating
identical data. Returning to the theme of \citep{GarDun01}, in later
work \citet{Rin08} introduced a fully constrained formulation with
elliptic gauge conditions and studied highly prolate Brill waves,
finding no evidence for violation of cosmic censorship. The best
degree of tuning for Brill wave initial data was around~$10^{-1}$ in
the studies discussed to this point.

In the era directly after the binary black-hole breakthrough of 2005
\citet{Pre05,BakCenChoi05,CamLouMar05} there was strong motivation to
work with the generalised harmonic or BSSN formulations. \citet{Sor09}
presented an axisymmetric generalised harmonic code and used it to
tune Brill wave initial data to the threshold of collapse in
\citep{Sor10}, but obtained results incompatible with earlier and
subsequent work.

To see what could be achieved using the moving-puncture gauge,
\citet{HilBauWey13} performed evolutions of both Brill and Teukolsky
waves using two BSSN codes in full 3D. One used Cartesian coordinates
and fixed mesh-refinement, whilst the other used a reference metric
approach to spherical polar coordinates. Comparable tuning was
achieved for families of Brill waves treated earlier in the
literature, and with compatible bounds. Teukolsky wave data were found
to be easier to evolve numerically, but a combination of a lack of
resolution, constraint violation and perhaps coordinate singularities
prevented precise bisection to the threshold for both Brill and
Teukolsky data.

In an attempt to overcome these issues, \citet{HilWeyBru15} prepared
a 3D multidomain pseudospectral code using a first order reduction
of the generalised harmonic gauge formulation with constraint
preserving outer boundary conditions, using the cartoon method for
symmetry reduction where possible and a bespoke axisymmetric
apparent horizon finder. In a first application \citep{HilWeyBru17}
they too evolved the family of Brill data from
\citep{Alcubierre_Brill}, and experimented with gauge source
functions for the evolution. For this single family they were able
to tune to an interval around~$10^{-6}$, observing tentative
evidence for power-law scaling with a wiggle in the maximum of the
Kretschmann scalar. They found on the subcritical side that close to
the threshold the maximum of the Kretschmann scalar occurred on the
symmetry axis but away from the origin. On the supercritical side
they found that disjoint apparent horizons formed around these
peaks. In their best-tuned supercritical data they found black holes
with $0.05$ of the ADM mass at the instant of formation.  Their
bisection search was ultimately hampered by an inability to classify
spacetimes very close to the threshold.

Independently, \citet{KhiLed18} investigated different gauge
conditions in the BSSN formulation, employing the cartoon method for
symmetry reduction, and using fixed mesh-refinement. For the purposes
of code validation they studied Brill waves, comparing quantitatively
with earlier results with a coarse bound for the threshold
amplitude. They found that by using gauge conditions that mimicked
maximal slicing they could avoid the problems reported in
\citep{HilBauWey13} with the canonical moving-puncture
gauge. Following on, \citet{LedKhi21} evolved two families of
Teukolsky wave initial data and two of Brill data. They made an
approach to the construction of nonlinear Teukolsky data very similar
to that of \citet{AbrahamsEvans}, and discovered that for certain
fixed amplitudes, their formulation of the constraints admitted two
solutions. Tuning to the threshold, they obtained bounds with an
interval of around~$|p-p_*|\sim 10^{-7}$ in their best case. They
invested less effort on Brill data, tuning to
the~$|p-p_*|\sim 10^{-1}$ level in their first family, and~$10^{-6}$
in the second. For individual families they found power-law scaling
(with superposed wiggles) of the Kretschmann scalar with
$|p-p_*|^{-4\gamma}$, but with the scaling exponent and ``wiggle''
period varying from family to family. Examining a scalar quantity
defined geometrically in axisymmetry, they found irregularly repeating
features in the vicinity of maxima within their best-tuned individual
spacetimes from each family and, remarkably, that local cross-sections
through these maxima appeared to be universal. More details can be
found in \citep{Khi21}.

Following up on this, \citet{RenCorHil23} presented an update to the
pseudo-spectral code of \citet{HilWeyBru15} to include adaptive
mesh-refinement. \citet{SuaRenCor22} used this tool to examine six
families of Brill waves. They found compatibility with earlier
published results and evidence for power-law scaling with
non-universal exponents. They found evidence for universal local
features in the Kretschmann scalar akin to those observed in
\citep{KhiLed18}.

Returning to the potential coordinate singularities observed in
\citet{HilBauWey13} with moving-puncture gauge, \citet{BauHil22}
investigated ``shock-avoiding'' \citep{Alc96,AlcMas97} slicing
conditions. These were then used by \citet{BauGunHil23} to treat two
families of nonlinear Teukosly wave data, built respectively from
pure $l=2$ and $l=4$ seed data. With a modest bracketing down to
three or four significant figures, they found that the threshold
solutions from both families displayed approximate discrete
self-similarity but with distinct parameters~$\Delta\simeq0.5$
and~$\Delta\simeq0.1$, in contradiction to the existence of a
universal critical solution.

\citet{BauBruCor23} then made a quantitative comparison of the data
from all three independent codes now achieving a similar degree of
tuning. They compared evolutions of two families of Brill data that
had been treated by all three groups and found good quantitative
agreement for the scaling of the Kretschmann scalar. Interpolating
spacetime scalars from their different evolution coordinates to the
same affinely parameterised outgoing null cones emanating from the
central wordline, they also found excellent agreement over an extended
region in their best tuned data.

In summary, although there is room for improvement in the degree of
tuning, all the evidence presently suggests that power-law scaling
(with a wiggle) is manifest for individual families, but the critical
exponent and echoing period are not universal across families. This
suggests that there is no universal threshold axisymmetric vacuum
solution. Threshold solutions may or may not exhibit approximate
DSS. Even if they do not, however, {\em local} curvature features do
repeat, albeit irregularly.
  
\citet{Ros25} reports on initial tests of a formulation of twist-free
axisymmetry that is closer to that of~\citep{AbrahamsEvans} than
previous ones, and in particular is fully constrained (the only
evolution equation is a single wave equation in first-order form). He
also suggests a resolution of the difficulty in recreating the initial
data of \citep{AbrahamsEvans} reported in \citep{LedKhi21,Khi21},
through a different interpretation of the notation of
\citep{AbrahamsEvans}.

\subsection{Axisymmetric electromagnetism}

Electrovacuum solutions are interesting for critical collapse because,
as in vacuum GR, there are no dynamical spherically symmetric
solutions. However, when the collapse is driven predominantly by the
electromagnetic field, it is plausible that the spacetime will be
easier to treat numerically than in vacuum collapse. Axisymmetric
electrovacuum spacetimes close to the threshold of collapse have been
investigated by~\citet{BauGunHil19}, \citet{MenBau21}, and
\citet{ReiCho23}.

In the first two of these papers the field equations are solved in
twist-free axisymmetry in spherical polar coordinates using a
reference metric version BSSN formulation. The source-free Maxwell
equations are reduced to a single wave equation
for~$A_\varphi=A_a\xi^a$, where here~$A^a$ is the vector potential
and~$\xi^a=(\partial /\partial \varphi)^a$ is the Killing vector.

\citet{BauGunHil19} evolved two families of initial data seeded by
dipolar~($l=1$) axisymmetric electromagnetic waves. Tuning to the
threshold, they found a single center of collapse and scaling of the
maximal energy
density~$\rho_\textrm{max}\simeq (p-p_\star)^{-2\gamma}$
with~$\gamma\simeq0.145$. Near the threshold they observed
qualitative similarity of the solutions from each family and {\it
approximate} DSS with~$\Delta\simeq 0.55$.

Following up on this, \citet{MenBau21} considered quadrupolar
$(l=2)$ seed data. Near the threshold of collapse, the quadrupolar
data still exhibit scaling and approximate DSS, but produce two
centers of collapse, and thus differ qualitatively from the dipolar
case. They found that quadrupolar initial data give rise to
different power-law exponents~$\gamma\simeq0.11$
and~$\gamma\simeq0.145$ (approximate) echoing
periods~$\Delta\simeq0.55$ and~$\Delta\simeq0.3$.

Working again in explicit axisymmetry but instead with a conformal
decomposition of the Z4 formulation, cylindrical polar coordinates,
and a formulation of the Maxwell equations with the electric and
magnetic fields evolved, \citet{ReiCho23} evolved the same families
of initial data as \citep{BauGunHil19,MenBau21}. They agreed with
the earlier results on all but the last point, finding instead
universality of the power-law exponents across all families. They
suggest that the disagreement is caused by the use of cylindrical
rather than spherical polar coordinates, which are better adapted to
fine features appearing away from the origin.

}

\subsection{Black string breakup}

\new{

A black string in 4+1 or higher spacetime dimension is subject to
the Gregory-Laflamme instability \citep{GreLaf93}, where any
perturbation that breaks the $z$-translation symmetry leads to a
pinch-off. \citet{HorMae01} conjectured that a new stable string
endstate is reached, but \citet{LehPre10} (see also
\citep{LehPre11}) showed by direct numerical time evolution of the
vacuum Einstein equations in 4+1 in $SO(3)$ symmetry with periodic
boundary conditions in $z$ that this is not so. Rather, the string
becomes a sequence of black holes joined smoothly by black string
segments. Each of these is subject to the Gregory-Laflamme
instability again, and the process starts again. Moreover, the
instability timescale is proportional to the black string radius, so
the process is self-similar, leading to arbitrarily large curvatures
(on the surface of arbitrarily thin black strings) in finite time.

Note that all (straight, infinitely extended) black strings are
unstable, so this process is an attractor. It is, however, not
universal: the spacetime structure depends sensitively on the
initial data.

\citet{FigFra23} carried out independent numerical time
evolutions. They confirmed the general results of \citet{LehPre10}.
In particular, they found that the minimum over $z$ of the string
area radius depends on time approximately as
\begin{equation}
R_\text{min}\simeq a(t-t_c).
\end{equation}
where $a$ is a dimensionless constant and $t=t_c$ is the singularity
(``pinch-off'') time. $a\simeq 0.005$ appears to be universal and
independent of the period $L$ in the $z$-direction.

As one would expect, the scenario of \citet{LehPre10} occurs also
for sufficiently thin closed black strings \citet{FigKun16}. A
similar instability applies to spinning Myers-Perry black holes in
5+1 and higher dimensions, and results in the same scenario. This
was shown by direct numerical time evolution in 5+1 dimensions in
$SO(3)$ symmetry in \citet{FigKun16}. \citet{BanFig19} extended this
to 6+1 in $SO(4)$ symmetry, and conversely showed that Myers-Perry
black holes are stable in 4+1 in $SO(2)$ symmetry.

}

\subsection{Two-body collisions}

\new{Type~I critical phenomena at the collapse threshold in the
  collision of two neutron stars were investigated by
  \citet{JinSuen,Wan10,Wan11}.} The matter is a perfect fluid with
``Gamma law'' EOS $P=(\Gamma-1)\rho_0\epsilon$, where $\Gamma\simeq 2$
is a constant, $P$ is the pressure, $\rho_0$ the rest mass density, and
$\epsilon$ the internal energy per rest mass, so that
$\rho:=\rho_0(1+\epsilon)$ is the total energy density. The initial data are
constructed with $P=k\rho_0^\Gamma$ for a constant $k$, which
corresponds to the ``cold'' (constant entropy) limit of the Gamma law
EOS. The initial data correspond to two identical stars which have
fallen from infinity. (The evolution starts at finite distance, with
an initial velocity calculated in the first post-Newtonian
approximation). The entire solution is axisymmetric with an additional
reflection symmetry that maps one star to the other.  The limited
numerical evidence is compatible with type~I critical phenomena, with
the putative critical solution showing oscillations with the same
period in the lapse and Ricci scalar.

Interesting numerical evidence for critical phenomena in the grazing
collision of two black holes has been found by
\citet{PretoriusKhurana}.  The threshold in initial data space is
between data which merge immediately and those which do not (although
they will merge later for initial data which are bound). The critical
solution is a circular orbit that loses 1 to 1.5\% of the total
energy per orbit through radiation. On both sides of the threshold,
the number of orbits scales as
\begin{equation}
  \label{Pretoriusn}
  n\simeq -\gamma\ln|p-p_*|
\end{equation}
for $\gamma\simeq 0.31\mbox{\,--\,}0.38$. The authors conjecture that
the total energy loss and hence the number of orbits is limited only
by the irreducible mass of the initial data, \new{which is much larger
  than the rest masses of the two black holes}. In particular, for
highly boosted initial data \new{where, intuitively}, the total energy
is dominated by kinetic energy, almost all can be converted into
gravitational radiation.

\new{The authors liken their evolutions to} the unstable circular
geodesics in the spacetime of the hypothetical rotating black hole
that would result if merger occurred promptly. These give orbital
periods, and their linear perturbations give a critical exponent, in
rough agreement with the numerical values for the full black hole
collision. \new{They conjecture} that these phenomena generalise to
generic initial data with unequal masses, and black hole spins which
are not aligned. \new{By contrast, \citet{Levin} claims, based on a
  2nd order post-Newtonian approximation,} that the threshold of
immediate merger is fractal if the spins are not aligned, and that the
system is therefore chaotic.

\citet{SperhakeCardosoPretoriusBertiHindererYunes} push the numerics
towards higher energies emitted and conjecture that the merger of
nonspinning black holes can in principle yield a black hole
arbitrarily close to extremal Kerr. \new{See also
  \citet{ShibataOkawaYamamoto} and \citet{GolBru12} for
  ultrarelativistic collisions of black holes, \citet{EasPre2012} and
  \citet{RezKen13} for perfect fluid stars, and \citet{ChoPre2009} for
  solitons.}

\new{

The problem has been addressed in the self-force approximation
in \citet{GunAkcBar12}. For related conjectures, see
also \citet{Pag22}.

A toy model for two-body collisions with non-zero impact parameter is
provided by two compact objects coupled to Einstein gravity in 2+1
dimensions. In contrast to 3+1 and higher dimension, the exterior
metric of an uncharged compact object in 2+1 depends only on its mass
and spin. \citet{BirminghamSen} considered the formation of a black
hole from the collision of two point particles in 2+1 gravity, but
without explicitly stating the family of initial data
considered. Addressing this, \citet{Gun24} gives explicit expressions
for the mass and spin of the resulting black hole as functions of the
rest masses and spins of the two compact objects and their relative
velocity and impact parameter.

}

\citet{Alvarez-Gaume0811} have calculated critical exponents for the
formation of an apparent horizon in the collision of two gravitational
shock waves.

\new{

  \section{Self-similar blowup in nonlinear dispersive PDEs and its
    stability}
\label{section:PDEblowup}

\subsection{Overview}

Self-similar PDE blowup on flat spacetime can serve as a toy model for
self-similar naked singularity formation in general relativity. More
specifically, a PDE that displays both blowup and dispersion,
depending on the initial data, separated by a self-similar critical
solution, can serve as a toy model for type-II critical collapse. The
existence and type of blowup depend on dimensionless parameters and on
the spacetime dimension.

In this subsection, we cover mainly rigorous mathematical results, but
also include numerical studies here if they are closely connected. The
mostly numerical work on nonlinear dispersive PDEs coupled to GR (and
so considered as ``matter''), however, was discussed above in
Sec.~\ref{section:matterobeyingwaveequations}, and rigorous
mathematical results on singularity formation in GR will follow below
in Sec.~\ref{section:mathematicalGR}.

In investigating self-similar blowup in a nonlinear PDE, and
establishing an analogy with self-similar naked singularity formation
in type~II critical collapse, one needs to go through a number of
steps: prove existence of a self-similar blowup solution, establish
its stability (mode stability, linear stability, nonlinear/orbital
stability) at least on a neighbourhood of the past of the singularity,
then prove critical behaviour (roughly, the phase space picture of
Fig.~\ref{figure:dynsim}). The choice of function space for these
steps is dictated by what can be proved. In the GR context, one would
naturally think of smooth functions, but this is not a natural
function space for PDE theory.

In this context, \citet{Don23} notes ``[The stability analysis]
consists of a `hard part' that proves the mode stability and a `soft
part' that embeds the mode stability problem into a proper
spectral-theoretic framework for the generator of the linearised
evolution. Once the linearised evolution is understood, the treatment
of the full nonlinear problem is routine. [This] is probably hard to
implement if the solution is not known in closed form.'' We note that
the Choptuik solution is known to be analytic, but is not known in
closed form.

A dispersive PDE is a time evolution equation whose phase velocity
$\omega(|{\bf k}|)/|{\bf k}|$ varies with $|{\bf k}|$ so that wave
packets spread out. We begin with an overview of widely investigated
dispersive PDEs, taking the nonlinear wave equation as our prototype,
and highlighting the similarities and relations between the different
PDEs as we go along.

\subsection{Nonlinear spherical wave equation}

The wave equation in $d$ space dimensions with power nonlinearity,
restricted to spherical symmetry, is
\begin{equation}
\label{nonlinearwaveequation}
-u_{,tt}+u_{,rr}+{d-1\over r}u_{,r}=\pm |u|^{p-1}u.
\end{equation}
This and the following examples have a conserved energy of the general
form
\begin{equation}
E=\int_0^\infty \left({1\over 2}(u_{,t}^2+u_{,r}^2)+V(u,r)\right)r^{d-1}\,dr.
\end{equation}
For the wave equation, obviously $V=V(u)$. For sufficiently simple
PDEs, if $u(t,r)$ is a solution with finite energy $E$, so is
\begin{equation}
\label{waveeqnsolnscaling}
u_\lambda(t,r)=\lambda^k\,u\left({t\over\lambda},{r\over\lambda}\right),
\end{equation}
with energy $E_\lambda=\lambda^sE$. We say a PDE is energy-critical
for $s=0$ and energy-supercritical for $s>0$. For the wave equation
with the power nonlinearity (\ref{nonlinearwaveequation}), we find
\begin{equation}
\label{waveeqnscaling}
k=-{2\over p-1}, \qquad s=d-2{p+1\over p-1}.
\end{equation}
Hence the wave equation with cubic nonlinearity is energy-critical in
$d=4$ space dimensions, and supercritical in $d\ge 5$. In the more
general case where $V=V(u)$ is a polynomial, there is no exact scaling
of solutions and their energy. However, if we assume that the
solutions scale approximately as (\ref{waveeqnsolnscaling}) with
$k<0$, then with decreasing $\lambda$ the potential $V(u_\lambda)$ is
dominated by the highest power $p$ in it. Hence in the limit
$\lambda\to 0_+$ the scalings (\ref{waveeqnsolnscaling}) and
(\ref{waveeqnscaling}) hold exactly.

Self-similar blow-up with finite energy can occur only in
energy-critical or super-critical PDEs, where the self-similar part of
the solution contains ever less energy as it shrinks. As we will see,
blowup in super-critical PDEs often occurs via a self-similar
solution, while in energy-critical PDEs it may occur via adiabatic
contraction of a scale invariant static solution.

We have introduced the absolute value and $\pm$ sign in
(\ref{nonlinearwaveequation}) to clarify another feature of nonlinear
wave equations. If spatial derivatives can be neglected, the PDE
becomes the ODE $-u_{,tt}=\pm u |u|^{p-1}$. For the upper sign, the
solutions of this ODE oscillate. This is consistent with the conserved
energy $E$ being positive definite. For the lower sign, if spatial
derivatives can be neglected, solutions experience ``ODE
blowup''. Consistently with this, the conserved energy $E$ is not
positive definite, and so cannot control any norm of the solution,
which would restrict the blowup. Such a nonlinear dispersive PDE is
called ``focusing''.

\citet{BizChmTab04} numerically study blowup in
(\ref{nonlinearwaveequation}) with $d=3$ and $p=3,5,7$ in the focusing
case. In all three cases, blowup approaches a spatially homogeneous
blowup solution that is known in closed form. Numerical blowup
simulations can be matched to this homogeneous solution and its linear
perturbations. For $p=5$ they find a critical solution at the
threshold of blowup that is static (with a free length scale), while
for $p=7$ they find that the critical solution is CSS. For $p=3$ a
critical solution could not be identified.

\citet{BizMaiWas07} prove the existence of a countable family of
self-similar solutions for all odd integer $p$. For $p=3$,
\citet{BizBreMaiWas09} show that the $n$-th solution has $n$ unstable
modes (not counting the time translation gauge mode).

\citet{KriSchla05} and \citet{NakSchla11} prove the critical behaviour
for $d=3$ with $p=5$ that was observed numerically in
\citep{BizChmTab04}: a codimension-one set of initial data approaches
a static solution, while initial data on one side blow up, and initial
data on the other side disperse. See also the review paper
\citet{Schla06}. \citet{KriNakSchla13} removes the restriction to
spherical symmetry.

\citet{BizChmSzpa11} numerically and analytically study
\begin{equation}
-u_{,tt}+u_{,rr}-u=-|u|^{p-1}u.
\end{equation}
(This is (\ref{nonlinearwaveequation}) with $d=1$ and odd integer
$p\ge 3$ in the focusing case, and with an additional mass term.)
Solutions at the blowup-threshold remain close to a static solution,
and for $p>5$ approach it. The approach rate is determined. This
complements the proof of critical behaviour in \citet{KriNakSchla10}
for $p>5$. This system can be seen as a toy model for type-I critical
collapse.

\citet{GloSch18} consider (\ref{nonlinearwaveequation}) with focusing
cubic nonlinearity ($p=3$), without symmetry restriction. In all
energy-supercritical space dimensions $d\ge 5$, they find an explicit
self-similar blow up solution. The particular choice of $d=7$
allows a rigorous solution of the spectral problem, and they prove
that the self-similar solution is a codimension-one
attractor. Specifically, for initial data in $H^4\times H^3$, they
show that the solution minus a multiple of the known growing mode
approaches the known similarity solution in the {\em homogeneous}
Sobolev space $\dot H^k \times \dot{H}^{k-1}$, for $k=1,2,3$, inside
the past lightcone of the blowup point, as a power of
$t_*-t$. Obviously, the blowup point and time is not given a
priori. \citet{GloMalSch20} investigate the same problem
numerically. For $d=5,7$, the self-similar blowup solution is an
attractor in the blowup-threshold. The perturbation spectrum of the
self-similar solution is computed numerically for general $d\ge 5$.

\subsection{Spherical Yang-Mills equation}

The Yang-Mills equation for the group $SO(d)$ also admits a
spherically symmetric ansatz in $d$ space dimensions. This gives rise
to the PDE \citep{BizonTabor}
\begin{equation}
\label{sphericalYM}
-w_{,tt}+w_{,rr}+{d-3\over r}w_{,r}=-(d-2){w(1-w^2)\over r^2},
\end{equation}
which looks like a wave equation\footnote{Notations vary: for
    example, \citep{BizBie15} uses $w=:1-u$, while \citep{Glo22} sets
    $w=:1-r^2u$.} in $d-2$ space dimensions with a potential $V(w,r)$.
    Solutions and their energy scale with $k=-1$ and $s=d-4$.

The Yang-Mills equation does not form singularities from smooth
initial conditions in $d=3$ space dimensions
\citep{EardleyMoncrief}. \citet{BizonTabor} numerically show
singularity formation in $d=4$ (the energy-critical dimension) and
$d=5$ (the first supercritical dimension). In $d=5$ there is a
countable family $W_n$ of CSS solutions with $n$ unstable modes, such
that $W_0$ is an attractor for blowup, and $W_1$ acts as a critical
solution separating blowup from dispersion, see also \citet{Biz02}. In
$d=4$ there are no self-similar solutions and the formation of
singularities seems to proceed through adiabatic shrinking of a static
solution. \citet{BizChma05} return to $d=5$, and numerically show
universal self-similar blowup, with the approach to the similarity
solution matched to its perturbation spectrum.

\citet{BizBie15} find explicit exactly CSS solutions for
(\ref{sphericalYM}), in the sense that $w=w(r/t)$, for $d\geq 5$. They
compute the linear perturbation spectrum to be the zeros of a special
function, which can be computed numerically to arbitrary
precision. All eigenvalues are negative (except for the trivial one
corresponding to an infinitesimal translation of any CSS solution in
$t$). They conjecture that they are attractors for blowup for $d\ge
5$, and give numerical evidence from time evolutions.  \citet{Glo22}
proves stable blowup for all odd $d\ge 5$.  This paper also gives a
nice summary of the literature on blowup in Yang-Mills and
corotational wave maps.

\subsection{Spherical wave map}
\label{section:sphericalwavemap}

Wave maps, under their alternative name sigma models, were defined in
Sec.~\ref{section:wavemapGR} above. If the target manifold $M$ has an
$SO(d)$ symmetry, the ``corotational'' (also called ``hedgehog'' or
``spherically symmetric'') ansatz is to identify the $d-1$ standard
angular coordinates in space with the corresponding angles in the
target manifold $M$, so the map reduces to one from time $t$ and the
radius $r$ in space to the radial coordinate in the target
manifold. For $M=S^d$ with full $SO(d+1)$ symmetry in particular, the
corotational wave map reduces to the PDE
\begin{equation}
\label{corotationalwavemap}
-u_{,tt}+u_{,rr}+{d-1\over r}u_{,r}=(d-1){\sin 2u\over 2r^2}.
\end{equation}
This looks like the nonlinear wave equation (\ref{nonlinearwaveequation}),
but now with a potential $V=V(u,r)$. Solutions $u$ and their energy
scale exactly with $k=0$ and $s=d-2$. The substitution
$u(t,r)=r\,v(t,r)$ gives 
\begin{equation}
\label{corotationalwavemapbis}
-v_{,tt}+v_{,rr}+{d+1\over r}v_{,r} =(d-1)\left({\sin 2rv\over
 2r^3}-{v\over r^2}\right). 
\end{equation}
The right-hand side is $O(v^3)$ as $v\to 0$, so to leading order in
$v$ we recover the nonlinear wave equation
(\ref{nonlinearwaveequation}) with defocusing cubic nonlinearity, but
in $d':=d+2$ space dimensions. Indeed, solutions $v$ and their energy
scale (exactly) with $k=-1$ and $s=d'-4=d-2$.

For 3+1 dimensions, \citet{Bizon1} has shown that there is a countable
family of smooth (before the future light cone of the
singularity) self-similar solutions labelled
by a nodal number $n\ge 0$, such that each solution has $n$ unstable
modes.  Numerical simulations in 3+1 in spherical symmetry in
\citet{BizonChmajTabor2} and \citet{LieblingHirschmannIsenberg}, and without
symmetry restriction in \citet{Liebling3D} show that $n=0$ is a global
attractor and the $n=1$ solution is the critical solution. See also
\citet{DonningerAichelburg1, DonningerAichelburg2} for numerical
computations of the largest perturbation eigenvalues of $W_0$ and
$W_1$. For the wave map in the critical 2+1 dimensions, generic blowup
proceeds through adiabatic shrinking of a static solution
\citep{BizonChmajTabor3}, while the critical solution is of the same
type.

In the paper \citet{BizBie15}, already mentioned in the context of
Yang-Mills, the authors also find explicit exactly CSS solutions for
(\ref{corotationalwavemap}), prove mode stability, and give numerical
evidence from time evolutions that they are attractors for blowup for
$d\ge 3$. \citet{BieBizMal17} show numerically and analytically that
the threshold of blowup in (\ref{corotationalwavemap}) is the
(codimension-1) attracting manifold of a self-similar solution for
$3\le d\le 6$.

\subsection{Non-relativistic spherical Euler and Navier-Stokes equations}
\label{section:fluidblowupnogravity}

Based on \citet{Guderley1942}, \citet{MerRapRod19} prove the existence
of smooth self-similar solutions of the Newtonian Euler equations
(without gravity) in arbitrary space dimension, in spherical
symmetry. The ideal gas equation of state is $P=(\Gamma-1)\rho e$
where $e$ is the internal energy per particle. The first law of
thermodynamics is $de=Pd(1/\rho)+Tds$, where $s$ is the entropy per
particle. The special case of constant $s$ gives the ``isentropic'' or
``barotropic'' ideal gas equation of state $P\propto\rho^\Gamma$. With
this barotropic equation of state, the spherically symmetric
self-similar solutions take the form (neglecting constant factors)
\begin{align}
\rho(t,r)&= e^{2{\beta-1\over\Gamma-1}\tau}\hat\rho(\tau,\x), \\
u(t,r)&= e^{(\beta-1)\tau}\hat u(\tau,\x),
\end{align}
where the similarity coordinates are 
\begin{align}
\tau&:=-{1\over \beta}\ln(t_*-t), \\
\x&:={r\over (t_*-t)^{1/\beta}}=e^{-\tau}r.
\end{align}
(We use the symbol $\beta$ for $r$ of \citet{MerRapRod19}, reserving
$r$ for the radius.) This ansatz reduces the equations to a system of
two autonomous ODEs in $x$, with $\beta$ a parameter.
For any space dimension $d\ge 2$, for any value of the state parameter
$\Gamma$ except a countable set of values, solutions exist for a
countable sequence $\beta_k$ of ``blow-up speeds'' accumulating from
below at some value $\beta_*(d,\Gamma)$. These solutions do
not have finite energy but decay at infinity.

(We note in passing that a more general CSS ansatz where $s$ is not
constant is also possible, parameterised by
$\rho=e^{-\kappa\tau}\hat\rho(x)$ with a free constant $\kappa$,
see~\citet{Jen25}. In this case, one still has a system of {\em two}
autonomous ODEs for $\hat u$ and $\hat c$, with
$c=e^{(\beta-1)\tau}\hat c(x)$ the sound speed, independent of
$\hat\rho$ \citep{Guderley1942}.)

Based on this existence result for smooth exactly self-similar
solutions, \citet{MerRapRod20} prove, in three space dimensions, that
for each blowup speed $\beta_k$ there is a finite-codimension manifold
in the space of spherically symmetric smooth initial data such that
the solutions blow up in the same way as the corresponding
self-similar solutions.

A similar result holds for the Navier-Stokes equations in $d=3$ space
dimensions, with the same equation of state and arbitrary {\em
  constant} shear and bulk viscosity parameters, but subject to the
sharp constraint $\Gamma<1+2/\sqrt{3}$: this follows simply from
demanding that the dissipation terms in Navier-Stokes, evaluated on
the self-similar solution of Euler, vanish as $\tau\to\infty$. As the
similarity solution shrinks, they can then be treated perturbatively,
even though they are principal.

A key step in proving finite-codimension stability of the
asymptotically self-similar blowup, inside and just beyond the past
soundcone of the blowup point, is the stability analysis of the
linearised Euler equations, given mainly in \citet{MerRapRod20b}. With
the linearised equations written as $X_{,\tau}=MX$, one wants to show
that after subtracting a finite number of unstable modes $X_i$, the
linear differential operator $M-\sum_{i=1}^NX_iX_i^\dagger$ is
stable. For this the smoothness of the background solution is
essential. The authors comment that counting the number of unstable
modes should be addressed numerically. We note here that a similar
count was carried out for the spectrum of (spherical and
non-spherical) linear perturbations of the fluid collapse critical
solutions, using numerical contour integration in the complex plane of
potential mode frequencies in \citep{Gundlach_nonspherical}.

\citet{Biasi2021} has numerically evolved initial data consisting of
the self-similar solution plus the unstable mode with small finite
amplitude (both constructed numerically). He finds that a shock always
forms. As the parameter space to explore is large it remains open how
many unstable modes there are and if any smooth self-similar solution
is stable.

The argument of \citep{MerRapRod20} requires a genericity condition
that does not hold for isolated values of $\Gamma$, such as
$\Gamma=5/3$. \citet{BucCaoGom22} apply methods 
similar to those of \citet{GuoHadJangSch21} to prove existence of
smooth self-similar solutions of Euler, and asymptotically
self-similar solutions of Navier-Stokes, for all
$\Gamma>1$. They also give a simpler proof of
linear stability of the Euler system, achieved by writing the full
equations in terms of Riemann invariants $u\pm c$ (where $u$ is the
radial fluid velocity and $c$ the sound speed) and similarity
variables $x$ and $\tau$.

\citet{Jen25} proves the existence of CSS solutions for
$\beta>1$ sufficiently close to $1$ that are continuous on all of
$\R^{3,1}$ except at the singularity, and fall off at
$r\to\infty$ sufficiently rapidly to have finite mass and energy. The
question of uniqueness for $t>0$ is also addressed, with a preliminary
statement that the solution is unique, at least in the absence of a
shock emanating from the singularity.

\subsection{Nonlinear Schr\"odinger equation}

The nonlinear Schr\"odinger equation with, for simplicity, polynomial
nonlinearity is
\begin{equation}
\label{NLS}
iu_{,t}+\Delta u=\pm |u|^{p-1}u,
\end{equation}
where $u$ is now complex. (Note we do not restrict to spherical
symmetry and $u$ is necessarily complex). The conserved energy
is \citet{MerRapRod20b}
\begin{equation}
E=\int \left({1\over 2}|\nabla u|^2\pm {|u|^{p+1}\over p+1}\right)\,d^dx.
\end{equation}
Separately, there is also a conserved particle number $N=\int
|u|^2\,d^dx$.  Solutions and their energy scale as
\begin{equation}
u_\lambda=\lambda^k u\left({t\over\lambda^2},{{\bf
 x}\over\lambda}\right), \end{equation} with $k$ and $s$ given by
 (\ref{waveeqnscaling}) above.

The nonlinear Schr\"odinger equation is related to the
Newtonian Euler equations by setting
\begin{equation}
\label{NLSEuler}
u=\sqrt{\rho}e^{i\phi/2},
\end{equation}
where $\rho$ and $\phi$ are now real. The real and imaginary part of
(\ref{NLS}) become
\begin{align}
\rho_{,t}+\nabla\cdot(\rho\nabla\phi)&=0, \\
\phi_{,t}+{1\over 2}|\nabla\phi|^2
+2\rho^{p-1\over 2}&=2{\Delta\sqrt{\rho}\over\sqrt{\rho}}.
\end{align}
Taking a gradient of the second equation, and writing ${\bf
v}:=\nabla\phi$, we obtain
\begin{align}
\label{Newtonianparticleconservation}
\rho_{,t}+\nabla\cdot(\rho{\bf v})&=0, \\
\label{almostEuler}
{\bf v}_{,t}+({\bf v}\cdot\nabla) {\bf v} +\nabla\left(2\rho^{p-1\over
2}\right)&=2\nabla\left({\Delta\sqrt{\rho}\over\sqrt{\rho}}\right).
\end{align}
We have recovered the Newtonian Euler equations for an {\em
irrotational} fluid with equation of state $P=2\rho^{p-1\over 2}$,
except for the term on the right-hand side of (\ref{almostEuler}),
which is not present in the genuine Euler equations.

\citet{MerRapRod20b} use the results of \citet{MerRapRod19} and the
relation (\ref{Newtonianparticleconservation},\ref{almostEuler}) to
prove a similar result on the existence of smooth self-similar blowup
solutions for the {\em defocusing} nonlinear Schr\"odinger equation
(\ref{NLS}). The proof holds explicitly for
$(d,p)=(5,9),(6,5),(8,3),(9,3)$, but may carry over for other
combinations. Like the viscosity terms in the Navier-Stokes equations,
the unwanted right-hand side of (\ref{almostEuler}) becomes less
relevant on smaller scales and the exactly self-similar Euler solution
provides the leading term for the nonlinear Schr\"odinger solution.
The condition for this to happen is $(p-1)(d-2\sqrt{d})>4$, which
requires $d\ge 5$. As in \citet{MerRapRod19}, there is a countable
sequence of blowup ``speeds'' $\beta_k$ approaching a limiting value
from below.

Using completely different methods, \citet{DonSchoe24} prove the
existence of a self-similar blowup solution for (\ref{NLS}) in the
{\em focusing} case in $d=3$ with $p=3$. (There is no suggestion of a
countable set of solutions.) The solution takes the form
\begin{equation}
u={1\over\sqrt{2\alpha}}(t_*-t)^{-{1\over
2}-{i\over2\alpha}}Q\left({r\over \sqrt{2\alpha(t_*-t)}}\right),
\end{equation}
where $\alpha\simeq 0.917$ is known. Note the different blow-up speeds
for the magnitude and phase.  The proof is computer-assisted, showing
convergence of an expansion of $Q(r)$ in Chebyshev
polynomials. (Compare also Sec.~\ref{section:ReitererTrubowitz}
below.)

\subsection{Relativistic Euler equation on flat spacetime}
\label{section:relativisticEuler}

Using similar methods to \citet{MerRapRod20}, \citet{ShaWeiZha24}
prove existence of smooth self-similar blowup solutions for the
special-relativistic Euler equations with the ultra-relativistic
equation of state $P=k\rho$ in $d\ge 2$ space dimensions, where $\rho$
is now the total energy density, rather than the rest mass density. In
contrast to the Newtonian case (where the equation of state was also
chosen differently), the similarity coordinate is $x=r/(t_*-t)$, and
existence is proved of only one solution for each $d$ and $k$. For
$d=2$ and $d=3$ the proof is for all $0<k<1$, and for $d>3$ it is for
$k_*(d)<k<1$ (but existence for smaller $k$ is not excluded).

An equivalence similar to that between nonlinear Schr\"odinger and
non-relativistic Euler also exists between the {\em relativistic}
Euler equations on Minkowski spacetime for an irrotational fluid and
nonlinear wave equations. To fix notation, the perfect fluid
stress-energy tensor is
\begin{equation}
T^{ab}=(\rho+P)u^au^b+Pg^{ab},
\end{equation}
where $u^au_a=-1$. As is well-known, $\nabla_aT^{ab}=0$ decomposes into
components parallel to $u^a$ (``energy equation'') and orthogonal to $u^a$
(``momentum'' equation). Assume $u^a$ is irrotational
in the sense that
\begin{equation}
\label{scalarfluidequivalence}
u_a=f^{-{1\over 2}}\nabla_a\phi, \quad f:=-\nabla_a\phi\nabla^a\phi,
\end{equation}
$P$ is a given function of $\rho$ only with $0\le P'(\rho)\le 1$, and
$\rho$ is a function of $f$ only with inverse $f(\rho)$. Then the
momentum equation is obeyed if and only if
\begin{equation}
{f'(\rho)\over f(\rho)}={2P'(\rho)\over P(\rho)+\rho}.
\end{equation}
The energy equation, in turn, is equivalent to the nonlinear wave
equation
\begin{equation}
\label{Fphieqn}
\nabla_a(F\nabla^a\phi)=0,
\end{equation}
where $F$ is understood to be $F(\rho)$, if and only if
\begin{equation}
{F'(\rho)\over F(\rho)}={1-P'(\rho)\over P(\rho)+\rho}.
\end{equation}
\eqref{Fphieqn} can also be written as
\begin{equation}
\left(-u^au^b+c_s^2\,h^{ab}\right)\nabla_a\nabla_b\phi=0,
\end{equation}
where $h_{ab}:=g_{ab}+u_au_b$ is the projector orthogonal to
$u^a$, and $c^2=P'(\rho)$ is the sound
speed. Hence $\phi$ obeys a wave equation in the acoustic metric.

For the linear equation of state $P(\rho)=k\rho$, these formulas give
us
\begin{equation}
\label{scalarfluidequivalencelinearEOS}
f=\rho^{2k\over 1+k}, \quad F=\rho^{1-k\over 1+k},
\end{equation}
and hence
\begin{equation}
\label{scalarfluidequivalencelinearEOSbis}
F=f^{1-k\over 2k}.
\end{equation}
In particular, if $P=\rho$, we have $f=\rho$ and $F=1$ and,
as is well-known, $\phi$ then obeys the wave equation.

\citet{ShaWeiZha24} note that their relativistic Euler result can be
used to construct {\em complex} self-similar blowup solutions for
(\ref{nonlinearwaveequation}) in the focusing case. Setting
$u=:\rho e^{i\phi}$, (\ref{nonlinearwaveequation}) with the upper sign
(the defocusing case) becomes the pair of real PDEs
\begin{align}
\label{urhophi1}
\rho^p+\rho\nabla_a\phi\nabla^a\phi&=\nabla_a\nabla^a\rho, \\
\rho\nabla_a\nabla^a\phi+2\nabla_a\phi\nabla^a\rho&=0.
\end{align}
Rescaling $\rho$ and $\phi$ by suitable powers of $\epsilon$, the
right-hand side of (\ref{urhophi1}) (the principal term!) can be made
to vanish as $\epsilon\to 0$, with the other terms unchanged. The
expectation is that this scaling will be realised dynamically in
asymptotically self-similar blowup, in a process called ``front
renormalisation'', as in \citep{MerRapRod20,MerRapRod20b}.  Setting
$\rho=F^{1/2}$, these PDEs become
(\ref{scalarfluidequivalencelinearEOSbis}) with $k=1/p$ and
(\ref{Fphieqn}), respectively. For a recent related proof of blowup in
the complex-valued defocusing septic wave equation on $\R^{4+1}$ see
\citet{BucChe24}.

\subsection{DSS blowup in a system of nonlinear wave equations}

\citet{Tao16} considers the coupled non-linear wave equations
\begin{equation}
\label{TaoPDE}
\square u=\nabla F(u),
\end{equation}
were $u:\R^{3+1}\to \R^m$ and the smooth potential $F:\R^m\to \R$ is
homogeneous of order $p+1$ for $|u|>1$ and defocusing in the sense
that $F(u)>0$. For $p>5$ and $m$ sufficiently large (in particular
$m=40$), Tao proves that there is a solution $u(\bx,t)$ that is smooth
on $|\bx|\le -t$, blows up as $t\to 0_-$, and is DSS in the sense that
there is a specific $\Delta>0$ with
\begin{equation}
\label{TaoDSS}
u(e^\Delta \bx,e^\Delta t)=e^{-{2\over p-1}\Delta}u(\bx,t),
\end{equation}
but {\em not} CSS.
The potential $F$ is in some sense constructed from the assumed solution,
using the Euler identity
\begin{equation}
(p+1)F(u)=\langle u,\nabla F(u)\rangle=\langle u,\square u\rangle,
\end{equation}
where the angle brackets denote the inner product in $\R^m$.

The construction assumes that, in contrast to $u$ itself, the
Euclidean norm $||u||$ (in $\R^m$) {\em is} CSS. The obvious possibility
of ``twisted CSS'', that is
\begin{equation}
u(\lambda\bx,\lambda t)=\lambda^{-{2\over p-1}}e^{J\ln\lambda}u(\bx,t),
\end{equation}
where $J\in so(m)$, for {\em all} $\lambda>0$ is ruled out explicitly \citep{Tao_pc}.

Tao also cites papers that construct {\em rough} CSS solutions, so
smoothness is essential for DSS. This result is analogous to the
spherical Einstein-scalar field system, where the Christodoulou
solutions are rough CSS solutions and the Choptuik solution is a
smooth DSS solution, although there is no proof yet of the
non-existence of smooth CSS solutions of (\ref{TaoPDE}).

\section{Rigorous results on self-similarity and naked singularity
  formation}
\label{section:mathematicalGR}

\subsection{CSS blowup in a spherical perfect fluid with Newtonian gravity}

As a preliminary step to full GR, we now move on to a perfect fluid
coupled to {\em Newtonian} gravity, described by the Euler-Poisson
system, with the polynomial barotropic equation of state
$P=\rho^\Gamma$. Here the spherical CSS ansatz is
\begin{align}
\rho&=(t_*-t)^{-2}\tilde\rho(\x), \\
u&=(t_*-t)^{1-\Gamma}\tilde u(\x), \\
\x&={r\over (t_*-t)^{2-\Gamma}}.
\end{align}
This gives
\begin{align}
m(\x,t)&\sim \int_0^{r(\x,t)}\rho\,r^2\,dr \sim (t_*-t)^{4-3\Gamma}, \\
E(\x,t)&\sim \int \rho u^2\,r^2\,dr
+\int {1\over
r}\left(\int \rho \,r^2\,dr\right) \rho \,r^2\,dr \nonumber \\ 
&\sim (t_*-t)^{6-5\Gamma}. \label{newtonianfluidgravity}
\end{align}
The two integrals in (\ref{newtonianfluidgravity}) represent the
kinetic and gravitational energy, and by construction they scale in
the same way. In contrast to the perfect fluid without gravity, the
CSS ansatz is now unique (with ``blowup speed'' $\beta=2-\Gamma$), and
the two ODEs in $x$ for density and velocity are no longer
autonomous. One might expect to obtain a third ODE representing the
Poisson equation but this can be solved in closed form.

The existence of self-similar collapse solutions with $1<\Gamma<4/3$
was shown by \citet{GuoHadJangSch21}.  No self-similar collapsing
solution with finite mass can exist for $\Gamma>4/3$, and for
$\Gamma=4/3$, collapsing solutions are quasistatic. The proof uses
computer-assisted interval arithmetic to control convergence of the
Taylor series in $\x$ about the sonic point, and to prove the
existence of a sonic point that can be matched to a regular centre.

The existence of the Larston-Penston self-similar spherical collapse
solution for the {\em isothermal} equation of state $P=k\rho$ (the
case $\Gamma=1$) was shown by \citet{GuoHadJang21a}. Physically, $k$
is the sound speed, and is proportional to the temperature. However,
in the Newtonian case, in contrast to the GR case, it can be absorbed
into a rescaling of variables. Subsequently, \citet{San23} showed
existence for the countable family of the Hunter solutions, which
differ from the Larston-Penston solution by an increasing number of
zeros in the velocity profile.

\subsection{CSS naked singularity formation in a spherical perfect
fluid in GR}
\label{section:mathfluidGR}

For exactly self-similar spherically symmetric perfect fluid collapse
in {\em general relativity}, the similarity coordinate has to be
$\x=r/(t_*-t)$ where, for example, $r$ is the area radius and $t$
proper time at the centre, and the equation of state has to be
$P=k\rho$ for $0<k<1$. Physically, $k$ is still the sound speed, but
in GR (in units $c=1$) it is dimensionless, and cannot be absorbed
into a redefinition of variables.

In similarity coordinates based either on comoving or polar-radial
coordinates, the fluid equations become again a system of two
non-autonomous ODEs. One would expect a third ODE for the
(Misner-Sharp, or Kodama) mass, but again this can be solved in closed
form. The limit $k\to 0$ of the CSS ansatz for $P=k\rho$ in GR is the
isothermal limit $\Gamma\to 1$ in the CSS ansatz for the Euler-Poisson
system \citep{HaradaMaeda2, HaradaMaeda3}.

For $k\ll 1$, \citet{GuoHadJang21b} show existence of the relativistic
Larston-Penston solutions of \citep{OriPiran,OriPiran2} as analytic
solutions of the field equation. Numerically, these solutions for
$k\ll 1$ are known to be attractors in spherical collapse, see
Sec.~\ref{section:verysoft}. Stability has not yet been proved
rigorously, and neither has existence or stability of the relativistic
Hunter A, B,... solutions which are believed to be codimension-1,
2,... stable.

By gluing the CSS solution to an asymptotically flat exterior outside
the past of the singularity, the authors show that the singularity is
naked in the sense that an outgoing null geodesic reaches infinity.

\subsection{Existence of the CSS critical solution for a corotational wave
map in GR}
\label{section:mathwavemapGR}

As already mentioned in Sec.~\ref{section:wavemapGR},
\citet{BizonWassermann2} have proved existence of a countable family
of spherically symmetric CSS solutions to the $SU(2)$ sigma model (or
wave map whose target manifold is the round $S^3$) coupled to GR that
are analytic inside the lightcone. They prove that for sufficiently
small value of the coupling $\eta$ in Eq.~(\ref{S3wavemapGRaction})
these solutions have a finite curvature on the future lightcone (the
wave map itself is only $C^0$) and so are naked singularities. One of
these was shown {\em numerically} to be a critical solution at the
black hole threshold by \citet{BizonWassermann}, making it the first
critical solution whose existence has been proved rigorously, and the
only CSS one so far. (The second existence proof of a critical
solution is the for the Choptuik solution \cite{ReiTru12}, and is the
only DSS one so far.) In this context it may be worth mentioning the
proof of \citet{BizWas10} that the 4+1-dimensional Einstein equations
with the Bianchi-IX ansatz of \citep{BizonChmajSchmidt} admits no
regular CSS solutions.

\subsection{Naked singularity formation in a spherical scalar field in
  GR}\label{section:mathscalarGR}

In the course of his examination of gravitational collapse in the
spherical Einstein-scalar system, Christodoulou has shown that naked
singularities can form from mildly singular initial data, and that
these spacetimes are non-generic, in the sense of admitting at least a
one-parameter family of perturbations that destroy the naked
singularity. See also Christodoulou's own review paper
\citep{Chr99}. This work, together with later contributions by others,
is reviewed in this subsection, as well as another key result for this
system, the existence proof for the Choptuik solution as an analytic
solution of the field equations.

\subsubsection{Sufficient criteria for collapse and dispersal}

The Einstein equations in spherical symmetry with a scalar field
$\phi$ can be stated in terms of a 2-dimensional quotient spacetime
with a timelike boundary $\Gamma$ representing that part of the centre
of spherical symmetry which is regular and timelike and an area radius
$r$ that vanishes on $\Gamma$. The mass function $m$ is defined as
$1-2m/r=|\nabla r|^2$.  Christodoulou considers initial data on an
outgoing null cone $C_0^+$. In spacetimes where $\Gamma$ ends at a
singularity $O$, the past lightcone of $O$ is denoted by $C_0^-$, and
its future lightcone (which may be empty) by ${\cal B}_0$.

\citet{Christodoulou2} gives a sufficient condition for a black hole
to form from initial data on $C_0^+$, in terms of the area radii
$r_1$, $r_2$ and masses $m_1$, $m_2$ at two points on $C_0^+$. The
only assumption about the function space is that the Einstein
equations hold in the strong sense, although in later papers it is
assumed that the results of \citet{Christodoulou2} also hold for BV
(bounded variation) data. By contrast, \citet{Christodoulou3} gives a
sharp sufficient condition for the evolution to remain regular
(neither a black hole nor a naked singularity forms and the scalar
field disperses) in terms of the total variation of the initial data
on $C_0^+$, assumed to be of BV.

\subsubsection{Construction of CSS naked singularities that are
  analytic except at the SH}

\citet{Christodoulou4} constructs CSS solutions with the most general
ansatz for the scalar field compatible with CSS in spherical
symmetry,
\begin{equation}
\label{Christodoulouansatz}
\sqrt{4\pi G}\,\phi(x,\tau)=\tilde\phi(x)+k\tau. 
\end{equation}
(See also \citet{Bra95} for an independent study of the same ansatz).
Here $x$ and $\tau$ are any similarity coordinates adapted to
spherical symmetry, such that the homothetic vector field is
$\xi=\partial_\tau$, and $k\ge 0$ is an arbitrary constant.
Specifically Christodoulou uses $x:=-r/u$ and $\tau:=-\ln(-u)$, where
$u$ is retarded time, normalised to be proper time on $\Gamma$ and
with $u=0$ at $O$, and $r$ is the area radius in spherical
symmetry. Solutions are obtained by solving the Einstein-scalar
equations under the CSS ansatz as a system of two autonomous ODEs in
$s:=\ln x$. In particular, $\Gamma$ is located at $x=0$ and $C_0^-$,
if it exists, at some $x=x_0>0$.

There is a one-parameter family of solutions that are analytic at the
centre $\Gamma$, parameterised by $k$. These solutions reach the past
light cone $C_0^-$ for $k^2<1$ only. $C_0^-$ corresponds to a critical
point of the dynamical system. The family of solutions has two
parameters, one of which, for given $k$, is fixed by connecting
$C_0^-$ to the analytic centre $\Gamma$. The other parameter
determines a non-unique continuation of the solution outside
$C_0^-$. Christodoulou proves that in an open set of such extensions,
$O$ is a naked singularity visible from infinity. Because of CSS,
these solutions are not asymptotically flat but can be glued to an
asymptotically flat (non-CSS) exterior on an ingoing null cone to the
future of $C_0^-$, thus preserving the naked singularity.

\subsubsection{\new{All CSS naked singularity examples are only
    $C^{1,\epsilon}$ at the past lightcone}}

Christodoulou has therefore constructed a 2-parameter family of
``solutions corresponding to regular asymptotically flat initial data
which develop singularities that are not preceded by a trapped region
but have future light cones expanding to infinity''
\citep{Christodoulou4}. However, the word ``regular'' is potentially
misleading here.

As the solution of an autonomous ODE system in $x$ with analytic
coefficients, the metric and scalar field are analytic in the
underlying Bondi coordinates $(u,r)$ from $\Gamma$ to $C_0^-$, and
again to the future of $C_0^-$, potentially up to, but not including,
${\cal B}_0$. However, pp.~623-624 of \citet{Christodoulou4}
explicitly show that they are not analytic in $s$, and hence not
analytic in $(u,r)$, at the critical point $C_0^-$. Generically, the
quantity $\theta:=r\phi_{,r}$, which represents free data on a surface
of constant $u$, has an expansion about the critical point starting
with $(s-s_*)^\epsilon$, where $0<\epsilon<1$ is the ratio of the two
Lyapunov exponents at the critical point. (In special cases, there are
$\ln(s-s_*)$ terms instead.) Hence the stress-energy is only
$C^{0,\epsilon}$ in $(u,r)$ on $C_0^-$, and the metric and scalar
field are $C^{1,\epsilon}$ there. \citet{Sin24} states that
$\epsilon=k^2/(1-k^2)$. This means that $\phi$ has the same regularity
as the null coordinate $V$ defined to vanish at the singularity
horizon and to be proper time on $\Gamma$.

In summary, the initial data for Christodoulou's examples all have a
breakdown of regularity precisely on the past lightcone of the naked
singularity that yet is going to form: they already ``know'' about the
singularity.

\subsubsection{\new{All naked singularity spacetimes are unstable under
perturbations that are only $C^{1,\delta}$  across the SH}}

\citet{Christodoulou5} shows that all naked singularity solutions are
nongeneric: in some sense, they have codimension. Let $\vartheta$ be
any initial data on $C_0^+$ that form a naked singularity. Then there
are choices of two functions $f_1$ and $f_2$ such that the data
$\vartheta+\lambda_1f_1+\lambda_2f_2$ form a black hole for any values
of $\lambda_1$ and $\lambda_2$ except for $\lambda_1=\lambda_2=0$. In
fact, both $f_1$ and $f_2$ represent classes of functions, so the
linear space $\lambda_1f_1+\lambda_2f_2$ that is attached to the
solution is not just two-dimensional but is a function space. See also
\citet{LiuLi18} for an alternative proof. Note that a black hole forms
for any non-zero value of $\lambda_1$ and $\lambda_2$, not just for
one sign. In this sense, the set ${\cal E}$ of naked singularity
solutions is convex, with $\lambda_1f_1+\lambda_2f_2$ a tangent plane
that lies entirely on the black-hole side. Hence the dimensionality of
this tangent space does not tell us anything about the codimension of
${\cal E}$ other than it is non-zero. It is known numerically that the
Choptuik solution contains a naked singularity and that there is one
perturbation mode which renders it regular, but the existence of
perturbations that render a given naked singularity solution regular
cannot be shown with the methods of \citep{Christodoulou4}.

The functions $f_1$ and $f_2$ vanish identically between $\Gamma$ and
$C_0^-$. This is essential for the proof, which requires the interior
of $C_0^-$ to be unchanged by the perturbation. $f_1$ is any function
that is absolutely continuous outside $C_0^-$, but is discontinuous,
jumping from $0$ to $1$, at $C_0^-$. $f_2$ is from a certain class of
functions that are absolutely continuous everywhere (AC) but, in
particular, not more regular than AC at $C_0^-$ itself. $f_2$ obeys a
limit condition given in terms of a function $\gamma$ that
characterises the gravitational blueshift along $C_0^-$.

A key observation is that the results of \citet{Christodoulou4} (naked
singularities can form in collapse) and \citet{Christodoulou5} (naked
singularities are not generic) are independent of each other. On the
one hand, \citet{Christodoulou5} shows that any naked singularity
solution, not only the ones of \citet{Christodoulou4}, {\em but also
  for example the Choptuik solution}, has some codimension {\em in the
  BV or AC function spaces}. On the other hand, the solutions of
\citet{Christodoulou4} are CSS by ansatz and are analytic except,
crucially, on $C_0^-$.

When the naked singularity spacetime is exactly CSS, one can show
(\citet{GunHilCos19}, compare (\ref{nuspherical}) with the definition
of $\gamma(t)/t$ in Lemma~1 of
\citep{Christodoulou5}) that $f_2$ can be chosen to be H\"older
continuous with regularity no better than $C^{0,\delta}$, and hence
the scalar field and metric perturbations can be chosen with
regularity no better than $C^{1,\delta}$ where
\begin{equation}
\label{deltaChr}
\delta={k^2\over 4(1+5k^2)}.
\end{equation}
We believe that this generalises to any CSS or DSS spherical
Einstein-scalar solution, such as the Choptuik solution, if one
replaces $k^2$ by $1-\nu$ of the similarity horizon, as defined below
in Sec.~\ref{section:surfacegravity}.

\citet{Sin24} has further investigated the blue-shift instability of
the solutions of \citet{Christodoulou4,RodShl19}, by examining the
stability of a {\em test} scalar field $\phi$ obeying
$\square_g\phi=0$ on the metrics of \citep{Christodoulou4}. He derives
\begin{equation}
\label{Singh1.4}
\left({\phi_{,v}\over R_{,v}}\right)_{,U}-{1+k^2\over (-U)}{\phi_{,v}\over
R_{,v}} =-{1\over R}\phi_{,U} \quad\text{(on the SH $v=0$)},
\end{equation}
along the similarity horizon, where $U<0$ is the retarded time that is
proportional to proper time at the centre of spherical symmetry and
hence obeys $\xi U=-U$. Consider in particular initial data where
$\phi=0$ for $v\le 0$ on $U=-1$, so that $\phi=0$ for $v\le 0$ for all
$-1\le U<0$. Then the right-hand side of (\ref{Singh1.4}) vanishes and
it integrates to
\begin{equation}
\label{blueshiftrate}
{\phi_{,v}\over R_{,v}}\sim (-U)^{-(1+k^2)}\quad\text{(on the SH
$v=0$)}.
\end{equation}
For generic $\psi$, we do not expect the right-hand-side of
(\ref{Singh1.4}) to cancel the second term on the left-hand side, and
Therefore, we expect blowup at the ``blue-shift rate''
$(-U)^{-(1+k^2)}$ as $U\to 0_-$, whereas in a self-similar solution
the blowup is as $(-U)^{-1}$, the ``self-similar rate''. A
straightforward generalisation of this argument to a spherically
symmetric test scalar field on an arbitrary spherically symmetric DSS
spacetime gives blowup as $(-U)^{\nu-2}$ (averaged over one period),
thus relating the surface gravity of Sec.~\ref{section:surfacegravity}
to the blue-shift instability.

\citet{Sin24} proves that a test field that is strictly
more regular than $C^{1,\epsilon}$ at the similarity horizon indeed is
polynomially in $U$ slower than the self-similar rate, one that is
$C^{1,\epsilon}$ at the similarity horizon blows up at the
self-similar rate (that is like the scalar field that generates the
metric), and one that is strictly less regular than $C^{1,\epsilon}$
blows up at a rate strictly between the self-similar and blue-shift
rates. These results hold for test field initial data both 
with support strictly outside the similarity horizon (those considered
in \citep{Christodoulou5}) and ones with support also inside the
lightcone.

Contrast this with Christodoulou's examples of unstable perturbations
of the full nonlinear equations, with the maximum allowed regularity
given by (\ref{deltaChr}). Either perturbations with H\"older
regularity between $\delta$ and $\epsilon$ are unstable as scalar test
fields but stable when coupled to the metric, or the maximum
regularity (\ref{deltaChr}) is not sharp. \citet{Sin24} conjectures
the latter.

It remains an open question if unstable nonlinear perturbations with support
in the interior can remove the naked singularity (rather than turning
it into a black hole, as in \citep{Christodoulou5}).

In complementary work, \citet{Nol06} investigates stability of a
smooth nonspherical scalar test field on an analytic spherical CSS
spacetime, up to the Cauchy horizon, and finds pointwise, $L^2$ and
energy bounds.

\subsubsection{\new{Non-generic stability of approximately CSS naked
    singularity solutions in spherical symmetry}}

\citet{Sin22} relaxes the assumption of exact CSS for the singularity
solutions of \citep{Christodoulou4}. Data are imposed at on $v=0$ (the
past light cone of the singularity) for $-1\le u\le 0$ and on $u=-1$
for $v>0$ with deviations from exactly CSS data for naked singularity
formation which fall off sufficiently rapidly as $u\to 0$ and as
$v\to 0_+$ and $v\to\infty$. These perturbations are non-generic. It
is shown that the resulting interior ($u<0$, $v<0$) solution
asymptotes to the self-similar naked singularity solution as
$u\to 0_-$, and that the perturbed exterior solution has an incomplete
future null infinity, that is, the naked singular persists under a
non-generic class of perturbations.

Intuitively, making the data decay as $u\to 0_-$ on $v=0$ suppresses
any growing perturbation in the implied data on ($u=-1$,
$v<0$). Similarly, the perturbations destroying the naked singularity
of \citep{Christodoulou5}, which have support only on $v>0$, are
suppressed by assumption of sufficiently rapid decay as $v\to 0_+$.

\subsubsection{Arbitrary codimension instability of CSS naked singularities
beyond spherical symmetry}

\citet{an2024naked} considers setting exact data for the Christodoulou
naked singularity solution on $v=0$ but perturbing the data on $u=-1$
for $v> 0$ non-spherically, with the perturbation bounded in some
weighted Sobolev space. The interior of $v=0$ is then obviously still
the Christodolou naked singularity. The solution on the exterior
exists up to $v=\delta>0$.

However, the naked singularity is unstable to non-spherical
perturbations with arbitrarily high codimension. For perturbations,
such that a certain integral
\begin{equation}
\int_0^v (\text{pert})^2\,dv'=f(\omega,v)\,v
\end{equation}
where $\omega\in S^2$, with 
\begin{equation}
f(\omega,v)\ge m, \quad \omega\in B_p(\epsilon)
\end{equation}
for a point $p\in S^2$, and arbitrarily small constants $m>0$ and
$\epsilon>0$, there is a MOTS embedded in each ingoing null cone of
constant $v>0$. Together these form a spacelike apparent horizon
emerges from the naked singularity and censors it, both locally and
globally. The perturbation that censors it can be arbitrarily small
but must be so that the integral above is proportional to $v$ in an
(arbitrarily small but finite) angular segment of $u=-1$.  Vector
spaces of perturbations that censor the naked singularity can be
constructed with arbitrary dimension.

The author mentions (Remark 7) that a similar instability proof may be
possible for the vacuum naked singularity solutions of Rodnianski and
Shlapentokh-Rothmann \citet{RodShl19}

\subsubsection{The Choptuik DSS solution is analytic}
\label{section:ReitererTrubowitz}

Numerically, the Choptuik solution can be constructed from an ansatz
of spherical symmetry, DSS, analytic centre $\Gamma$ and analytic past
light cone $C_0^-$ \citep{Gundlach_Chop2}. In particular, the
assumption of analyticity is represented in the numerical construction
by expanding in truncated power series about $x=0$, corresponding to
$\Gamma$, and about $x=1$, corresponding to $C_0^-$. DSS is imposed as
periodicity in $\tau$ with some yet unknown period $\Delta$. The
scalar field then takes the form $\phi=\phi_*(x,\tau)$ with
$\phi_*(x,\tau+\Delta)=\phi_*(x,\tau)$, and similarly for suitable
metric coefficients.

Based on this construction, \citet{ReiTru12} have rigorously proven
that the Choptuik solution exists as an analytic solution from
$\Gamma$ through to somewhat to the future of $C_0^-$. The proof is
based on an expansion of the DSS ansatz in terms of sines and cosines
in $\tau$ and Chebyshev polynomials in $x$. Starting from an
approximation with a specific finite (very large) number of such basis
functions, and using a contraction map argument, it is eventually
proved that the infinite series converges at a sufficient rate with
Fourier and Chebyshev index that the Choptuik solution is
complex-analytic in a strip surrounding the real line in $\tau$ and an
ellipse surrounding the interval $-(1+\delta)<x<1+\delta$ for an
explicit small finite $\delta$, thus implying real analyticity through
$\Gamma$ and $C_0^-$.

Crucial ingredients of the computer-aided proof are that the reference
approximate solution is found in rational, rather than the usual
floating point, arithmetic, that the formulation of the field
equations is polynomial (in fact, quadratic) in the dependent
variables (essentially a tetrad and connection) and thus compatible
with rational arithmetic (quadratic terms are implemented as
convolution in Fourier space), and that the operator norms in the
contraction map arguments are in $l^1$, again compatible with rational
arithmetic.

It is a remarkable fact that a CSS ansatz for the spherical Einstein
scalar system only gives solution that are not smooth across the
similarity horizon, whereas a DSS ansatz gives rise to a solution that
is analytic.

While we believe that the Choptuik solution is a naked singularity
spacetime, and is analytic up to (but not across) its Cauchy horizon,
there is no rigorous proof of this yet. It might be possible to give a
brute force proof using the method of \citet{ReiTru12}, based on the
numerical results of \citet{critcont}.

\subsection{Asymptotically CSS vacuum solutions}

We now move from the Einstein-scalar system to the Einstein equations
in vacuum. \citet{RodShl17} and \citet{RodShl19} have investigated
CSS and asymptotically CSS solutions, and the naked singularities they
give rise to, without symmetry assumptions. Both papers make use of
double-null coordinates with zero $v$-shift, in which the metric takes
the form
\begin{equation}
ds^2=-4\Omega^2\,du\,dv+\
\cancel{g}_{AB}(d\theta^A-b^A\,du)(d\theta^B-b^B\,du),
\end{equation}
in which the curves of constant $(u,\theta^A)$ are the null geodesics
generating the null surfaces of constant $u$.  Both papers consider
the initial value problem on the intersecting null cones $u=-1$ and
$v=0$, such that the singularity $O$ is at $u=v=0$. $\Omega$ and the
conformal class $\hat{\cancel{g}}$ of $\cancel{g}$ can be specified
freely on the two cones, with continuity at the corner, and $b^A$ can
be freely specified on $v=0$.  (Their homothetic vector field $K$
points in the opposite direction to the one called $\xi$ throughout
this review, $K=-\xi$.)

\subsubsection{Asymptotically CSS vacuum solutions with $\kappa=0$}

In \citet{RodShl17}, all spacetime dimensions $\ge 4$ are considered,
but for simplicity we restrict here to $4$ dimensions, where
$\cancel{g}$ is a 2-metric. Only a narrow wedge to the future of the
similarity horizon is considered, namely $-1\le u<0$ and
$0\le v<c(-u)$ for some small constant $c$.

\citet{RodShl17} assume that the homothetic vector field takes the
form (\ref{xiuvnaive}), thus restricting to solutions with geometric
invariant $\kappa=0$. (We retrospectively use the notation of
\citet{RodShl19}.) The data on $v=0$ are assumed to be compatible with
exact CSS. With the gauge choice $\Omega=1$ on $v=0$ and $\kappa=0$,
this implies $\cancel{g}(u)=u^2\cancel{g}_0$ on $v=0$. Regularity then
requires the further gauge choice $b^A=0$ on $v=0$. There are two
groups of theorems, concerning exactly CSS solutions and
asymptotically CSS ones.

Theorem~1.3 states that the 2-metric $\cancel{g}$ and (in 4 spacetime
dimensions) $h={\rm tf}\,\cancel{g}_{,v}$, both evaluated at the
corner, uniquely determine an exactly CSS solution. The analogy of $h$
in the spherical Einstein scalar system is $\phi_{,v}$ at the corner.
Theorem~1.4 is an equivalent of Theorem~1.3 in the analytic case,
previously obtained by Fefferman-Graham in very different notation.
Theorem~1.6 states that if $h=O(\epsilon^{-1/2})$ the solution exists
for $v/(-u)<B\epsilon^{1/2}$ but a trapped surface forms for
$v/(-u)>b\epsilon$, and so the singularity is not naked.

Theorem~1.1 then states that for exactly CSS data on $v=0$ and a
certain class of ``admissible data'' on $u=-1$, for which
$\hat{\cancel{g}}$ changes sufficiently slowly away from the corner, a
unique weak solution of the Einstein equations for those data exists
in some Sobolev space on the wedge $-1\le u<0$ and $0\le v<c(-u)$. In
the case of 4 spacetime dimensions, this construction forces the
Hawking mass on $v=0$ to be zero.  (This is surprising but, in
hindsight, is due to the restriction $\kappa=0$.)  Theorem~1.2 states
that the combination of exactly CSS data on $v=0$ and admissible data
approach a unique exactly CSS solution as $u\to 0$. Theorem~1.5 states
that Theorems~1.1 and 1.2 can be extended to the case where the data
on $v=0$ approach exactly CSS data sufficiently rapidly as $u\to 0$.

\subsubsection{Asymptotically CSS vacuum naked singularity solutions
  with $\kappa>0$}

\citet{RodShl19} introduce a new type of self-similarity where, in
contrast to previously known CSS or DSS solutions, including those of
\citet{RodShl17}, the homothetic vector field is not tangent to the
similarity horizon $v=0$, but winds around it, and so becomes
spacelike on the horizon. It is this generalisation that allows {\em
  CSS vacuum} solutions to contain naked singularities and, in the
terminology we introduce in Sec.~\ref{section:surfacegravity}, allows
the SH to have non-trivial surface gravity.

The main result is Theorem~1, which states the existence of vacuum
solutions in four spacetime dimensions on a manifold given by
$0\le v<\infty$, $-1\le u<0$, such that $u=v=0$ is a naked curvature
singularity, the solution is asymptotically flat as $v\to\infty$, and
$v=\infty$ is an incomplete future null infinity, ending at $u=0$.

In this construction, the outgoing null data on $v=0$ are compatible
with exact CSS with some $\kappa>0$, in the sense that
\begin{align}
\Omega&=(-u)^\kappa \tilde\Omega(\theta^A), \\
\cancel{g}_{AB}&= (-u)^2\cancel{g}_{AB}(\theta^C), \\
b^A&=(-u)^{-1}\tilde b^A(\theta^B).
\end{align}
(Here $u$ is the preferred null coordinate $U$, and $\kappa$ is
$1-\nu$ of Sec.~\ref{section:surfacegravity}). The presence of the
shift vector $b^A$ means that, on $v=0$, the homothetic vector field
$\xi=-u\partial_u$ and the null generators
$\partial_u + b^A\partial_A$ wind around each other a number of times
that diverges with $\ln(-u)$. The ingoing data on $u=0$ are not
explicitly specified, but are not assumed to be compatible with exact
CSS. However, they must not deviate from exact CSS too rapidly away
from the corner.

Solutions are characterised by a a small parameter $\epsilon$ such
that along $v=0$ the ratio of the Hawking mass over the area radius is
$\sim \epsilon^2$. The outgoing shear is $\sim \epsilon^{-1}$. The
Riemann tensor component $R(e_4,e_A,e_4,e_B)$ diverges as
$v^{-1+c\epsilon}$, where $e_4=\Omega^{-1}\partial_{v}$. In this
sense, these solutions are strong even for small $\epsilon$. Indeed,
as $\epsilon\to 0$, $g$ converges to the Minkowski metric in
$H^1_{\rm loc}$ but not in $C^1$.

The metric in coordinates $(u,v)$ is $C^N$ (or potentially $C^\infty$
in a future extension of the theorem), but only $C^{1,c\epsilon^2}$
across the past light cone $v=0$, thus indicating where the naked
singularity will be already in the ingoing null data on $u=-1$.

\citet{Shl22} gives the extension to the past of $v=0$ to a regular
``centre'' at $u=v<0$, with the metric there in $C^N$ in suitable
double-null coordinates. On the future lightcone $(u=0,v>0)$ of the
singularity, the metric is H\"older-continuous in suitable
coordinates. In this sense, these solutions have a naked
singularity. (The $\kappa=0$ solutions of \citet{RodShl17} do not.)
\citet{Shl22a} considers twisted self-similarity more generally and in
particular the relation between Fefferman-Graham formal expansions and
actual solutions.

\subsection{DSS exterior naked singularity solutions in the spherical
Einstein-scalar system}

\citet{CicKeh24} construct approximately DSS 
solutions of the spherical Einstein-scalar system on the null
rectangle bounded by to the past by the similarity horizon $v=0$ and
the regular null surface $u=-1$, and to the future by the Cauchy
horizon $u=0$ and future null infinity $v=\infty$. The DSS singularity
is at $u=v=0$. The initial data are exactly DSS only on $v=0$ but not
on $u=-1$, already in order to make the solution asymptotically flat,
but the authors state that one should be able to construct also
exactly DSS, non-asymptotically flat, solutions. 

The compactness $2m/r\ll 1$ everywhere in the solution, but undergo an
infinite number of oscillations as $u\to 0_-$. This also means that
the surface gravity of the similarity horizon and Cauchy horizon obey
$1-\nu\ll 1$ (see Sec.~\ref{section:surfacegravity}). The assumed
smallness of $2m/r$ means that the (exterior) Choptuik solution is not
one of the solutions constructed. (Indeed, $2m/r$ is small on the
Cauchy horizon of the Choptuik solution, but large on its similarity
horizon.) The authors state that any smooth interior ``fill-in''
solution would have $2m/r$ {\em not} small in the interior.

The authors note the similarity of their construction method to that
of \citet{RodShl19}, where vacuum, non-spherical, asymptotically CSS
solutions are constructed (which are exactly CSS on the similarity
horizon $v=0$), but stress differences: while the spherical scalar DSS
solutions are smooth everywhere except on the Cauchy horizon $u=0$,
and in particular across $v=0$ (similar to the Choptuik solution
of \cite{ReiTru12}), the vacuum CSS solutions of \citet{RodShl19} are
only $C^{1,\delta}$ across $v=0$, similar in that regard to the
spherical scalar CSS solutions of \citet{Christodoulou4}.

\subsection{Surface gravity of the similarity and Cauchy
horizons of CSS and DSS spacetimes}
\label{section:surfacegravity}

\new{

Assume that on a CSS or DSS spacetime we have a preferred worldline
$\Gamma$ and null coordinates $(U,V)$, such that the surfaces of
constant $U$ are future null cones emanating from $\Gamma$, the
surfaces of constant $V$ are past null cones emanating from $\Gamma$,
$U$ and $V$ are normalised to be proper time along $\Gamma$, and the
homothetic vector field in CSS, or a DSS-adapted vector
field in DSS, is given by
\begin{equation}
\label{xiuvnaive}
\xi={\partial\over\partial \tau}=-U{\partial\over\partial
  U}-V{\partial\over\partial V}.
\end{equation}
This is always possible in spherical symmetry, where $\Gamma$ is the
central worldline. $V=0$ is then the similarity horizon and $U=0$ the
Cauchy horizon, and $\xi^a$ is tangential to both. $\Gamma$ is given
by $U=V$. 

However, even if the spacetime itself is regular across the Cauchy
horizon $V=0$, the coordinate $V$ is in general not regular
there. Instead, there is a unique power $\nu\ne 1$ such that
\begin{equation}
\label{nuvdef}
\hat v:=V^{\nu}
\end{equation}
is regular across $\hat v=0$.

We define a similarity coordinate $x$ such that the similarity horizon
is at $x=0$ and $x$ is regular across the horizon. Then to leading
order $x$ must be proportional to $\hat v$, not $V$. If we further
want to define $x$ as the ratio of $\hat v$ to another null
coordinate, then this must be $x=-\hat v/\hat u$, where
\begin{equation}
\hat u:=-(-U)^{\nu}.
\end{equation}
The homothetic or DSS-adapted vector field is now
\begin{equation}
\label{xinuunuv}
\xi=-\nu\left(\hat u{\partial\over\partial\hat u}
+\hat v{\partial\over\partial\hat v}\right).
\end{equation}
These $(\hat u,\hat v)$ are the double null coordinates used at the end
of \citet{Christodoulou4}, in \citet{ReiTru12}, in \citet{RodShl19}
and in \citet{CicKeh24}. However, the introduction of $\hat u$ is not
essential. An alternative choice of $x$ in spherical symmetry, made
for example in most of \citet{Christodoulou4}, is $x=R/U$, where $R$
is the area radius.

To characterise $\nu$ without introducing null coordinates, we recall
that 
\begin{equation}
n^a:=\left.-\nabla^a\hat v\right|_{\hat v=0}
\end{equation}
is the tangent vector to an affinely parameterised family of null
generators of the horizon, that is, it obeys $n_an^a=0$ and
$n^a\nabla_an^b=0$. 

The vector $-\nabla^ax$ is also tangential to the
horizon generators, but non-affinely parameterised. (This is because
$|\nabla\hat v|^2=0$ everywhere, but $|\nabla x|^2=0$ only on the
horizon.) Specifically, with
$x=-\hat v/\hat u$ we have
\begin{equation}
\label{gradxn}
\left.-\nabla^ax\right|_{x=0}=An^a,
\end{equation}
where, from $x=-\hat v/\hat u$, 
\begin{equation}
A:=-{1\over \hat u} = e^{\nu\tau}\tilde A,
\end{equation}
where $\tilde A>0$ is constant (in CSS), or periodic in $\tau$ (in
DSS). Here and in the following a tilde denotes a function that is
periodic in $\tau$ at constant ($x,\theta^a)$.  Hence we have
\begin{subequations}
\label{Ascaling}
\begin{equation}
\label{AscalingCSS}
{\cal L}_\xi A=\nu A \quad \text{(for CSS)},
\end{equation}
and
\begin{equation}
\label{AscalingDSS}
\Phi_*A=e^{\nu\Delta}A \quad \text{(for DSS)}.
\end{equation}
\end{subequations}
Note that in DSS $\xi^a$ and hence $A$ are not unique, and in CSS they
are unique only up to a constant factor, but the scalings
(\ref{Ascaling}) are unique. Hence we can now define $\nu$ from $n^a$
and $x$ only. 

Going further, we now eliminate $x$ from the definition in favour of
$\xi^a$. As by assumption $\xi^a$ is tangential to the horizon, we
decompose it there as
\begin{equation}
\label{xiBb}
\xi^a=Bn^a+b^a \quad \text{on the SH},
\end{equation}
made unique by imposing $b^a\nabla_a\tau=0$. Working in adapted
coordinates $(\tau,x,\theta^A)$,
we find
\begin{align}
B&=-{\tilde A\over \tilde g^{\tau x}}e^{(\nu-2)\tau}, \\
b^A&=-{\tilde g^{\tau A}\over \tilde g^{\tau x}}, \\
b^\tau&=0, \\
b^x&=0.
\end{align}
where the $\tilde g^{\mu\nu}$ are constant (in CSS) or periodic in
$\tau$ (in DSS). Hence $B$ obeys
\begin{subequations}
\label{Bscaling}
\begin{equation}
\label{BscalingCSS}
{\cal L}_\xi B=(\nu-2) B \quad \text{(for CSS)},
\end{equation}
and
\begin{equation}
\label{BscalingDSS}
\Phi_*B=e^{(\nu-2)\Delta}B \quad \text{(for DSS)}.
\end{equation}
\end{subequations}
$B$ and $b^a$ depend on the choice of $\tau$ and $\xi^a$, but the scaling
(\ref{Bscaling}) does not. Hence we can now define $\nu$ from $n^a$
and $\xi^a$ only. 

In this way, we can define the geometric invariants $\nu_\text{CH}$
for the Cauchy horizon and $\nu_\text{SH}$ for the similarity horizon,
either from $\hat u$ and $\hat v$, or from $n^a$ and $x$ (with the
horizon at $x=0$ and $n^a$ the tangent vector of its affinely
parameterised generators), or from $n^a$ and $\xi^a$ (a DSS-adapted
vector field tangent to the horizon).

From (\ref{xiBb}) we read off
\begin{equation}
\xi^b\nabla_b\xi^a={d\ln B\over d\tau}\xi^a+\xi^b\nabla_bb^a.
\end{equation}
We can re-express (\ref{Bscaling}) as $d\ln B/d\tau=\nu-2$ in CSS, or
$\overline{d\ln B/d\tau}=\nu-2$ in DSS, where here and in the
following the overbar denotes the average over one period $\Delta$ in
$\tau$, at constant $(x,\theta^A)$. The constant $\nu-2$ therefore
parameterises the failure of $\xi^a$ to be affinely parameterised, and
$b^a$ parameterises its failure to be geodesic. This suggests an
interpretation of $\nu-2$ as the surface gravity of the similarity
horizon (or Cauchy horizon). We now assemble calculutions of $\nu$
from the literature.

In spherical symmetry (but not in general!) a DSS-adapted vector field
$\xi^a$ that is tangential to the horizon is also tangential to its
null generators. We then have
\begin{equation}
\xi^b\nabla_b\xi^a={d\ln B\over d\tau}\xi^a.
\end{equation}
From this and (\ref{Bscaling}) we then have
\begin{equation}
\nu-2=\overline{{\Gamma^\tau}_{\tau\tau}}
\quad \text{ (spherical symmetry only)}.
\end{equation}
Computing ${\Gamma^\tau}_{\tau\tau}$ in the similarity coordinates
(\ref{x_tau}) based on the polar-radial coordinates (\ref{tr_metric}),
we find
\begin{equation}
\nu-2=-\overline{a^2_0} \quad \text{ (spherical symmetry only)},
\end{equation}
}where $a$ is the metric coefficient defined in (\ref{tr_metric}), and
here and below the suffix 0 denotes either the similarity horizon or
the Cauchy horizon. With
$a^{-2}=1-2m/r$, where $r$ is the area radius and $m$ the Kodama mass,
we can write this as
\begin{equation}
\label{nuspherical}
\nu-1=-\overline{\left({2m_0\over r_0-2m_0}\right)}
\quad \text{ (spherical symmetry only)}.
\end{equation}

Eq.~(\ref{nuspherical}) also shows us that $\nu\le1$, at least in
spherical symmetry. Independently, $\nu>0$ follows from the definition
of $\nu$ in terms of null coordinates, so we have $0<\nu\le 1$.

The quantity $\epsilon:=1-1/\overline{a^2_0}$ is
defined in \citet{critcont}, equivalent to
\begin{equation}
\nu=:1-{\epsilon\over 1-\epsilon}\quad \text{($\epsilon$ of \citep{critcont})}.
\end{equation}
\citep{critcont} also gives the identity
$\overline{a^2_0}=1+2\overline{V^2_0}$, where
$V_0=\sqrt{2\pi G}\phi_{,\tau}$ on the horizon. Hence in the spherical
scalar field system we have 
\begin{equation}
\label{sphericalscalarnu}
\nu=1-4\pi G\overline{\phi_{,\tau}^2}
\quad \text{ (spher. symm. scalar field only)},
\end{equation}

The Choptuik solution is not analytic at its {\em Cauchy} horizon, and
so there is no preferred null coordinate $\hat u$ there, but $m/r$ is
continuous there. The value of $\epsilon$ on the Cauchy horizon is
given to 8 significant digits in
\citet{critcont}. This translates into
\begin{equation}
\nu_\text{CH,Choptuik}\simeq 1-1.4710461(4) \cdot 10^{-6}.
\end{equation}

Using (\ref{sphericalscalarnu}), the spherical CSS ansatz of
(\ref{Christodoulouansatz}) for the Einstein-scalar system gives
\begin{equation}
\nu_\text{SH,Christodoulou}=1-k^2.
\end{equation}
(This agrees with Eq.~(4.13) of \cite{Christodoulou4}.)

\citet{ReiTru12} parameterise
$\nu=4\pi\mu/\Delta$ for the DSS Choptuik solution, where
$\Delta\simeq 3.44$ is the scale-echoing period. They find values of
$\mu$ and $\Delta$ to 80 significant digits which imply
\begin{equation}
\nu_\text{SH,Choptuik}\simeq 0.6138...
\end{equation}

In their study of vacuum CSS solutions, \citet{RodShl19} define the
parameter $\kappa$ as, in our notation,
\begin{equation}
\nu_\text{SH,RodS-R}=1-2\kappa_\text{RodS-R}.
\end{equation}
In their study of spherical scalar DSS solutions, \citet{CicKeh24} also
introduce a parameter $\kappa$ as
\begin{equation}
\nu_\text{SH,CicKeh}=1-\kappa_\text{CicKeh}.
\end{equation}

\subsection{Comparison of naked-singularity solutions}

The vacuum CSS naked singularity solutions of \citet{RodShl19} have
many features in common both with the CSS naked singularity solutions
of \citet{Christodoulou4} for the spherical Einstein scalar system,
and the Choptuik solution \citep{Choptuik92,critcont,ReiTru12}. All
have the same spacetime structure: the singularity is a point. In all,
the centre is analytic. In both the scalar and vacuum CSS
solutions, the similarity horizon is only H\"older-continuous, while
in the DSS scalar solution it is analytic.

One might think that a DSS solution can be more regular than a CSS
solution because the solution space is infinitely larger, and this can
be made explicit in terms of shooting between a regular centre and a
regular similarity horizon. But the proof of \citet{BizWas10} that the
critical solution for an Einstein-wavemap system is analytic, even
though it is only CSS, makes that intuition questionable.

However, we know numerically that the {\em Cauchy} horizon of the
Choptuik solution is only H\"older-continuous \citep{critcont}, in a
very similar way to the similarity horizons of the CSS
solutions. Considerations of the available solution space then suggest
to us the following conjecture: Any naked singularity solution, in
vacuum or with matter whose characteristics are light cones,
regardless of symmetry, that is regular everywhere to the past of the
CH cannot also be regular through the CH. However, for spherical CSS
solutions with perfect fluid matter, for the purpose of imposing
regularity the roles of the similarity and Cauchy horizons in the CSS
ansatz are played by the past and future {\em sound} cones of the
singularity, which are timelike. In such a solution the Cauchy horizon
precedes the future sound cone of the singularity, and there is no
reason why it should not be regular.

The unstable perturbations of the Christodoulou solutions are only
$C^{1,\delta}$ across the SH, like the solutions themselves, which are
$C^{1,\epsilon}$ there. By contrast, the single unstable perturbation
of the Choptuik solution is analytic, like the solution
itself. Perturbations of the Choptuik solution with sufficiently low
H\"older continuity across its SH will also be unstable
\citep{Christodoulou5,Sin24} but these would not form from smooth
initial data, whereas the analytic unstable mode does.

There is another difference between the CSS vacuum naked singularity
solutions of \citet{RodShl19}, and any hypothetical DSS vacuum, or
Einstein-scalar, naked singularity solutions.  In the most general DSS
Einstein-scalar solution, the shift and a scalar field in similarity
coordinates take the form
\begin{align}
b^A&=e^\tau\,\tilde b^A(\tau,x,\theta^B), \\
\phi&=\tilde\phi(\tau,x,\theta^B)+k\tau,
\end{align}
where in DSS $\tilde b^A$ and $\tilde\phi$ are periodic in $\tau$ with
period $\Delta$, and $k$ is uniquely defined as the $\tau$-average of
$\phi_{,\tau}$. $\kappa>0$ in CSS requires at least one of
$\tilde b^A(x,\theta^B)\ne 0$ or $k\ne 0$. By contrast, the spherical
DSS Choptuik solution has both $b^A=0$ (it is spherically symmetric)
and $k=0$, which is possible because $\tilde\phi(\tau,x)\ne 0$.

Similarly, there may be DSS vacuum solutions in which
$\tilde b^A[\tau,x,\theta^B]\ne 0$ but its average vanishes, so that a
DSS-adapted vector can be chosen parallel to the horizon
generators. In particular, in twist-free axisymmetry
$\tilde b^\varphi$ vanishes and $\tilde b^\theta$ has vanishing
average, so there are no CSS naked singularity solutions, but
(potentially) DSS ones, with the DSS-adapted vector field chosen
parallel to the horizon generators, are possible. Indeed approximately
DSS twist-free axisymmetric vacuum solutions appear to have been found
numerically at the threshold of collapse, see
Sec.~\ref{section:axi_vacuum}.

\subsection{The extremal critical collapse conjecture}
\label{section:extremalcriticalcollapse}

Previously known critical solutions, both type~I (static, stationary
or time-periodic) and type~II (CSS or DSS) are highly compact but have
no trapped surfaces. In two influential papers, Kehle and Unger have
noticed that in an extremal Kerr-Newman {\em exact} solution the
horizon is {\em marginally} outer-trapped but there are no fully
trapped surfaces. It may be possible that regular initial data at the
threshold of black hole formation settle down to an extremal black
hole which has the same property, even though it is created in
collapse. If so, it may further be possible that such solutions act as
critical solutions. (In hindsight, there may be a precedent for a
critical solution being marginally trapped, namely in spherical
massless scalar field collapse in 2+1 dimensions, see
\citet{JalGun15}.)

\citet{KehleUnger24} have proved (Thm.~1) the existence of
one-parameter families of solutions of the spherical
Einstein-Maxwell-charged Vlasov system (from now, sEMcV) with the
following properties: All solutions in the family are geodesically
complete in the past, and in particular admit regular one-ended Cauchy
surfaces at arbitrarily early times.  The solution with $\lambda=0$ is
Minkowski, solutions with $0<\lambda<\lambda_*$ disperse, the solution
with $\lambda=\lambda_*$ forms an extremal Reissner Nordstr\"om (from
now, eRN) black hole with no trapped surfaces anywhere in the
spacetime, and solutions with $\lambda>\lambda_*$ form a sub-extremal
Reissner-Nordst\"om (RN) black hole. In other words, these families
interpolate between dispersion and collapse, with the threshold
solution settling down to an eRN black hole. They call this ``extremal
critical collapse'' (Def.~1.5).

They conjecture that the exterior of eRN is codimension-one
nonlinearly orbitally stable, that is, the exterior of a critical
black hole formed from collapse settles down to eRN for a
codimension-one set of data (Conj.~1 for massless and Conj.~2 for
massive particles). The conjectures are complicated by the Aretakis
instability of extremal RN, and the possibility of Vlasov ``hair''
remaining outside the horizon.

In this context, see also~\citep{DafermosHolzegelRodnianskiTaylor21}
for a proof of nonlinear stability of the Schwarzschild exterior and
the conjecture that eRN is codimension-four stable. (Fixing the
angular momentum to remain zero and the charge/mass ratio to remain
one imposes four conditions on the initial data). Less is known about
the stability of extremal Kerr. The two closest results are
\citet{ShlTei20,ShlTei23} for a proof of the stability of the
Teukolsky equation on Kerr with any $|a|<M$, and \citet{Tei20} for
mode stability of the Teukolsky equation on $|a|=M$.

\citet{KehleUnger24} further conjecture (Conj.~3) that extremal
critical collapse is stable in the sEMcV system, in the sense that the
black hole threshold in solution space contains an open set of black
holes that settle down to eRN but do not contain trapped surfaces. The
black-hole region of solution space is then foliated by hypersurfaces
of constant final values of $|e|/M$, and the hypersurface $|e|/M=1$ is
locally identical to the collapse threshold. The topology on the space
of solutions is left open. The interpolating families of Thm.~1 would
then be special only in that the threshold solutions are isometric to
exact eRN outside the event horizon after finite advanced time, while
generic threshold solutions asymptote to it.

Crucially, if Conj.~3 holds the black hole threshold solution along
any interpolating one-parameter family of solutions not only has an
extremal RN exterior, but does not contain trapped surfaces anywhere
in the interior. The absence of trapped surfaces in a neighbourhood of
the event horizon of any eRN would be implied by the conjectured
stability, as any proof of exterior stability would have to control
also a region inside it. However the absence of trapped surfaces
everywhere in the interior is definitely not true for all eRN black
holes formed in collapse, see below. The argument that it is
nevertheless true for eRN black holes at the black hole threshold
appears to be as follows, and appears to us to be tied to spherical
symmetry.

First, in sEMcV and other spherical Einstein-matter systems, if there
are no spherically symmetric trapped surfaces there are none at all,
see Prop.~B.1 of~\citep{KehleUnger22}. A spherically symmetric
two-surface is trapped if and only if $R_{,v}<0$, where $R$ is the
area radius of the symmetry two-spheres. The Raychaudhuri equation
implies that if $R_{,v}$ is negative at a point $(u,v)$, it remains so
at all larger $v$. Consider now the graph of the function
\begin{equation}
f:=1-{2m\over R}=g^{uv}R_{,u}R_{,v}\propto R_{,v},
\end{equation}
where $m$ is the Kodama mass, $R$ is the area radius, and $u$ and $v$
are null coordinates, against $u$ at constant $v$. (We assume $f$ is
at least $C^1$). In exact subcritical RN, $f<0$ between the inner and
outer horizon, with a trapped region in between, but in exact eRN, $f$
only touches zero from above. In a generic spacetime that settles down
to eRN, generically $f$ may then become negative again further inside
the horizon, but this is not the case in the threshold solutions
constructed in Thm.~1. By continuous dependence of the metric, and in
particular $f$, on the initial data, it can therefore also not be the
case in an open neighbourhood in solution space. (By contrast, because
an apparent horizon is the outermost level set $f=0$, it does {\em
  not} in general depend continuously on the function $f$, and the
same holds for the event horizon.)

That, within the (codimension-one) set of black holes that settle down
to eRN, there is an open set of eRN black holes that {\em do} contain
trapped surfaces in their interior is illustrated by an earlier
paper~\citep{KehleUnger22} that constructs solutions for the spherical
Einstein-Maxwell-charged scalar field (from now on, sEMcSF) system by
characteristic gluing techniques. The main thrust of that paper is
actually to provide counterexamples to the third law of black-hole
thermodynamics, by constructing $C^k$ solutions (for arbitrary $k$)
which evolve from a regular one-sided Cauchy surface, settle down to
subcritical RN for advanced time $v_1<v<v_2$, and then are charged up
to extremality, settling down again to exact extreme RN for
$v>v_3$. In this case, there is a trapped region in $u\ge
u_\text{AH}$, where $u=u_\text{AH}$ is the apparent horizon during the
exact subextremal RN phase, while the event horizon (the apparent
horizon of the final extremal RN phase) is at some
$u=u_\text{EH}<u_\text{AH}$.

Going beyond sEMcV, Kehle and Unger conjecture that ``extremal
critical collapse'' occurs also in sEMcSF (Conj.~4) and in vacuum
(Conj.~5). In contrast to Conj.~3 for the sEMcV these conjectures do
not explicitly include stability (local genericity).

As a matter of terminology, note that in~\citep{KehleUnger24}
``critical solution'' denotes the threshold solution of a
one-parameter family, not an attractor in the black-hole threshold in
the space of solutions (such as the Choptuik solution, or as the eRN
solution is conjectured to be). Similarly, ``extremal critical collapse'' is
defined as the existence of threshold solutions that asymptote to an
extremal black hole, with stability (local genericity) not part of the
definition.

As already noted in \citep{KehleUnger22,KehleUnger24}, the formation
of extremal RN black holes in collapse is unlikely unless the local
charge to mass ratio can be large. \citet{Rea24} has given a proof of
the third law with matter that obeys a local charge/mass relationship
\begin{equation}
T_{00}\ge \sqrt{\sum_iT_{0i}^2+J_0^2+\tilde J_0^2}
\end{equation}
with respect to an orthonormal 3+1 basis, where $T_{ab}$ is the
stress-energy tensor and $J^a$ and $\tilde J^a$ are the electric and
magnetic charge density. The proof uses the fact that the eternal
extremal RN solution is a classical solution of supergravity with
vanishing gravitino, that is, a solution of Einstein-matter that
admits a Maxwell-covariantly constant Dirac spinor. If an extremal RN
black hole forms from regular initial data with charged matter in
finite time, the resulting spacetime must admit a {\em regular}
spacelike 3-surface $\Sigma$ bounded by a cross-section $S$ of the
extremal RN horizon. $S$ is ``supersymmetric'' in the sense that the
entire spacetime metric on $S$ is isometric to a cross-section $\tilde
S$ of the horizon of an eternal extremal RN. It is then shown that if
$\Sigma$ with $\partial\Sigma=S$ is smooth, $S$ cannot have electric
or magnetic charge.

In another related result \citet{AngKehUng24} have shown the nonlinear
stability of extremal RN black holes as solutions of the spherically
symmetric, Einstein-Maxwell-{\em neutral} scalar field model. In other
words, the (nonlinear) perturbations considered are spherical and do
not carry charge. As noted in \cite{KehleUnger24}, the extremal
critical collapse conjecture in spherical symmetry with charge assumes
such a stability result (but with the perturbations carrying charge).

}

\begin{acknowledgments}

  For the 2007 version, CG and JMM would like to thank David Garfinkle
  for a critical reading of the manuscript and the Louisiana State
  University for hospitality while this work was begun. JMM was
  supported by the I3P framework of CSIC and the European Social Fund,
  and by the Spanish MEC Project FIS2005-05736-C03-02.

  For the 2010 update of the 2007 version, JMM was supported by the
  French A.N.R.\ (Agence Nationale de la Recherche) through Grant
  BLAN07-1\_201699 entitled ``LISA Science'', and CG was supported by
  the A.N.R.\ Grant 06-2-134423 entitled ``Mathematical Methods in
  General Relativity''. CG would also like to thank the IAP and LUTH
  for hospitality.

  \new{For the 2025 version, CG and DH are grateful to Jo\~ao Costa,
    Mihalis Dafermos, Eugene Trubowitz, Michael Reiterer, Igor
    Rodnianski, Yakov Shlapentokh-Rothmann, Christoph Kehle, Ryan
    Unger, Irfan Glogic, Birgit Sch\"orkhuber, Roland Donninger, Mahir
    Had\v zi\'c, Matthew Schrecker, Harvey Reall and Piotr Biz\'on for
    helpful conversations about the content of
    Secs.~\ref{section:PDEblowup} and \ref{section:mathematicalGR}. CG
    was supported by the ``GWverse'' European Union COST action for
    visits to Lisbon. CG and DH are grateful to the Oberwolfach
    Research Fellows programme for support in 2019, 2022 and 2025. DH was
    supported by the FCT (Portugal) IF Program IF/00577/2015,
    PTDC/MAT-APL/30043/2017, FCT (Portugal) projects UIDB/00099/2020
    and UIDP/00099/2020 and PeX-FCT (Portugal) program
    2023.12549.PEX.}

\end{acknowledgments}

\bibliography{critphen_lrr_revtex.bbl}
      
\end{document}